\documentclass[11pt]{article}
\usepackage[utf8]{inputenc}
\usepackage[american]{babel}
\usepackage[style=apa,natbib=true,doi=false,isbn=false,url=false,citestyle=authoryear,maxnames=3, minnames=1,uniquelist=false,date=year,alldates=year]{biblatex}
\DeclareDelimFormat*{finalnamedelim}
  {\ifnum\value{liststop}>2 \finalandcomma\fi\addspace\bibstring{and}\space}

\DeclareDelimFormat[bib,biblist]{finalnamedelim}{%
  \ifthenelse{\value{listcount}>\maxprtauth}
    {}
    {\ifthenelse{\value{liststop}>2}
      {\finalandcomma\addspace\bibstring{and}\space}
      {\addspace\bibstring{and}\space}}}

\DeclareDelimFormat*{finalnamedelim:apa:family-given}{%
  \ifthenelse{\value{listcount}>\maxprtauth}
    {}
    {\finalandcomma\addspace\bibstring{and}\space}}

\usepackage{csquotes}
\DeclareLanguageMapping{american}{american-apa}
\usepackage{physics}
\usepackage{graphicx}
\usepackage[hang]{subfigure}
\usepackage{bbm}
\usepackage{blindtext}
\usepackage{enumitem}
\usepackage[letterpaper, top=1.25in, bottom=1.25in, margin=1.25in]{geometry}
\usepackage{caption}
\usepackage{comment}
\usepackage{float}
\usepackage{booktabs}
\usepackage[usenames, dvipsnames]{color}
\usepackage{amsmath}
\usepackage{xcolor}
\usepackage{changepage}
\usepackage{pdflscape}
\usepackage[normalem]{ulem}
\usepackage{soul}
\usepackage{titlesec}
\usepackage[hang,flushmargin]{footmisc}
\titlespacing\section{0pt}{12pt plus 4pt minus 2pt}{0pt plus 2pt minus 2pt}
\titlespacing\subsection{0pt}{12pt plus 4pt minus 2pt}{0pt plus 2pt minus 2pt}

\usepackage{setspace} 
\usepackage[justification=centering]{caption}
\usepackage{lscape}
\usepackage[flushleft]{threeparttable}
\usepackage{float}
\usepackage{hyperref}
\usepackage{afterpage}
\usepackage{mdframed}
\usepackage{comment}
\usepackage{amssymb}
\usepackage{amsthm}
\usepackage{bbm}
\usepackage{datetime}
\usepackage{bbold}

\newdateformat{monthyeardate}{%
  \monthname[\THEMONTH] \THEYEAR}

\makeatletter
\g@addto@macro\appendix{\setcounter{figure}{0}}
\makeatother
\makeatletter
\g@addto@macro\appendix{\setcounter{table}{0}}
\makeatother
\makeatletter
\def\blfootnote{\xdef\@thefnmark{}\@footnotetext}
\makeatother

\DeclareCaptionLabelFormat{table}{TABLE~#2}
\DeclareCaptionLabelFormat{figure}{FIGURE~#2}
\title{\vspace{-35pt}\Large{\sc{Can the Black Lives Matter Movement Reduce Racial Disparities? Evidence from Medical Crowdfunding}}}

\author{\vspace{-15pt}Kaixin Liu$^{\dagger\diamond}$  \quad Jiwei Zhou$^{\ddagger\diamond}$ \quad Junda Wang$^\pm$}

\date{\vspace{-15pt} \monthyeardate\today }

\begin{document}

\maketitle
\addtocounter{page}{0}%
\vspace{-30pt}
\begin{abstract}
\noindent  Using high-frequency donation records from a major medical crowdfunding site and careful difference-in-difference analysis, we demonstrate that the 2020 BLM surge decreased the fundraising gap between Black and non-Black beneficiaries by around 50\%. The reduction is largely attributed to non-Black donors. Those beneficiaries in counties with moderate BLM activities were most impacted. We construct innovative instrumental variable approaches that utilize weekends and rainfall to identify the global and local effects of BLM protests. Results suggest a broad social movement has a greater influence on charitable-giving behavior than a local event. Social media significantly magnifies the impact of protests.

\noindent \textbf{Keywords:} Social Movement; Black Lives Matter; Racial Disparity; Medical Crowdfunding; 
\newline 
\noindent\textbf{JEL Classification Numbers:} D74; I14; J15; P32

\end{abstract}
\vfill
\blfootnote{\scriptsize{We wish to thank David Autor, Hanmin Fang, Shuaizhang Feng, Elaine Hill, Zibin Huang, Josh Kinsler, Narayana Kocherlakota, Cong Liu, David Ong, Ande Shen, John Singleton, Yalun Su, Kegon Tan and Ne\c{s}e Yildiz for their suggestions for this paper. We are especially thankful to our advisors Travis Baseler, Lisa Kahn, and Ronni Pavan for their fantastic guidance on this project. We especially thank Junda Wang for his generous help in data access and technical support.\vspace{10pt} \\
$^\diamond$ These authors contributed equally. 
\begin{flushleft} 
$\dagger$ Institute for Economic and Social Research, Jinan University, email: kaixinliu@jnu.edu.cn or liukaixin888@gmail.com \\
$^\ddagger$ Department of Economics, University of Rochester, email: jiwei.zhou@rochester.edu \\
$^\pm$ Department of Computer Science, UMass Amherst, email: jundawang@umass.edu\\
\end{flushleft}
}
}
\thispagestyle{empty}

\newpage
\setcounter{page}{1}

\section{Introduction}

Beyond the ideological support and political reform it garners, what tangible impact does a social movement bring about to the economic racial disparity? Moreover, how can we interpret the nature and the process of this impact: Which demographics does the movement primarily influence given the polarization of public opinion? In an era of extensive social media and widespread news coverage, what is the scale of local gathering effects when compared to the overall global influences of the movement? These questions are important to answer but missing from the literature.

This paper endeavors to illuminate these critical questions by analyzing the impact of the ``Black Lives Matter" (BLM) movement. The BLM movement, founded in 2013, gained unprecedented national attention following the murder of George Floyd in May 2020, which triggered wide-reaching discussions on racial injustice. From June to August 2020, over 450 major protests were held across the United States, with an estimated 15-21 million participants. Recent studies suggest that this movement prompt changes in public empathy, emotional support, and some anti-racism actions towards Black people (\cite{reny2021opinion}; \cite{campbell_black_2021}; \cite{agarwal_antiracist_2022}; \cite{engist2022_BLM_local_voteregistra}). Yet, it remains unclear whether or how this movement contributes to an actual improvement of Black people's economic conditions or to a vast reduction of racial inequality.

Leveraging high-frequency donation records from an online medical crowdfunding platform, this paper provides innovative research designs and insights into these questions. Specifically, we focus on the impact of the BLM movement on charitable donations, and examine how the movement alters an individual’s propensity to donate to the medical crowdfunding projects of Black beneficiaries.

First, we show that the surge of the BLM movement leads to an increase in donations to projects organized by Black individuals, and that this rise is not due to a rising demand for financial support from Black individuals. Second, we demonstrate that it is non-Black rather than Black donors, who primarily contribute to this donation surge. Third, utilizing geographical variations in the BLM protests, we observe that residents in areas with a moderate number of gatherings are the most responsive to the movement. We present evidence showing that the effect of massive protests spill over across geographics, and affecting every corner of the U.S. Finally, to quantify such a spillover channel that significantly amplifies the effect of protests,    we develop two novel instrumental variable methods to identify the causal effects of the national protests and the local protests. Using holidays and weekends as an instrumental variable, we estimate the global causal effect of protest intensity on reducing the racial donation gap. Using the daily rainfall of each county as an instrument, we identify the effect of local protests  on donation giving. By comparing the relevance  of local rallies and the broader national movement, we highlight the substantial role that social media and news coverage play in shaping the impact of social movements. 

We make the following key empirical findings. First, the BLM movement reduces the racial disparity in the final raised funds by about half, from an initial baseline gap of 20\% down to 10\%. Second, we exploit daily-level donation records, and find that following the beginning of BLM, the daily number of donors to Black beneficiaries increases by 10 percent (from 4.0 to 4.4), which is the main driver behind the reduction in the final sum of money raised. The effect on donor numbers is found to fade away three months after the murder of George Floyd. Third, an analysis of detailed donor name records reveals that 95\% of the extra donations to Black beneficiaries come from non-Black donors. Lastly, our weekend IV approach shows that an additional 100 protests occurring in the U.S. can reduce the racial gap in the daily number of donors by 0.27 heads, amounting to 30 percent of the racial disparity in raised funds. The rainfall IV approach shows that one local rally can lead to a reduction of 0.8 heads for projects launched in that county. Since a local protest only happens with a probability of 0.03 in a day and there are 101 national rallies daily, the impact of local rallies is only one-tenth as influential as the global impact of BLM protests altogether. Social media and general media (e.g., Facebook, Google search, and newspapers) coverage on racial disparity play a critical role in disseminating the influence of the local rallies  beyond their counties of origin.

High frequency medical crowdfunding records provide us with a rare set-up to fulfill our goal for its unique features. First, crowdfunding for medical issues is purely philanthropic and reflects donors’ altruistic concern for the adverse health conditions of others. Some categories of crowdfunding operate more like microfinance tools, where fundraisers must repay their funds.\footnote{Thus, the gap in donations between Black individuals and non-Black individuals reflects a lack of empathy towards Black people, rather than a biased opinion over the ability of Black people to earn money or their tendency to default.}   Second, access to detailed information regarding crowdfunding posting and donation giving allows us to separate the demand behavior from the supply behavior. 

Crucially, these high-frequency donation records allow us to implement research designs that precisely identify the causal role of the BLM movements on the donations to Black people and to discuss the underlying mechanisms. The simultaneous occurrence of the Covid-19 pandemic and associated policy responses could confound the identification for all BLM-related research. Solving this confounding problem would not be feasible with monthly or even weekly data, since the temporal proximity (less than 7 days) of these potential confounding factors to May 27, 2020 is close. Moreover, the construction of weekend instruments and rainfall instruments to understand the nature of protest gatherings also requires daily-level measurements of the outcome variables.   Our data overcome all of these challenges to answering our questions.  

This paper contributes to three strands of existing literature. 

First, our paper expands on the understanding of the economic impact of social movements. Social movements strive to alter unsatisfactory conditions in society by means of achieving particular societal or political objectives. Understanding how these movements reshape society has been a long-standing goal among social scientists. However, previous research has primarily focused on political or ideological outcomes, and has failed to explore the influence of social movements on the economic behavior of individuals.\footnote{Chapters and books in social science such as \cite{snow_cultural_2018}; \cite{andrews_legitimacy_2016}; \cite{giugni_how_1999}; \cite{bosi_consequences_2015} confirm the literature gap and highlight the future agenda for this field of study. Until very recently, works by economists, such as \cite{levy_effects_2019} and \cite{luo_scandal_2022} start to understand the effect of social movements on aspects other than political outcomes.}  Some existing research (\cite{reny2021opinion}; \cite{campbell_black_2021}; \cite{agarwal_antiracist_2022}; \cite{engist2022_BLM_local_voteregistra}) provides insights that are relevant to our area of study on the BLM movement. However, existing research largely focuses on the political and ideological outcomes of BLM movements such as, police use of lethal force, demand for books with anti-racism themes in schools, public opinion towards Black communities, and voter registration. Our study stands as pioneering work that demonstrates how BLM movements can not only enhance people’s awareness of equality and racial preferences, but importantly yield tangible economic benefits, which alleviate adverse health, insurance, or financial conditions for Black individuals. 

Second, our paper contributes to existing literature on the causal identification of the effect of social movements, and to the methods to study the mechanism through which these movements achieve impact. Methodologically, establishing a causal link between social movements and their subsequent outcomes presents a significant challenge, and so forms a substantial part of future research agendas in social movement literature(\cite{giugni_how_1999}; \cite{bosi_consequences_2015}). We rise to this challenge by advocating for the use of high-frequency panel data to establish a persuasive causal effect of social movements. By leveraging daily-level outcomes, we effectively distinguish the BLM’s effect from potential confounders such as Covid-19 policy shocks.\footnote{The identification strategy used in previous studies on the BLM movement (\cite{campbell_black_2021};\cite{agarwal_antiracist_2022})) fail to establish the causal role of the George Floyd protests due to insufficient temporal variations to distinguish them from other possible confounders (e.g.., Covid policies). \cite{campbell_black_2021} uses weekly level reports; \cite{agarwal_antiracist_2022}) only uses monthly level data; most other literature uses either monthly level data or yearly level data.}  Therefore, we emphasize the need for the gathering of more detailed high-frequency data in future research in order to enhance our understanding of the causal relationships between social movements and their outcomes. Furthermore, we design a novel instrument approach  - a weekend instrument that exploits the spike in the protests during the weekend - to recover the causal impact of global protests on economic outcomes. This approach, join with the rainfall instrument by \cite{madestam2013qje_teaparty_gather} to recover the local impact of a protest assembly, provides a direct way to understand the relation between a massive social movement and the local assemblies. Further, our investigation of how social medias interact with this relation offers a novel way of identifying how social media spillovers and amplifies the effect of local assembly.

Lastly, our paper contributes to the literature on racial disparities in health care and health insurance. Studies show that the health status and life expectancy of Black individuals are generally poorer compared to other racial groups (\cite{yearby_racial_2018}; \cite{williams_racial_2015}; \cite{carratala_health_2020}; \cite{orsi_blackwhite_2010}). Many studies (\cite{lillie-blanton_role_2005}; \cite{nelson_unequal_2002}; \cite{buchmueller_effect_2016}) have identified the underlying cause of this issue as the disproportionately limited access that Black individuals have to health care and health insurance.\footnote{In 2013, 25.8\% of Black individuals were uninsured. Black individuals are 1.5 times less likely than white individuals to be covered by any health insurance.} We extend on this broad literature by first uncovering a novel source of health disparity that is prevalent and resilient on the medical crowdfunding platform, a type of informal health insurance and social safety net.\footnote{Related works include \cite{younkin_colorblind_2018} and \cite{jenq_beauty_2015}. Much of the literature here consists of opinions, commentaries, and perspective pieces.} We find that there is a fundraising gap of approximately 20\% between Black and non-Black beneficiaries, even after we adequately control for observed characteristics of crowdfunding posting information. This constitutes a significant source of healthcare disparity, implying the importance of introducing future initiatives to address this problem.

The structure of the paper is as follows: Section \ref{sec_Background} provides background information and data on the 2020 surge of the BLM movement, and the landscape of medical crowdfunding; Section \ref{sec_GapDecrease} delivers a descriptive analysis of the racial fundraising gap, focusing specifically on the time frames before and after the emergence of the BLM movement; Section \ref{sec_Identification} exploits daily-level donation records to assess the causal impact of the BLM movement on mitigating the fundraising gap; Section \ref{sec_race_decompose} scrutinizes the racial composition of the donors who were mobilized by the BLM movement; Section 6 explores the geographical distributional effect of the BLM movement, highlighting the non-exclusive role of local protest rallies; Section \ref{sec_global_local} presents the weekend IV approach and the rainfall IV approach to estimate the local effect of protest rallies and the global effect of protests;  Section \ref{sec_Conclusion} concludes the paper. We present additional discussions in Online Appendix \ref{sec:online_app_1}, and additional Figures and Tables in Online Appendix \ref{sec:online_app_2}.

\section{Background and Data}\label{sec_Background}
\subsection{Black Lives Matter movement and George Floyd protest}
The Black Lives Matter (BLM) movement is rooted in the long history of social activism and the struggle against racial injustice in the U.S. The movement began in earnest in 2013 after the acquittal of George Zimmerman, the neighborhood watch volunteer who fatally shot Trayvon Martin, an unarmed African-American teenager. The hashtag \#\textit{BlackLivesMatter} was created in response to the verdict, and quickly became a rallying cry for people protesting against systemic racism and police brutality against Black individuals. Over the years, the BLM gained prominence as a decentralized movement addressing a wide range of issues related to racial inequality.

On May 25, 2020, in Minneapolis, Minnesota, George Floyd, a 46-year-old Black man, was killed during an arrest when a police officer knelt on his neck for over nine minutes. Floyd repeatedly told the officers that he could not breathe, but his pleas were ignored. The incident was captured on video by a bystander and quickly went viral on social media, sparking widespread outrage and protests. 

While the protests were sparked by the specific incident of George Floyd’s murder, they share a common goal with the BLM movement: the addressing of racial injustice and police brutality against Black individuals. During the George Floyd protests, many protesters chanted ``Black Lives Matter” and carried signs with BLM slogans, showing solidarity with the movement. These protests helped amplify the BLM movement’s message and contributed to an increased global awareness of its goals. In summary, though the George Floyd protests were a distinct series of events, they formed part of the larger BLM movement due to their shared focus on racial justice and police reform. In this paper, we use the BLM movement or  George Floyd protests to represent the surge of the BLM following the murder of George Floyd. 

More importantly, the Floyd protests represent a more pivotal point in the BLM campaign in the U.S. than any preceding events. The protests soon became the largest social movement in the history of the U.S. since the 1960s. An estimated 15 to 26 million people participated in the 2020 BLM protests.\footnote{Source: \url{https://www.nytimes.com/interactive/2020/07/03/us/george-floyd-protests-crowd-size.html}} Further, the protests attracted a wide range of participants, including people from diverse racial, ethnic, and socio-economic backgrounds. 

\begin{figure}[htb]
    \centering
    \caption{George Floyd/BLM: Protest, Search Behavior and Media Coverage}
    \includegraphics[width = \textwidth]{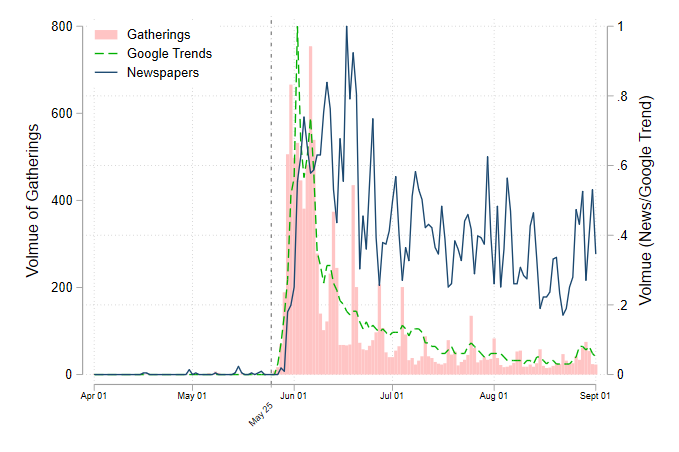}
    
    \label{fig:protests_news_google}
    \vspace{1em}
    \begin{minipage}[0.1cm]{1\textwidth}
    \small \textit{Note:} Figure \ref{fig:protests_news_google} presents the time pattern of the national wide Floyd/BLM related protest gatherings, google search behavior and news coverage. y-axis on the left side is the number of protests at each date. y-axis on the right side is the normalized scale of: (i) Google search trends for related keywords of George Floyd or BLM; (ii) newspaper trends for these two topics. As one can observe, before the death of George Floyd, the scale of both the protest and public attention are negligible. After May 27, 2020 the protest and public attention surges significantly. As one can observe, the peak of the both the protest and public's attention appear in June to July. Though the protests and search/media coverage are significant increased comparing to the pre-Floyd period, their scale also shrink largely comparing to the peak during the June and July. 
    \end{minipage}
\end{figure}

This surge in the number of protests and enlarged public attention to the BLM is reflected through multiple data resources. Figure \ref{fig:protests_news_google} shows the temporal pattern of the protests’ intensity, media coverage, and Google search trends around the time of George Floyd’s death. 

We use the Crowd Counting Consortium\footnote{See webpage \url{https://github.com/nonviolent-action-lab/crowd-counting-consortium}.} to collect the protest data. The data set contains records on political crowds reported in the U.S. obtained from newspapers and Twitter, including the date, claim, and location of the protest/demonstration. We select BLM-related records. As shown in Figure \ref{fig:protests_news_google}, there is a small number of BLM protests before the circulation of the video, but immediately after May 25, 2020, the protests surge at the national level. 

Data on public attention shows a consistent pattern. We use the Google search index and media coverage on BLM as the metrics of public attention. Prior to the death of George Floyd and subsequent protests, the BLM did not gain much public attention from social media. Figure \ref{fig:protests_news_google}  shows that public attention to the BLM surges after May 28, 2020, the day following the Floyd incident on May 27. Prior to May 28, there were barely any Google searches for “BLM” or newspaper reports related to BLM. Thus, we expect the full treatment effect of the BLM movement to be initiated and reflected in the data after May 28, which we choose as our cut point later in our research design.

\subsection{Medical Crowdfunding Records}
We obtain access to the medical crowdfunding project records on GoFundMe. It is the largest crowdfunding platform for personal fundraising, especially for medical fundraising. More than \$15 billion have been raised since its launch in 2010. 

A typical medical crowdfunding project on GoFundMe adheres to the following process. First, the fundraiser must set up a project on the platform (crowdfunding posting). The donors then have access to the project title, text description, profile photos of the beneficiary(es), and the goal(target amount to be raised). Figure  \ref{fig_eg} shows an example of what a donor will see when accessing a medical crowdfunding project on GoFundMe. After seeing this information, the donors then make their decision on whether to give money and if so, how much they wish to donate. 

As econometricians, we observe the following information for each fundraising project: title; start date of the project; text description (usually a paragraph describing why they need the money); beneficiary(es)’s name; zip code where the beneficiary(es) live; photo of the beneficiary(es); goal(\$); amount of money(\$) so far raised; total number of donations(\#) so far received. 

More importantly, under each project, we have visibility into each donation record, capturing up to the most recent 100 donations. The information on a donation record includes the donor’s name, the amount of the donation, and the timing of the donation on the day. These records enable us to track the donation process on a high-frequency basis and identify the impact of medical crowdfunding in a granular time setting. 

For projects that have received more than 100 donations throughout their life-cycle, although we lose track of the earlier donations beyond the most recent 100 , we can impute the average daily flow for those records we know its launch date and the total number of donors giving money to the project. For projects with more than 100 donations, the imputation is to allocate the invisible donation to dates between the project launch date and the date when the last 100th donation appears.\footnote{As an illustrative example, consider a project that was launched on May 29, 2020, and has received a total of 150 donations. The most recent donation was made on July 1, 2020, and the 100th most recent donation was made on June 20, 2020. Our observed data consists of donations made between June 20 and July 1, capturing the latest 100 donations. To complete the dataset, we would then need to allocate the remaining 50 earlier donations to the time period spanning from May 29 to June 19.}  We assume that the weight used to allocate the unobserved donation records follows the donation flow distribution of the fundraising projects that have aggregate donations of the entire life cycle less than 100 donations.\footnote{Here is an example to illustrate our allocation rule: for a project that has extra 10 donations (beyond 100), we have to allocate them to 2 dates (first date and second date). We observe that on average, for projects that have less than 100 donations, they receive 20 donations on their first day and 10 donations on their second day. We calculate a weight $(2/3,1/3)$  from this observation. Using this weight, we assign $10\times 2/3$ to the first date and $10\times 1/3$ to the second date. 

Besides this weighting strategy, to check robustness, we also use an equal weighting strategy when it comes to estimating the effect of BLM. Changes in allocation rules barely change any main result developed further.}

Our original data consists of around 550,000 medical crowdfunding projects worldwide from January 1, 2019, to July 31, 2021. Our data provide a complete snapshot of all medical crowdfunding projects on July 31, 2021. Namely, at the end of July 2021, we observed the status of all crowdfunding projects launched from January 1, 2019, to July 31, 2021. To focus on the impact of BLM movements within the U.S. and exclude the impact of ongoing fundraising, we only keep records in the U.S. that started before March 31, 2021 (the average life-cycle of the project is about three months), which provides us with 390,000 medical records. 

Panel A of Table \ref{tab_su_GoFundMe} summarizes the characterizing variables of all crowdfunding projects. On average, for one project, the beneficiaries want to raise around 13 thousand dollars, yet only 40 percent of the goal is fulfilled. Only 10 percent of medical crowdfunding projects are entirely or over-funded (i.e., raised more money than the goal). There are around 38 donors for each project, and each donates around \$80.

We determine the beneficiary’s race based on the uploaded photo in the project. Specifically, we utilize the Baidu Research API, which identifies racial information  (the number of faces and their races) from faces in the images. This approach is remarkably accurate when distinguishing the faces of Black and non-Black people (\cite{serengil_lightface_2020}; \cite{yang_benchmarking_2021}). For each photo, the approach reports the number of faces and the number of Black faces. Around 90\% (362,953 records) of the projects have photos of beneficiaries.  Panel B of Table \ref{tab_su_GoFundMe} reports the racial information about the photos. On average, each photo has two faces. Around 10 percent of projects have photos including Black faces, and we identify those projects with at least one Black face as projects raised by Black people.\footnote{As a robustness check, we also change the classification rule by using the number of faces.}  

Previous literature shows that the text description on the project’s profile may influence the outcomes of crowdfunding projects (\cite{gorbatai_gender_2015}). Following \cite{younkin_colorblind_2018}, we utilize text-analysis software, notably LIWC (Linguistic Inquiry and Word Count) (\cite{pennebaker_development_2015}) to measure the key attributes of the text description in the crowdfunding records.  Panel C of Table \ref{tab_su_GoFundMe} reports those attributes. On average, the description of these projects comprises around 1600 words, though some have no text descriptions. Among all the words, twenty of them show authenticity, about four of them deliver positive attitudes/emotions, and two of them express negative attitudes. We do not see a large gap in the number of gender-related words. On average, all project descriptions have around two male- and female-related words. We identified those donors’ racial information from their names using the ``\textit{ethnicolr}" package, which is trained by the Census and Florida voting registration data.\footnote{\url{https://github.com/appeler/ethnicolr}}  On average, 10 percent of the donors are Black as based on their name classification.


\section{Racial Disparity in Medical Crowdfunding}\label{sec_GapDecrease}

This section demonstrates the gap in the outcomes of Black and non-Black beneficiaries at the end of the project. In particular, we are interested in documenting the evolution of the fundraising gap from two angles -- money raised and total number of donors.

\subsection{ Gaps and the Convergences Since 2020}

\begin{figure}[htb]
    \centering
    \caption{Amount of money eventually raised by launch time}
    \includegraphics{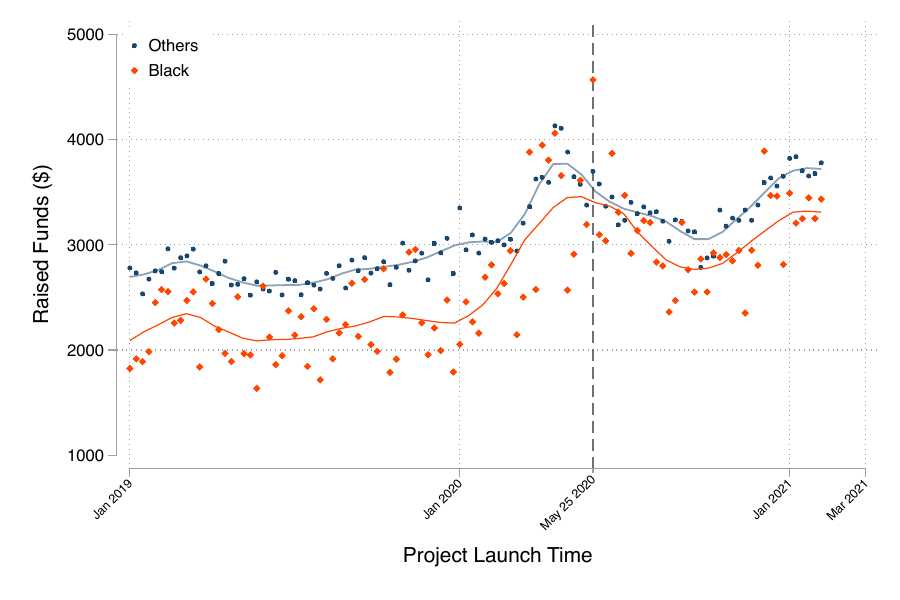}
    
    \label{fig:raisedfunds_t}
\vspace{1em}
\begin{minipage}[0.1cm]{1\textwidth}
    \small \textit{Note:} Figure \ref{fig:raisedfunds_t} depicts the total funds raised for projects launched at various times. The x-axis denotes the project launch date, and the y-axis signifies the total funds raised in dollars. Each data point reflects the average amount raised for projects initiated in that week. We use orange dots to represent projects involving Black beneficiaries and navy dots for all other projects. Non-parametric fitted lines indicate trends for each group. There is a notable initial gap between Black and non-Black beneficiaries for projects launched before January 2020. This gap begins to narrow for projects launched from February 2020 onwards, reaching its smallest in May 2020. Though the gap slightly widens after this point, it largely remains stable in subsequent months.
\end{minipage}
\end{figure}

Figure \ref{fig:raisedfunds_t} plots the money eventually raised against the project launch time  for Black and non-Black as well as the respective fitted lines using a local polynomial. Our time index on the x-axis represents the launch date of the fundraising project. Regarding the total money eventually raised, we first demonstrate a significant fundraising gap between Black and non-Black beneficiaries, but this gap decreases for projects with launch dates after January 2020. We find that for projects that start between January 2019 and January 2020, a significant gap persists in the raised funds across Black and non-Black beneficiaries. On average, Black beneficiaries raise 12\% fewer funds (average of \$3667) than non-Black beneficiaries (average of \$4097), and this gap is especially pronounced (23\%) for projects that start prior to January 2020 (average of \$2949 for Black v.s. average of \$3635 for non-Black people). 

As Figure \ref{fig:raisedfunds_t}  show, the reduction in the gap of raised funds starts to occur for projects launched after January 2020. This reduction cannot be explained by the channel that Black beneficiaries systematically change the style of how they edit their project profile (e.g., setting a higher goal or changing the writing style of the text description). Table II verifies our conclusion using a difference-in-differences framework and identifies a significant 12-15\% reduction in the racial gap after January 2020. Additional control in Columns (2) and (3) of Table II show that project goal, characteristics variables for text descriptions and profile picture, as well as state/date fixed effects, barely change our estimates. 

Next, we focus on the gap reduction in the total number of donors for each project. Figure \ref{fig:donors_t} presents the total number of donors over the project launch time for Black and non-Black people. The x-axis remains the launch date of the fundraising project.

Consistent with the pattern we have shown for the money raised, with regards to the total number of donors, we also observe an initial gap between Black and non-Black beneficiaries for those projects launched before January 2020. We find that on average Black beneficiaries receive 5\% fewer donors (average of 35 donors) than non-Black beneficiaries (average of 37 donors). This gap starts to decrease for projects launched after February 2020, and at the end of these projects’ donation life cycle, Black people’s projects have collected more donors than those of non-Black people. This inverse in the eventual number of donors remains stable in later months. 

Another outcome that relates to both the final raised money and the total number of donors is the average amount of donation. We define the average amount of donation to a project as the money raised in that project divided by the total number of donors. In Online Appendix \ref{sec:online_app_2}, Figure \ref{fig:avgdonation_t} plots the average donation for Black and non-Black people over the project launch dates. The average amount of donations for Black beneficiaries (\$70) remains 17\% percent smaller than that of non-Black beneficiaries (\$82). Unlike the number of donors or the final amount of money raised, the gap in the average amount of donations remains stable. This shows that the reduction in the gap in funds raised after 2020, as shown in Figure III, is due to more donors participating in donations to Black beneficiaries, rather than each donor donating a greater sum of money.  Table \ref{tab4} from Online Appendix \ref{sec:online_app_2} reports the estimates from the corresponding difference-in-difference (DID) regression.

In summary,  the analysis of this section reveals that the fundraising gap between Black/non-Black beneficiaries starts to decrease after January 2020, and the reduction reaches its maximum level (gap itself reaches its minimum level) at around the start time of this wave of the BLM movement triggered by George Floyd’s death. This finding motivates us to ask whether the surge of the BLM movement following the death of George Floyd could cause the convergence in the fundraising outcome across Black and non-Black beneficiaries. 

An explanation of BLM movements driving the reduced gap is consistent with our empirical findings: On average, a project lasts around three months. Thus, a project whose launch time is February 2020 would start experiencing the potential impact of BLM on the public’s charitable giving behavior at the end of its life cycle. Projects launched between February and May 27, 2020, would be partially impacted by BLM protests (partial treatment). The closer a project’s launch time is to the George Floyd event, the greater the impact on the project. Projects launched after May 27 are fully affected by the BLM (full treatment), as the life cycles of these projects are covered by the period of the BLM. Therefore, this section concludes, with reserve, that BLM movements may be a possible explanation for the reduction in the fundraising gap.

\subsection{Stable Temporal Pattern of Fund Demand}
The previous subsection highly suggests that the increment in the intensity of the BLM may cause more people to donate to the projects of Black individuals. However, before moving to our research design, which examines the causal effect of BLM on donors, we first need to exclude the possibility that this racial gap reduction is not resultant of change to the money demand side. 

The crowdfunding platform is a typical two-sided market: we have the money demand side – the people who launch projects (beneficiaries), and the money supply side – the people who browse the profiles of the projects and make donations (donors). The BLM can drive the reduction in the racial gap through both sides of the platform. Through the supply side, the BLM movement directly influences the behavior of donors, making them more willing to give money to Black individuals. 

However, through the demand side, even in an extreme case where the movement has no impact on the supply side, the surge of the BLM movement may still drive the same racial gap reduction by inducing more donations to Black people. Black individuals may strategically upload or edit their profiles during the Floyd protests. In this case, even if the donors’ tendency to give money to Black people holds constant, Black beneficiaries may still attract more donations by means of launching more projects or embellishing their profiles. 

We first address the concern that Black people may launch more projects on the platform during the Floyd protests period. An increase in Black people’s projects during the surge of the BLM, would expose donors to more Black people’s profiles, hence more donor money would then be given to Black people.  Figure \ref{fig:num_project_date}  from Online Appendix \ref{sec:online_app_2} plots the time pattern of the number of newly created projects on each day by race group. As can be observed, the ratio between Black and non-Black people’s newly created projects remains stable over time. During the surge of the BLM movement (after May 28 2020), Black people launch no more projects on the platform. Namely, donors do not therefore observe more Black people’s profiles during the Floyd protests, and it is unlikely that the extra donations observed during this period are due to an increment in the number of Black people’s profiles. 

We then test the hypothesis that Black people may strategically edit their project profiles to make them more attractive during the surge of the BLM movement. Under this hypothesis, donors still provide more donations to Black people even if the Black people do not launch more projects, and the tendency of donors to give money to Black people remains constant but is conditional on the attractiveness of characteristics of Black people observed on the profile. Table \ref{tab_observed_charcts}  from Online Appendix \ref{sec:online_app_2} tests this hypothesis by means of DID regressions that use the observed profile characteristics as the outcomes. Our observed characteristics are the length of the text description, the goal of the project, and the likelihood that the text description is about a certain topic. The regression results show that with regards to projects launched during the Floyd protest, Black fundraisers do not systematically change the style of how they edit their profiles. Namely, it is not likely that donors make more donations to Black individuals as the Black fundraisers upload more attractive profiles.

\section{The Overall Causal Impact of BLM Movements}\label{sec_Identification}

This section shows that the reduction in the fundraising gap is due to the surge of the BLM movement immediately after the spread of the news of the murder of George Floyd. 

\begin{figure}[htb]
    \centering
    \caption{Number of eventual donors for projects by launch time}
    \includegraphics{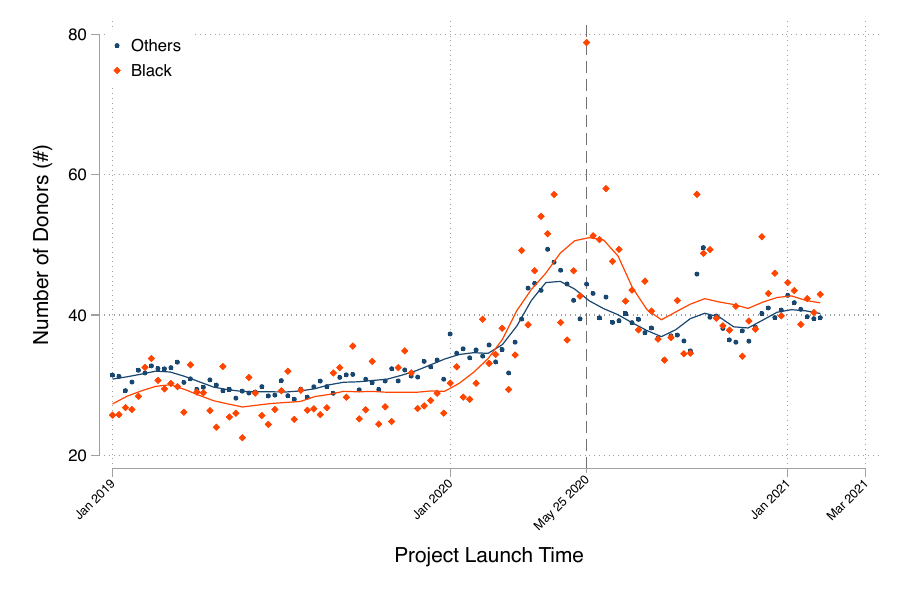}
   
    \label{fig:donors_t}
    \vspace{1em}
    \begin{minipage}[0.1cm]{1\textwidth}
    \small \textit{Note:} Figure \ref{fig:donors_t} plots the average number of donors for projects launched at different times. The x-axis represents the project's launch date, and the y-axis represents the number of donors. Each data point denotes the average number of donors to projects launched in that week. We use orange dots for projects involving Black beneficiaries and navy dots for other projects. The lines represent the non-parametric fitted lines for each group. An initial gap exists between Black and non-Black beneficiaries for projects launched before January 2020. However, this gap begins to close for projects launched after February 2020, and Black beneficiaries' projects start to attract more donors. This reversal in donor numbers remains consistent in the subsequent months.
    \end{minipage}
\end{figure}

The empirical findings in Section \ref{sec_GapDecrease} highly suggest that the surge of the BLM or the Floyd protests drives the reduction in the racial gap in medical crowdfunding projects by attracting more donors to Black people’s projects. However, it remains a challenge to argue the causal role played by the surge of the BLM. As Figure \ref{fig:raisedfunds_t} and \ref{fig:donors_t}  show, the gaps in the money raised and number of donors begin to decrease in February 2020, which is four months earlier than the surge of the BLM. Although the projects started between February and early May 2020 may be partially affected by the BLM given the average 3-month lifetime of the projects\footnote{For example, it is possible that a project started in March 2020 remains available for donors who browse the platform in June 2020. }, other shocks during this 3-month period, such as the Covid pandemic and pandemic-related policy shocks may also contribute to this racial gap reduction. Thus, we need a more rigorous strategy to identify the causal effect induced by the BLM and exclude other confounders.   

\subsection{Identification Strategy}
To solve the identification challenge, we leverage the daily-level donation inflow records of each project and develop a difference-in-differences (DID) strategy. 

This section uses the daily number of donations to each project to construct the identification strategy. As Section \ref{sec_GapDecrease} identifies that the reduction in the racial gap of final raised money is driven by an increase in the number of donors giving money to Black people, we focus on the daily number of donations received by one project.\footnote{We also examine whether the amount of donation (\$) is affected by the BLM movement. Our results show that there is a zero impact. See Figure \ref{fig:donation_volume} from Online Appendix \ref{sec:online_app_2} for evidence.} 

The detailed cash flow daily panel data for each project provides two critical advantages for identifying the causal impact of BLM movements. First, tracking the daily level enables us to identify whether or not one project is affected by the surge of the BLM movement. We can deduce which projects have a donation life cycle overlapping with our treatment period. Moreover, within one project, we can distinguish the dates when the project were impacted by the surge of the BLM movements. Solely relying on the project-level ultimate outcomes means we are unable to identify whether and to what extent the project is affected by the BLM event (start from which time point), especially concerning those projects launched before late May 2020.\footnote{For example, we are unable to identify whether projects started in April 2020 were influenced by the later BLM in June 2020; If we only have access to the project-level data, some of the projects could have a very short life-cycle, and could potentially have already ended before the George Floyd event.}

Second, this daily data provides a measurement that immediately responds to the social movement. The high-frequent cash inflows enable us to trace the instantaneous donations (if any) on fundraising platforms back to the changes in the BLM movements. This is particularly advantageous to consolidating the causal impact of social movements when there are other contemporary events that may contaminate our estimations. 

\subsubsection{The Treatment Definition}

We are interested in understanding the effect of the ``surge of the BLM movement" or the ``Floyd protests" on donation flow. The term ``Floyd protests" refers to the complex changes that unfolded throughout the protest period. As a treatment period,   the ``Floyd protests" period encapsulates the multifaceted societal transformation triggered by the event, including amplified public discourse and cultural evolution. This broader interpretation acknowledges that the impact of BLM extends beyond the physical protests and encompasses the intricate, systemic changes that transpired during this time, and contributed towards the ongoing pursuit of racial justice. In this section to estimate the causal impact of the BLM movement, we will use the Floyd protests period as our treatment.

We use May 28, 2020, as the starting point of the treatment period. Despite Floyd’s death occurring on May 25, it was not until May 26 that the cellphone footage of the incident became public, leading to protests erupting in Minneapolis. By May 28, demonstrations expanded throughout the U.S., resulting in a significant increase in media attention, as illustrated in Figure \ref{fig:protests_news_google}. Consequently, we anticipate that the complete impact of the protests will be evident in public opinion data starting from May 28, which we select as our cut point. 

It is noticeable that the pinnacle of both the protests and public focus on George Floyd and the BLM take place during June and July 2020. While the protests and search/media coverage still exhibit a significant increase relative to the pre-Floyd period three months after the event, their scale diminishes substantially in comparison to the peak observed in June and July. Hence, we include the donation inflow data from April 1 to September 1, 2020  in our main analysis.

\subsubsection{The Identification Assumption and Difference-in-differences (DID) Regression}

\begin{figure}[htb]
    \centering
    \caption{The Daily Average Number of Donors Per Project}
    \includegraphics[width = \textwidth]{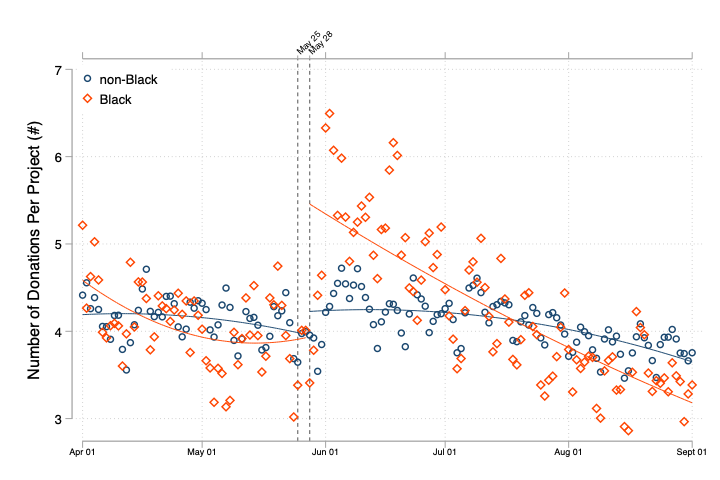}

    \label{fig:daily_num_donors_t}
    \vspace{1em}
    \begin{minipage}[0.1cm]{1\textwidth}
    \small \textit{Note:} Figure \ref{fig:daily_num_donors_t} depicts the average number of donations received per active project per day. Each data point signifies the average number of donors per project on a given date $t$, calculated as the total number of donations made that day divided by the total number of active projects on that day. Remarkably, prior to the onset of the Floyd protests, no significant difference exists between Black and non-Black beneficiaries' accounts. However, during the height of the Floyd protests, we observe a notable surge in daily donations to projects associated with Black beneficiaries (rising from approximately 3.5 per day to around 5 per day), while the daily number of donations to non-Black beneficiaries remains relatively stable (around 4 donations per day).
    \end{minipage}
\end{figure}

Figure \ref{fig:daily_num_donors_t} non-parametrically illustrates the DID estimation of the effects of the Floyd protests on the number of daily donations received by a project. We use the projects of the non-Black individuals as the control group and those of Black individuals as the treatment group. Each data point represents the daily average number of donations received across all active projects on that day. 

It can be observed that before the George Floyd protests, there was no significant gap between Black and non-Black people’s projects in respect of daily donation numbers. On average, every project received around 3.5 donations per day. During the surge of the Floyd protests, there was a dramatic increase in the number of donations supporting projects of Black beneficiaries: around 90\% surge within a week, and 40\% surge over two months. On the other hand, the number of donors for non-Black beneficiaries remained stable after the event, revealing no crowding-out effect associated with the increase in the number of donations for Black beneficiaries. 

\textbf{Regression specification}

To formally quantify the impact of BLM movements, we employ the following DID regression by utilizing the timing discontinuity around the George Floyd event. Project $j$ is active (be available for donors) at date $t$, which falls
\begin{equation}\label{eq:main}
\begin{split}
    y_{j,t} = \gamma_0 Black_j \times \textbf{1}(t\geq t_{May28}) + \gamma_1 Black_j + X_{j}\Gamma + \tau_{t}+\mu_{State}+\kappa_{CreateDay}+\varepsilon_{jt}    
\end{split}
\end{equation}
in the range of April 1 to September 1, 2020. $y_{j,t}$ denotes the number of donations that fundraising project $j$ receives on date $t$.  $Black_j$  indicates whether or not the beneficiary of project $j$ is identified as a Black person. $\textbf{1}(t\geq t_{May28})$ denotes the indicator for whether the donation happens after May 28, 2020. $X_{j}$ represents a set of time-invariant control variables at the project level, including the goal of the project, and project description variables (length, emotions, language use). $\tau_{t}$ is the time fixed effect for date $t$, which absorbs the nationwide common trend in donations. $\mu_{State}$ is the state fixed effect of where the beneficiaries live. $\kappa_{CreateDay}$ is the fixed effect of the date when project $j$ is created, which absorbs the project's natural pattern of donation diminishing over time.  $\gamma_0$ returns our estimated effect of BLM on the donation.

To analyze the duration of the surge of the BLM's effect, we also introduce the event study framework of Equation \eqref{eq:main}:
\begin{equation}\label{eq:main_eventstudy}
\begin{split}
    y_{j,w} = & \sum_{w=-8}^{-1}\eta_{pre,w}\times Black_{j}+ \sum_{w=0}^{w=15}\eta_{post,w}\times Black_{j} + \gamma Black_j + X_{j}\Gamma + \\
    & \delta_{w}+\mu_{State}+\kappa_{CreateDay}+\varepsilon_{jw}    
\end{split}
\end{equation}
To have a clearer read from the event study, we pool the data into week level. $y_{j,w}$ is the number of donations received by project $j$ in week $w$. We define May 28, 2020, as the 1st day of week 0, and, to be consistent with our data range from April 1 to September 1, 2020, we include $w\in [-8,15]$ in the event study. $\delta_w$ is the week fixed effect absorbing the common trend. Other variables have the same interpretations as Equation \eqref{eq:main}.

We have two identification assumptions: (i) If there is no surge of BLM, the daily donation number provided to projects of Black and non-Black individuals should still share the parallel time trend after May 28, 2020; (ii) No other shocks can have a heterogeneous influence across races during our treatment period. From Figure \ref{fig:daily_num_donors_t}, we learn that before the Floyd protests, there was no significant difference in time pattern across Black and non-Black people. This fact suggests that assumption (i) is plausible. As for assumption (ii), other shocks in spring and summer 2020, such as the pandemic, stay-at-home order (SAH), end of the stay-at-home order, and so on might be threats. We develop a framework to rule out these possible threats based on our daily level records.

\textbf{Main empirical results}

Column (1) to Column (4) in Table \ref{tab:reg_mainDID} present the estimates using Equation \eqref{eq:main}. For project $j$ on date $t$ in both regression specifications, we control for the log value of the goal setting of the project, the characteristics of the text description (including the length of the description, the emotion index of the description, the gender tendency index, and the authenticity index), the fixed effect of the beneficiaries’ living states, the fixed effect of date $t$, and the fixed effect of the start date of the project. 

To address the concern that Black people excessively raise more projects for Covid-related reasons, we also control for the dummy that indicates if the project is asking for money due to Covid as shown in Column (2). To mitigate the potential confounding effects of varying SAH across states and over time, we introduce additional controls in Column (3) of our analysis. Specifically, we control for an interaction term between an indicator for Black individuals and two time dummy variables for the SAH shocks: one for dates prior to the enactment of SAH orders and another for dates after the termination of such orders. Column (4) controls for the project's fixed effect, which absorbs all the project-level time-invariant confounders. 

 Our results from Columns (1) - (4) show that the surge of the BLM causally initiates an average increase in donations to Black people’s projects of 0.406 to 0.419 donations per day. Compared to the mean value of the non-Black control group before the treatment period, it is a significant 10\% increase in the daily donation number. 

\begin{figure}[htb]
    \centering
    \caption{Event study of the effect of the surge of BLM}
    \includegraphics[width = \textwidth]{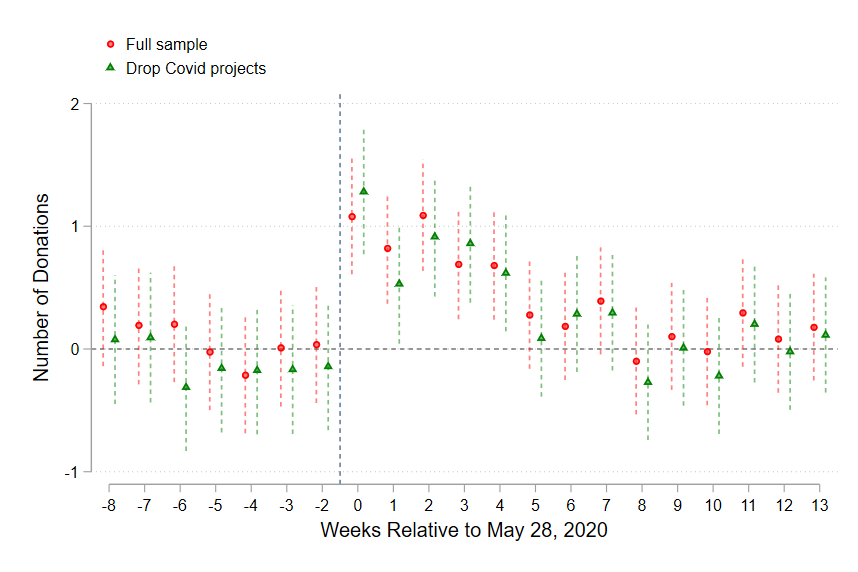}
    \label{fig:reg_eventstudy_dailynumdonation}
    \vspace{1em}
    \begin{minipage}[0.1cm]{1\textwidth}
    \small \textit{Note:} Figure \ref{fig:reg_eventstudy_dailynumdonation} illustrates the estimation results of Equation \eqref{eq:main_eventstudy}. In an effort to address concerns that the pandemic may disproportionately impact Black individuals, we present results both with and without projects soliciting funds due to COVID-19. Additionally, we account for the interaction between Black individuals and state-level pandemic-related policy shocks. As observed, before the onset of the BLM movement, the daily number of donations to Black and non-Black individuals maintain a parallel trend. Moreover, the influence of the BLM movement on Black individuals is sustained for approximately 12 weeks following the nationwide dissemination of the campaign.
    \end{minipage}
\end{figure}

To diagnose the temporal duration of the effect, we also estimate the event study in Equation \eqref{eq:main_eventstudy}. Figure \ref{fig:reg_eventstudy_dailynumdonation} presents the estimation results. To address the possible threats from the pandemic, we report the estimates from both the full sample and the non-Covid-related projects. For the estimates from the full sample, we also control for the dummy of Covid. It is evident that there is no significant difference between the estimates from the full sample and the restricted sample.

The x-axis represents the relative weeks to May 28, 2020, which is the start date of our treatment period. The negative tick marks represent the weeks prior to the treatment period. It is obvious that before the treatment period, there was no significant pre-trend for projects organized for Black people, but in the immediate aftermath of the start of the treatment period, their projects experienced a significant positive impact compared to those of the non-Black people. This pattern suggests that it is very likely that without the surge of the BLM, the projects of Black and non-Black beneficiaries would hold a parallel trend in the number of daily donations. 

The second observation from the event study is that despite both the statistical and economic significance observed, the effect of the surge of BLM on Black individuals is short-lived. The first five weeks of the spread of the BLM movement see Black people receive about one more donation per day compared with non-Black people, which is about a 30\% increase, compared to the pre-treatment period. However, this effect only lasts for at most twelve weeks and shrinks to negligible scales in the following weeks. 

Overall, Table \ref{tab:reg_mainDID} and Figure \ref{fig:reg_eventstudy_dailynumdonation} indicate that using the periods after May 28, 2020 as the treatment period, the surge of BLM induces a significant increase in the number of daily donations made to Black people’s projects. In our treatment period from May 28 to September 1, 2020, the surge of BLM induces on average  0.419 more daily donations to Black people. According to the event study, this effect appears in the first three months of the spread of the BLM movement. In the first five weeks of our treatment period, Black people receive on average of one more donation per day and this effect shrinks in the following weeks. 

\textbf{Robustness check}

We conduct two additional exercises on top of our main regression specification as robustness tests. We briefly discuss our main takeaways here and leave the details in Appendix \ref{app_robustness}.

First, we address the concern that Black fundraisers might strategically react to the surge of BLM movement by restricting our sample to projects that were started from May 1 to May 25. Although Section \ref{sec_GapDecrease} argues that there are no significant changes in the characteristics of projects launched both before and during the Floyd protests, there might be unobserved differences between the two groups.

Given the average 3-month life of the projects, these projects from restricted sample are very likely to experience both the control and treatment period. Besides, the random timing of the murder of George Floyd ensures that fundraisers have no advanced information about the surge of BLM when editing and launching new projects in that period. Hence, the DID estimates based on the restricted sample is purely down to changes made by the donors.

Second, we attempt to understand the causal effect that the surge of BLM has on the donations by using a regression discontinuity in time (RDiT) design on Black and non-Black individuals’ projects. The RDiT design helps us to estimate the treatment effect within each racial group. Besides, the estimation within the non-Black people’s projects also helps us see if the positive effect of BLM on Black people is partly down to its negative effect on non-Black people. 

Given the unexpected and unplanned nature of George Floyd’s killing, we are able to mitigate any concerns about “anticipation effects” in our RDiT estimation (\cite{reny2021opinion}). The RDiT design estimates local causal effects by leveraging as-if-random variation around an arbitrary cutoff, which is comparable to randomized controlled trial (RCT) treatment effects (\cite{wing2013strengthening}).

\subsubsection{Testifying Other Confounding Events and Why High Frequent Data Help}\label{sec_subsec_valid_assumption_strategy}

In order to identify the causal role of the surge of the BLM movement on the donations, we must assume that during our treatment period (the peak of the Floyd protests), there were no other social-economic events that could contemporaneously affect fundraising processes disproportionately for Black and non-Black beneficiaries. The major concern comes from the shock of the pandemic and the related policies in 2020. Previous literature has shown that both Covid-19 and public health policies have heterogeneous effects across Black and non-Black people. These heterogeneous effects based on race could then be threats to our identification. 

This section briefly discusses the empirical strategy used to provide evidence that both the pandemic per se and public health policy shock would not contaminate our estimation of the effect of the BLM movement by leveraging our daily-level data.\footnote{Do the events we selected to test sufficiently support our identification? We use Google Search Trend API to report us queries with the biggest increase in search frequency from April 1 to August 31, 2020. We find the top 10 search keywords are: ``coronavirus" ``stimulus check" ``coronavirus symptoms" ``popular google doodle games" ``thank you coronavirus helpers" ``coronavirus tips" ``coronavirus news" ``george floyd" ``kobe bryant" ``n95 mask". The popularity of these keywords increased by more than 1000\%. In particular, the keyword on ``george floyd" increased by 3400\%. Therefore, it is unlikely that other major events except for Covid-related issues during those periods could significantly affect the fundraising gap. } The detailed manipulation and results discussion can be found in Appendix \ref{app_assumptiontest}.

Our strategy primarily aims to examine whether there exists any discontinuous effect for the other confounding events. Our results show that immediately before and after the Covid reopening policies, there was no sudden change in the fundraising disparity; and there was a smooth evolution of the number of Covid-19 cases. We run RDiT regressions with a changing bandwidth from large to small and plot our estimates to investigate the existence of discontinuity. This allows us to focus on sharp local neighborhood days around these confounding events, owing to the daily-level feature of our data. Thus, we conclude that the impact of the BLM movement on our estimation is not confounded by either the pandemic itself or the shock to public health policies.

Besides, the framework we use to rule out the possible confounders highlights the power of using daily-level data to evaluate the effect of the BLM movement. As some potential contaminators (shocks) occur in the very close time window alongside the surge of BLM (10 days away from the Covid reopen policies), it is difficult to exclude their possible statistical effects by using monthly or even weekly level data, which the previous literature base on.

\subsubsection{The Impact of the Floyd Protests Is Unprecedented in the History}
The ``Floyd" movement had an unprecedented effect in comparison to previous BLM movements. In Figure \ref{fig:previous_killings}, we plot the donation pattern to Black community over time, ranging from May 2019 (our earliest possible data) to November 2020. Fifteen instances of police violence against Black people are covered during this period. While earlier incidents led to limited changes in charitable giving, the George Floyd movement resulted in a significant, structural shift in donations to the Black community. This is consistent with its standing as the largest social movement in U.S. history, a status likely elevated by extensive protests and broad social media coverage.  This sets the ``Floyd" movement apart from earlier BLM efforts in terms of its impact on charitable engagement.

\section{Racial Decomposition of Donors}\label{sec_race_decompose}
The previous section demonstrates that the BLM movement assists in procuring more donations for projects related to the Black community. Given the racial context and the widespread influence of the movement in 2020, a consequential question surfaces: Which racial group of donors is primarily responsive to the surge of the BLM movement? 

This question is vital for comprehending the mechanisms underlying the impact of the BLM movement on the number of donations to Black people and for gaining insight into the social outcomes resulting from this wave of the movement. The overall effect of the rise of the BLM may be primarily driven by either Black or non-Black donors. On the one hand, it is reasonable to assume that donors might be more inclined to support beneficiaries of their own race. Consequently, one potential mechanism through which the BLM movement influences donors is by amplifying within-race preferences (altruism). In other words, the 2020 surge in BLM activity may primarily encourage Black donors to offer greater support to their Black peers. On the other hand, however, the extensive reach of the BLM movement in 2020 also affects a substantial number of non-Black individuals, suggesting that donations to Black beneficiaries could primarily originate from non-Black donors. If the first mechanism prevails, the rise of the BLM movement may not necessarily reduce racial discrimination within society. Conversely, if the second mechanism dominates, it could indicate a partial move towards a more integrated society. 

Therefore, this section aims to address the question by examining the racial composition of donors to determine which racial groups respond to the BLM movement’s surge. This analysis enables us to differentiate between within-race altruism and cross-race altruism.

The initial step involves extracting the racial information about donors. To estimate race, we employ the Python package ``\textit{ethnicolr}.”\footnote{\url{https://github.com/appeler/ethnicolr}} The package exploits the U.S. census data, the Florida voting registration data, and the Wikipedia data to predict race and ethnicity based on first and last name or just the last name and was widely used in existing literature for racial classification (\cite{chilton2020political}). In practice, when provided with the first and last names of a donor, the ethnicolr package returns the probability of that individual being classified as Asian, Hispanic, non-Hispanic White, or non-Hispanic Black. We define the first three categories as non-Black donors and the last category as Black donors.  For each donation $i$ to project $j$, we denote the probability of donation  $i$ coming from a Black donor as $p_j^{i\in B}$, and from a non-Black donor as $p_j^{i\in nB}$. We calculate the expected number of donations from Black/non-Black received at date $t$ as:
\begin{align*}
     y_{j,t}^{B} \equiv \sum_{i} p_j^{i\in B} \times \textbf{1}(i \text{ happens at } t) \\
     y_{j,t}^{nB} \equiv \sum_{i} p_j^{i\in nB} \times \textbf{1}(i \text{ happens at } t)
\end{align*}
Between April 1, 2020, and September 1, 2020, our primary analysis encompasses 2,413,873 donations. Utilizing the racial classification package, the expected number of donations from Black donors is 252,549, while the expected number of donations from non-Black donors stands at 2,161,324. The proportion of Black donors is 11.7\%, which aligns with the overall proportion of Black donors in crowdfunding, as reported by  \cite{smith2016shared}.

To break down the effect of the surge of the BLM on Black and non-Black donors separately, we estimate regression Equations \eqref{eq:race_Black_effect} and \eqref{eq:race_nonBlack_effect}: 
\begin{equation}
        y_{j,t}^{B} = \gamma_0^{B} Black_j \times \textbf{1}(t\geq t_{May28}) + \gamma_1^{B}Black_j + X_{j}\Gamma+ \tau_{t}+\mu_{State}+\kappa_{CreateDay}+\varepsilon_{jt}    
        \label{eq:race_Black_effect}
\end{equation}
\begin{equation}
        y_{j,t}^{nB} = \gamma_0^{nB} Black_j \times \textbf{1}(t\geq t_{May28}) + \gamma_1^{nB} Black_j + X_{j}\Gamma+ \tau_{t}+\mu_{State}+\kappa_{CreateDay}+\varepsilon_{jt}
        \label{eq:race_nonBlack_effect}
\end{equation}
The estimated $\hat{\gamma}_0^{B}$ and $\hat{\gamma}_0^{nB}$ represent the effects of the BLM movement on the expected number of donations from Black and non-Black donors, respectively. Mathematically, we can express this relationship as:
$\hat{\gamma}_0 = \hat{\gamma}_0^{B} + \hat{\gamma}_0^{nB}$,
where $\hat{\gamma}_0$ is the overall effect estimated from Equation \eqref{eq:main}. The estimated $\hat{\gamma}_1^{B}$ and $\hat{\gamma}_1^{nB}$ denote the initial disparities in the racial composition of donors who give to projects associated with Black and non-Black individuals. These values capture the differences in the expected number of donations from Black and non-Black donors for projects owned by Black individuals.

 

Table \ref{tab:reg_race_decomposse} presents the estimates of Equation \eqref{eq:main} in Column (1) and the estimates of Equations \eqref{eq:race_Black_effect} and \eqref{eq:race_nonBlack_effect} in Columns (2) and (3). 
We first affirm the initial difference in the racial composition of donors across projects related to Black and non-Black individuals. In comparison to projects associated with non-Black individuals, those associated with the Black community receive an additional 0.175 donations per day from Black donors, while they experience 0.119 fewer donations per day from non-Black donors. This is consistent with the notion that a person prefers donating to their peer races.

\begin{figure}[htb]
    \centering
    \caption{Event study of the effect of the surge of BLM by donors' races}
    \includegraphics[width = \textwidth]{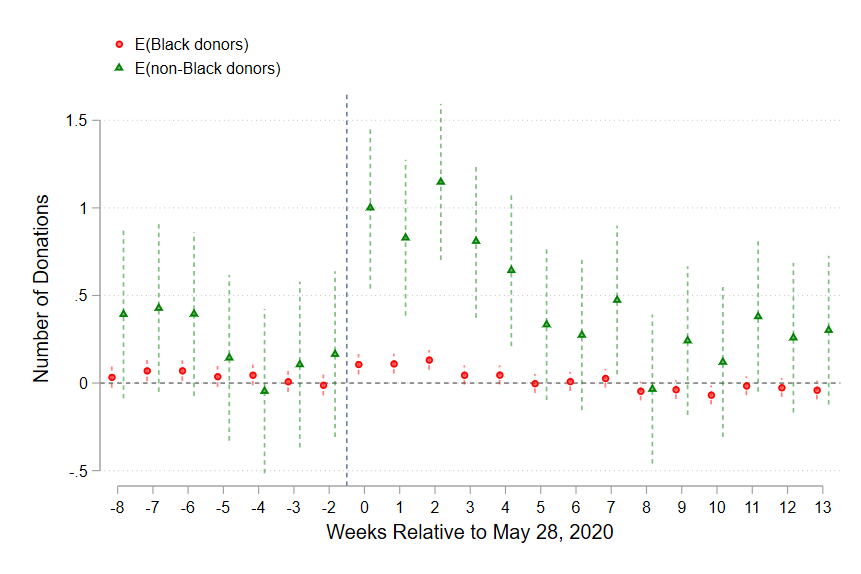}
    \label{fig:reg_eventstudy_donor_race}
    \vspace{1em}
    \begin{minipage}[0.1cm]{1\textwidth}
     \small \textit{Note:} Figure (\ref{fig:reg_eventstudy_donor_race}) presents the results of the event studies for Equations \eqref{eq:race_Black_effect} and \eqref{eq:race_nonBlack_effect}. Both Black and non-Black donors do not show any pre-existing trend of donating to Black beneficiaries. It can be observed that it is predominantly non-Black donors who contribute to the increase in donations to Black beneficiaries' projects following the surge of the BLM movement.
    \end{minipage}
\end{figure}

More critically, the primary takeaway pertains to the differential impact of the BLM movement on various groups of donors. The overall effect of the BLM movement on the daily increase in donations to projects associated with the Black community primarily originates from non-Black donors. As reported in Column (1), the BLM surge results in an additional 0.413 donations per day to projects related to the Black community. This effect is composed of an additional 0.016 donations from Black donors and 0.397 donations from non-Black donors. In other words, over 95\% of the BLM effect is attributable to non-Black donors. Consequently, the observed overall effect is not primarily driven by an influx of Black donors supporting projects related to Black beneficiaries during the treatment period. Figure \ref{fig:reg_eventstudy_donor_race} displays the event study of the donors’ racial composition, which reinforces the conclusion that non-Black donors predominantly account for the overall effect of the BLM treatment. 

This observation remains consistent even when considering that non-Black individuals constitute the primary source of donations for projects associated with Black beneficiaries. Before the treatment, projects related to the Black community receive 0.572 donations per day from Black donors and 3.383 donations per day from non-Black donors. The BLM treatment leads to an 11.7\% increase in non-Black donations and only a 2.8\% increase in Black donations, suggesting that the BLM surge exerts a more substantial effect, both economically and statistically, on non-Black donors.

\section{Heterogeneous Effects Across Geographics}\label{sec_geo_HTE}

This section deepens our understanding by exploring the geographical distributional effects of the BLM movement. Our main focus is on answering whether protest gatherings matter in reducing the racial disparity by exploiting geographical variations in the protest gatherings associated with the BLM movement and the residential information about fundraisers.

We first investigate how the BLM movements impact on donation behavior according to the characteristics of local residential counties. We consider six aspects: (1) pre-BLM racial disparity in fundraising; (2) prejudice against Black people (measured from 2019's Implicit Attitude Test); (3) the Black population ratio; (4) average income of Black people; (5) Pennsylvania v.s other states; (6) political tendency (2019’s voting outcome: Republican or Democratic). 

Figure \ref{fig:heterogeneity_BLM} presents the counterpart graphs of Figure \ref{fig:daily_num_donors_t} according to county characteristics. The figure reveals the following four conclusions: First, there is a significant path dependence – counties with higher racial disparity and higher prejudice against Black people tend to benefit most from the BLM movements; Second, the local economic level and population composition do not matter; Third, non-surprisingly, the  impact of the BLM movement is most profound in Pennsylvania, where the treatment effect is three times that of the average treatment effect across the U.S. Lastly, Republican counties are slightly more motivated to donate to Black beneficiaries than democratic counties. 

We next exploit geographical variations in the protest gatherings and examine whether protest gatherings are important.

\subsection{Geographical Distribution of Protests}
The crowdfunding projects in our analysis encompasses 2,612 counties across the entire country. Figure \ref{fig:geodist_protest} presents the geographical distribution of the total occurrence of protest gatherings in each county and state in the U.S. The Figure shows quite a large geographical variation in protest gatherings for us to explore.

Among these, both Black and non-Black beneficiaries initiated their projects in 1,070 counties; there were no counties in which only Black individuals launched projects. It is note-worthy that both Black and non-Black fundraisers are disproportionately located in areas with higher occurrences of BLM-related rallies that were sparked by the murder of George Floyd. This observation is particularly pronounced for Black beneficiaries. 

Figure \ref{fig:protest_cum_project} illustrates the distribution of projects across counties with varying levels of BLM-related gatherings between May 26 and September 1, 2020. The x-axis represents the ranking of the counties based on the total number of BLM-related gatherings. The y-axis indicates the cumulative number of projects. The vertical dashed lines depict the number of gatherings at each rank. The red curve corresponds to all projects, while the blue curve represents projects associated with the Black community. 

This figure illustrates a notable contrast in the distribution of gatherings and projects. Although most counties, in terms of sheer numbers, only experience a few gatherings during the treatment period, they only account for a small proportion of the initiated projects. Conversely, a small number of counties experience a high number of gatherings and see the launch of a significant proportion of projects. Of the 2,612 counties, approximately 2,200 experience no more than four gatherings, with only a quarter of the projects relating to the Black community being launched in these counties. On the other hand, only 38 counties experience more than 42 gatherings, yet these counties see the launch of a quarter of the projects related to the Black community. This reveals that the majority of projects were concentrated in a small number of counties where a massive number of gatherings occurred.

\subsection{Do Counties with More Gatherings See Larger BLM Effects?}\label{sec:heter_DDD}
In reference to Figure \ref{fig:protest_cum_project}, we categorize our sample into four quantiles, based on the number of gatherings in each project’s associated county. By doing so, we account for the uneven distribution of projects across different counties, thus ensuring an equal number of projects associated with Black beneficiaries in each quantile. 

Consequently, we have the following quantiles:\footnote {The empirical findings remain robust even when the cut-off points for each group are marginally adjusted.}
\begin{itemize}
    \item \textit{Quantile} 1: Projects initiated in counties with 0 to 3 gatherings. This group contains 1,515 projects led by Black people.
    \item \textit{Quantile} 2: Projects initiated in counties with 4 to 18 gatherings. This group includes 2,165 projects led by Black people.
    \item \textit{Quantile} 3: Projects initiated in counties with 19 to 41 gatherings. This group comprises 1,856 projects led by Black people.
    \item \textit{Quantile} 4: Projects initiated in counties with more than 41 gatherings. This group consists of 2,024 projects led by Black people.
\end{itemize}
In order to estimate the heterogeneous effects of the BLM on each quantile group, we run the following regression for project $j$ initiated in county $c$:
\begin{equation}\label{eq:DDD_county}
\begin{split}
    y_{jct}  = &  \sum^{3}_{k=1}\beta^k_0 \textbf{1}(t\geq t_{May28})\times Black_j \times Q_k + \beta_1 \textbf{1}(t\geq t_{May28})\times Black_j \\
    & +\beta_2 Black_j + X_j\Gamma + \tau_{t}+\mu^1_{County}+ \mu^2_{County\times Black}+ \mu^3_{County\times \textbf{1}(t\geq t_{May28})} \\
    & + \kappa_{CreateDay}+\delta_{SAH\times Black}+\varepsilon_{jct}
\end{split} 
\end{equation}
$y_{jct}$ remains the number of donations to project $j$ at date $t$. $Q_k$ (where $k = 1,2,3$) is a series of dummies indicating which quantile group the project is located. We use \textit{Quantile} 4 as the base level. 

In addition to the fixed effects in our main analysis, we control for the following: county fixed effects $\mu^1_{County}$; interaction between Black people's projects and counties $\mu^2_{County\times Black}$; interaction between $\textbf{1}(t\geq t_{May28})$ and counties $\mu^3_{County\times \textbf{1}(t\geq t_{May28})}$. These account for time-invariant county characteristics, the existing racial gap in each county, and varying trends among counties during the treatment period, respectively.

The estimated $\Hat{\beta_1}$ is the effect that the BLM has on projects in \textit{Quantile} 4. The estimated $\Hat{\beta^k_0}$, where $k = 1,2,3$, is the extra effect that BLM has on projects belonging to other groups.

\begin{figure}[htb]
    \centering
    \caption{Event study of the effect of the surge of BLM by counties}
    \includegraphics[width = \textwidth]{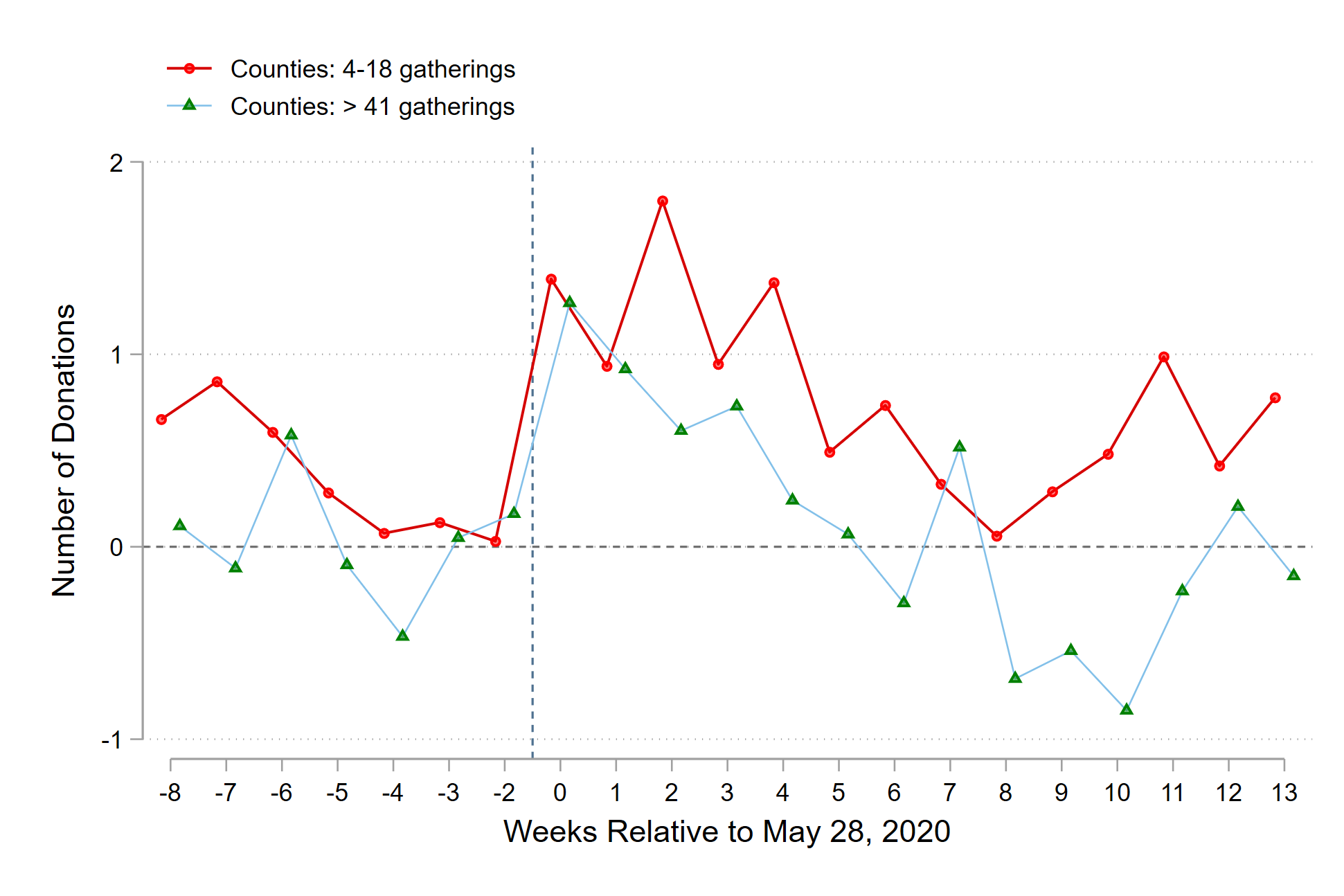}
    \label{fig:reg_eventstudy_heter_county}
    \vspace{1em}
    \begin{minipage}[0.1cm]{1\textwidth}
     \small \textit{Note:} Figure \ref{fig:reg_eventstudy_heter_county} presents the event study estimate of the effect of BLM movement, in analog to Figure \ref{fig:reg_eventstudy_dailynumdonation}, separately for Quantile 2 (counties that witnessed 4-18 gatherings) and Quantile 4 (counties that witnessed over 41 gatherings).
    \end{minipage}
\end{figure}

Table \ref{tab:reg_county_DDD} presents the estimates of Equation \eqref{eq:DDD_county} under various specifications. Column (3) represents the primary specification, which reveals that after accounting for the additional effects that the BLM movement has on each group, the BLM contributes to an increase of approximately 0.46 donations per day to projects associated with the Black community. Notably, projects in \textit{Quantile} 2, initiated in counties that witnessed a moderate number (4 - 18) of BLM gatherings, benefit from an additional 0.39 donations per day during the treatment period. Columns (1) and (2) feature specifications with different controls for the county-related fixed effect. The finding that the projects in \textit{Quantile} 2 experience the most significant effect from the BLM remains consistent across alternative specifications from Column (1) to (3).  Furthermore, the significant estimates of the coefficients for $Black\times Q_k$ and $\textbf{1}(t\geq t_{May28})\times Q_k$ suggest differences in the pre-existing racial gap across counties and varying trends among counties during the treatment period. Figure \ref{fig:reg_eventstudy_heter_county} further illustrates the event studies, separating projects in \textit{Quantile} 2 and \textit{Quantile} 4. This distinction clearly indicates that the effect of the BLM movement is more pronounced effect on projects in \textit{Quantile} 2.


The heterogeneous effects revealed by Equation \eqref{eq:DDD_county} suggest that residents living in the ``marginal counties" of the Floyd protests, where a moderate number of BLM gatherings took place, were most influenced by the movement. The impact of the social movement in these areas was more significant than in counties with either sparse or numerous gatherings.

\section{Local and Global Effect of the BLM Protests}\label{sec_global_local}

The findings from Section \ref{sec_geo_HTE}  demonstrate that the effects of BLM movements vary across regions with different levels of protest gatherings, but non-linearly. This motivates us to explore whether donations made to Black beneficiaries at one region could be affected by the occurrences of rallies at some other regions.

This section discusses the effect of the overall BLM protest gatherings versus the effect of the local rallies under the movement on donation behavior. Our goal is to elucidate the interplay between the overarching movement and its localized events, particularly in an era where social media serves as a powerful force in magnifying the reach and impact of social movements.

We start from a simple regression that examines the effect of national aggregate protests and local rallies. 
\begin{equation}\label{eq:reg_localgather_OLS1}
\begin{split}
    y_{jct}  = &  \beta_0 Black_j \times protest_{ct} + \beta_1 Black_j \times Protest_{-ct} + \beta_2 Black_j\times\textbf{1}(t\geq t_{May28}) \\
    & + \beta_3 Black_j + X_{jt}\Gamma_0  + \tau_{t}+\varepsilon_{jct}
\end{split} 
\end{equation}
$protest_{ct}$ represents the number of gatherings that occurred in county $c$ at day $t$. $Protest_{-ct}$ is the total number of gatherings with that occurred in the U.S. other than county $c$ at the same time at day $t$. We leave the detailed discussions of specifications and control variables in Appendix \ref{sec_app_olsprotest}. 

We run regressions of the Equation \ref{eq:reg_localgather_OLS1} for full sample and by four quantiles. Table \ref{tab:reg_localprotest_OLS} reports the estimates $\hat{\beta}_0$ and $\hat{\beta}_1$, representing the effect of local rallies and global gatherings. Although the estimates from Columns (1)-(5) hardly have any causal interpretation, they suggest one conclusion -- the BLM movement mainly exerts its influence through its broad global impact, especially those massive movements in counties with a lot of protest gatherings. More specifically, 
we find that \textit{Quantile} 3 and 4 are influenced by the BLM movements because of the role of their own local protest gatherings while \textit{Quantile} 1 and 2 are surprisingly influenced because of the role of protest gatherings that happened nationally and away from \textit{Quantile} 3 and 4. 

With the above motivating evidence, we proceed to causally explore the role of protests and spillover effects in this movement. 

\subsection{The Global Effect of Protests}

We begin by estimating the ``global effect" of BLM protests. ``Global effect" refers to the influence that the nationwide trend of protest occurrences would have on donation gaps. Conceptually, the ``global effect" encompasses a range of impacts, not limited to the influence of witnessing local rallies in one's own residential county. It also includes the effects stemming from exposure to news reports and online information about large-scale protests occurring across the U.S.

Estimating the elasticity of the donation gap with respect to national protest occurrences presents a significant challenge. A simplistic comparison between dates with more protest gatherings v.s. fewer protest gatherings could lead to endogeneity problems. Donors with a greater empathy towards the Black community may donate more and earlier. As the overall national protest trend diminishes over time, an OLS regression would compare early donors, who may have less racial bias, with later donors who may harbor more racial prejudice. This could result in an overestimation of the elasticity. 

To navigate this challenge, we employ an instrumental variable approach, comparing the dates of holidays and weekends with regular workdays. This approach is informed by the observation of remarkable spikes in protest numbers occurring on every weekend following the George Floyd incident (as seen in Figure \ref{fig:protests_news_google}). The underlying idea is that a holiday or weekend day exogenously increases people’s free time for protest participation. By promoting protest activities on weekends or holiday dates,  the fundraising gap is further narrowed. As indicated in Table \ref{tab:iv_1st_week}, the first-stage regression shows that being a holiday or weekend significantly boosts the number of protest gatherings by 116, as supported by the F statistics of 100. 

Our exclusion restriction assumption is that being a non-workday does not impact the racial disparity (or donation behavior) between Black beneficiaries and non-Black beneficiaries, other than to increase donors’ engagement with the BLM movement. A crucial concern is that non-workdays differ from workdays, particularly because people generally spend more time online during weekends or holidays and are therefore more exposed to fundraising advertisements. If this exposure effect was to vary across races, then directly comparing weekdays with weekends could bias our results.\footnote{Another substantial concern is that on a non-working day or holiday, people’s free time to read and absorb news or social discourse on BLM significantly increases. We provide a heterogeneity analysis in Section 7 demonstrating that internet or digital device access, acting as a proxy for information digestion, does not significantly affect our IV estimates when using weekends as an instrument.}

To address this concern, we incorporate an additional layer of comparison—the difference in the fundraising gap between workdays and non-workdays prior to the BLM event. By using this pre-BLM difference as a baseline, we can account for unobserved differences between workdays and non-workdays. Therefore, we construct a triple- differences framework to capture the causal elasticity of protests. The following is the regression specification for the reduced-form effect: 
\begin{equation}\label{eq:iv_weenkend}
    y_{j,t} = \alpha_0 Black_j \cdot 1(t\geq t_{May28}) \cdot \textbf{\textit{h}}_t + \alpha_1 Black_j \cdot 1(t\geq t_{May28}) + \alpha_2 Black_j \cdot \textbf{\textit{h}}_t 
\end{equation}
\[ +\alpha_3 1(t\geq t_{May28}) \cdot \textbf{\textit{h}}_t +\gamma_4 Black_j \times \mu_{week} + \gamma_5 1(t\geq t_{May28}) + \gamma_6 \textbf{\textit{h}}_t + \beta X_{jt} + \mu_{week}+ \epsilon_{jt}\]
with a first-stage regression specified as:
\begin{equation}\label{eq:iv_weenkend_1st_stage}
    Protest_{t} = \theta_0 \textbf{\textit{h}}_t \cdot 1(t\geq t_{May28}) + \theta_1 \textbf{\textit{h}}_t + \theta_2 1(t\geq t_{May28}) + \mu_{week} +e_{jt}
\end{equation}
$\textbf{\textit{h}}_t$ denotes the indicator for whether time $t$ is a non-workday (including Saturdays, Sundays, and other national holidays), and $\mu_{week}$ represents optional week dummy variables. The week fixed effect can be included when comparing holidays versus non-holidays within the same week. The coefficient $\alpha_0$ captures the change in the racial donation gap on non-workdays compared to workdays after the George Floyd event.

Table \ref{tab:iv_reduced_week} presents our reduced-form estimates from Equation \eqref{eq:iv_weenkend}. From Column (2), we observe that weekends or holidays after the George Floyd incident see a 0.32 increase in the daily number of donors for Black beneficiaries relative to non-Black beneficiaries. This holds true regardless of whether week fixed effects are incorporated.

Columns (3) - (6) provide a separate breakdown for Black and non-Black beneficiaries. We find that being a non-workday increases the daily number of donors for Black beneficiaries by 0.24, while, intriguingly, donations to non-Black beneficiaries see a significant decrease of 0.12. This suggests that one effect of the BLM protests is a redistribution of resources from other racial groups to the Black community.

Table \ref{tab:iv_main_week} presents the 2SLS estimates for the causal effect of an additional protest on the nation’s donation flow. We find that an extra protest significantly reduces racial disparity (measured in the daily number of donors) by 0.0027. This means that an additional 100 protests occurring in the U.S. can decrease our racial gap in daily number of donors by $0.27$. In comparison to the OLS estimate from Column (1) of Table \ref{tab:reg_localprotest_OLS}, a simple OLS regression evidently overestimates the global causal effect, which aligns with the prediction of endogeneity whereby those individuals who donate more also tend to respond earlier to the BLM protests. 

\noindent\textbf{How much does the protest gatherings explain the overall impact of the BLM movement?} The BLM movement is multifaceted, and encompasses aspects such as protests, social media coverage of the protest gatherings, societal reflections on U.S. racial issues, and political debates around police misconduct. By identifying the global effects of protests, the BLM movement’s impact can be partitioned into two components: the first is responsive to national protest intensity and the second is independent of it (such as societal reflections on racial issues and political debate on police misuse of power). 

Given that there was an average of 101 protests per day across the U.S, the effect of national protests in reducing racial disparity approximates to 0.0027 × 101 = 0.273 donors, which constitutes about 70 percent of the overall BLM effect (0.4 donors as in Table \ref{tab:reg_mainDID}). Therefore, while protest gatherings account for a significant 68 percent of the impact of social movements, the remaining 32 percent can be attributed to a fixed effect arising from the initiation of the BLM movement and independent of the protest intensity.

\subsection{The Local Effect of Protests}

This section proceeds to explore whether the presence of local rallies causally affects people’s donation behavior regarding Black beneficiaries.


The main challenge in measuring the effect of experiencing a local protest rally within one’s own county is that the unobserved culture of empathy or racist taste of a county is deeply correlated with the intendancy to protest against racial inequality. For example, New York City, known for its diversity and substantial support towards the Black community, experienced the highest number of protests during the BLM movement but BLM affects less on donors in NYC. Consequently, a straightforward OLS comparison across regions could potentially lead to an underestimate of the effects of local protests. 

To solve such an issue, we adopt a rainfall IV approach. The idea is similar to \cite{madestam2013qje_teaparty_gather}, which observes that a rainy day decreases the propensity for a crowd of people to gather for protests. We rely on the coincidence of a rainy day, deviating from the local rainy propensity and global date-level precipitation. \footnote{We differ with \cite{madestam2013qje_teaparty_gather} in that we do not rely on a cross-region comparison of rainy counties v.s. no-rainy counties. This is due to the fact that rainfall in a region is deeply connected with the local economic and societal development, migration-in pattern, and the local racial composition.} Given the likely randomness variation of the daily rainfall incidence, it acts well as an exogenous shifter that influences the racial disparity or donation behavior only by affecting that region’s protest occurrences.

More specifically, our main strategy considers the following first-stage regression:
\begin{equation}\label{eq:iv_rain_1st_stage}
    protest_{ct} = f(rain_{ct})\theta' + \theta_c + \mu_{t} + \epsilon_{ct}
\end{equation}
For the dates after May 27, 2020, where $protest_{ct}$ is the number of protests that happened in county $c$ on date $t$\footnote{ In our case, we're primarily examining the extensive margin of protesting numbers, where the impact of rainy days is anticipated to be relatively mild. This is in contrast to the study by \cite{madestam2013qje_teaparty_gather}, which considers the number of participants at protests, a context where weather conditions could have a more substantial effect.}, $rain_{ct}$ is the measure of the rainfall (in inches) of county $c$ on date $t$, and $f(.)$ is a flexible function we choose to fit our first stage regression. $\theta_c$ represents the local rainy propensity, proxied by the county-average precipitation during this period. $\mu_t$ represents the global date-level precipitation. $\epsilon_{ct}$ denotes other unobserved terms that affect the occurrence of protests.

Table \ref{tab:iv_1st_rain} presents the first-stage estimates with different specifications. It illustrates that increased precipitation typically reduces the occurrence of protests.  Columns (1)-(4) present the estimates for when the treatment specification is the log of rainfall volume, actual rainfall volume, log of rainfall at the counties' population center, and actual rainfall at the population center, respectively. Our estimates in Columns (1)-(3) suggest that rainfall significantly reduces the occurrence of protests ($t$-stats is about 7 for the first specification, satisfying strong instrument standard). For instance, Column (1) shows that increasing rainfall volume by 1 percent will decrease the number of protests on each date in each county by about 0.0037$\times 0.01$. Given that on the baseline, an average county witnessed 0.03 protests every day during this period, such an effect is equivalent to a 0.12 percent decrease in protest gatherings.

The significance of our first-stage estimates is sensitive to the choice of the functional form of the instrumental variable. The exact data generation process determining how rainfall affects protest gatherings remains unknown. Therefore, we also try all these instruments in Columns (1)-(4) of  Table \ref{tab:iv_1st_rain} combined together to maximize the relevance of the first stage, when we perform IV estimations.

\textbf{Exogeneity Check:} The key identification assumption is that precipitation is uncorrelated with other determinants of fundraising outcomes for Black people, such as the local economic and social conditions. To validate this assumption, we use the pre-BLM donation records to examine whether rainfall occurrence per se systemically affects donations received by Black people. Table \ref{tab:iv_exogeneity_rain} reports estimates of regressions examining whether precipitation predicts local fundraising gaps, conditional on county rainy propensity and date-fixed effects. Our results show that rainfall occurrence does not explain the variations in fundraising, fundraising for Black people, nor the racial disparity.\footnote{On the other hand, counties with larger rainfalls on average tend to be associated with a larger racial gap in fundraising. }

Next, we consider the second stage regression that examines how local protest rallies causally affect the fundraising disparity between one county’s Black and non-Black beneficiaries. 
\begin{equation}\label{eq:iv_rain_main}
    y_{jct} = \gamma_0 \cdot protest_{ct}\times Black_j   + Black_j \times \theta_c + Black_j \times \mu_t+ \gamma_2 protest_{ct} +
\end{equation}
\[\theta_c + \mu_t + B X_{jt}+\epsilon_{jct}\]
We estimate the Equation using dates after May 27, 2020.   Here we are interested in $\hat{\gamma}_0$, which represents the effect of protest rallies on the donation gap. We include controls of local rainy propensity $\theta_c$ and global rainy trends $\mu_t$ as well as their interaction with the Black indicator. The interaction terms are essential to capture the unobserved local heterogeneity in biased fundraising, as well as the effect of the global protests.\footnote{We show that without such interaction terms as controlling covariates, there is an overestimation of the true causal impact of local protests.} Similar to previous sections, we also include project-level covariates (goals, project text descriptions, picture descriptions) as a control.

Table \ref{tab:iv_main_rain} presents the second-stage IV estimates  $\hat{\gamma_0}$ as well as the OLS regression. Column (1) shows the estimate using the log of the average rainfall precipitation of each county as an instrument, and Column (2) combines all instrumental variables. Both results show that the effect of experiencing any protest occurrence within a county has an insignificant effect on the fundraising received by Black beneficiaries. The point estimate suggests 0.87 extra daily donors for Black beneficiaries per local protest rally. Compared with the OLS estimate in Column (3), the IV estimate presents a much larger estimate. This aligns with the endogeneity problem whereby the counties with more protest rallies have already made substantial donations to Black beneficiaries, and therefore residents there respond less to the protest rallies.

\subsection{Spillover Amplifier and the Role of Social Media}
In an era where social media wields an increasingly pivotal role, comparing the relevance of local rallies and global protest gatherings can help us understand how social media shapes the effect of social movements. Our proposed identification methods from previous sections provide a good way of examining the role of social media in the spread of social movements.

Our previous results show that experiencing one local protest increases the fundraising for Black beneficiaries by 0.8 donors, while an extra global protest gathering increases the fundraising by 0.0027 donors. This implies that ``seeing" a protest gathering is more effective than ``hearing about" a protest gathering from somewhere else in the U.S. This is reasonable given the large geographical distance and information barrier that limits the degree of spillover. 

However, this does not mean that experiencing local protest rallies is any more relevant than experiencing nationwide protests when it comes to explaining how the BLM movements affect donation behavior. To calculate the relevance, we must acknowledge that an average county witnesses about 0.03 local protest gatherings per day. Thus, the additional donations received by Black beneficiaries owing to local protests amount to 0.87 times 0.03, approximately 0.026. On the other hand, there are an average of 101 nationwide protests each day during our period of interest. Therefore, the additional donations received by Black beneficiaries due to aggregate protests across the U.S.  amount to 101 times 0.0027, which is approximately 0.273. Subsequently, the local rallies account for 0.026/0.273= 10\% of relevance while the global effects make up the remaining 90\%.

To understand the degree of spillover, we define a spillover parameter as the following wedge between global effects and local effects.

\[Spillover = \frac{\# Protests \ in \ U.S \times \mathit{Effect} \ \mathit{of} \ Global \ Protests}{\# Local \ Protests \times \mathit{Effect} \ \mathit{of} \ Local \ Protests} - 1\]
This parameter reflects what the relative size of the overall impact of BLM protests v.s. local rallies would be if each county was a disconnected island from the rest. A measurement of 0 indicates that there is no spillover effect. Under our estimates, we find that the spillover parameter is 10. Therefore, for an average county, the impact on donation giving of protests elsewhere across the U.S. is ten times as large as the impact of local rallies. 

\subsection{The role of social media}
 This section explores the importance of different social media platforms in conveying the effects of the BLM protests beyond one's original location. 
We split counties into two groups by the median of social media exposure intensity, based on three different measurements:(1) Google search index on “BLM” or racial issues\footnote{We also include ``protest" ``Floyd" ``racial justice" as part of our keywords.}; (2) newspaper coverage on “BLM” or racial issues; (3) constructed county exposure to global BLM protests using Facebook connectivity.\footnote{Except for unavailability of Twitter data, these are main-stream media channels that people can obtain news from.}   In dividing our groups, we condition on deciles of the county average racial gap (prior to BLM) in order to exclude the possible path dependence or income effects. We illustrate more details on these measures in our Appendix \ref{sec_app_socialmedia}. Here, we detail the process of constructing the county exposure to BLM protests using Facebook connectivity.

We use the social connectivity index $SCI_{c,c^{'}}$ from \cite{bailey2018social}, which measures the probability of a Facebook friendship link between a given Facebook user in county $c$ and a given user in county $c^{'}$. The local exposure to global protests $Fcp_c$ is then constructed by using this connectivity index as a weight to calculate the exposure of a county to protests from all other counties.
$$Fcp_c = \sum_{c^{'}}  SCI_{c,c^{'}} \cdot protest_{c^{'}}$$
This measure is de facto a Bartick instrument and reflects whether socially connecting to places where more protests happened will increase the protest effect. 

Table \ref{tab:iv_socialmedia} presents the estimates of the weekend IV approach, the corresponding spillover multiplier, and the relevance of global gatherings, by the measure of social media exposure. Overall, results show that counties exposed to higher social media reports on ``BLM" will experience a significantly higher response to global protests, and be associated with a much larger spillover multiplier. A comparison between Column (3) and Columns (1) and (2) implies that both online media and traditional newspapers are important. 

Moreover, in Column (4), we examine whether counties with higher internet/digital devices accessibility\footnote{This comes from ACS 2020, which is imputed for each zip code based on the survey question ``has one or more types of computing devices?".} will experience a larger spillover effect from the global protests. In contrast to social media reports on ``BLM", we find that internet/digital device accessibility barely makes any difference. Instead of contradicting our previous findings, this implies that the ``BLM" movement information and the county reports actively broadcast by social media, are of greatest relevance. Although the ability to access information on a given matter is important, without exposure to specific ``BLM" information or racial justice discussion, people will not have a higher level of motivation to donate to Black charities. 


\section{Conclusion} \label{sec_Conclusion}
This study shows that racial justice movements can economically help disadvantaged races by motivating charitable donations from other races, and increasing inclusiveness of the social safety net. This indicates a strong discrimination-debiasing role of the social movements in the set-up of the U.S. where extremely polarized views on race and inequality persist. 
 
Protests or rallies per se matter during this movement. Especially when occurs massively over the country, the effectiveness of a single local protest rally will be amplified.  Social media plays an essential role in amplifying and conveying the meaning of a decentralized assembly. This indicates that as central pillars of democracy, freedom of speech and assembly are complementary to one another. 

The effectiveness of social movements fades over time as the protests cease. It is unsurprisingly consistent with the fact that numerous affirmative and equal opportunity legislations introduced since 1964 have had a mixed and limited effect. People’s racial stereotypes towards Black people, rooted in the legacy of prejudice, segregation, and discrimination, cannot be drastically altered. While it sounds discouraging, a potential hope to address racial disparity might be to consistently keep the public informed and aware of the issue of racial discrimination. 

To reach these lessons, the high-frequency features of data are needed. It not only helps establish compelling causality, but also provides grounds for implementing a variety of novel methodologies to study the nature of protest gatherings. We believe it to be of general use to the field study of social movements. 

Our findings lead to several research questions: (1) How do protest gatherings perpetuate themselves and spill over from the large metropolitan regions to marginal regions? (2) How do we quantify and model the simultaneity between social media coverage and protests? (3) Given that protest gatherings might be detrimental to the economy and safety of society while anti-racism discussions and articles are beneficial, how could government and social planners exploit the power of social media and minimize the side effects of protest rallies?  Answering these questions will deepen our understanding of the functions and institutions that both sustain democracy and lay the constructive foundation for a climate of equal opportunity.

\newpage
\printbibliography

@article{williams_racial_2015,
  title={Racial bias in health care and health: challenges and opportunities},
  author={Williams, David R and Wyatt, Ronald},
  journal={Jama},
  volume={314},
  number={6},
  pages={555--556},
  year={2015},
  publisher={American Medical Association}
}

@article{lillie-blanton_role_2005,
	title = {The {Role} {Of} {Health} {Insurance} {Coverage} {In} {Reducing} {Racial}/{Ethnic} {Disparities} {In} {Health} {Care}},
	volume = {24},
	issn = {0278-2715, 1544-5208},
	number = {2},
	journal = {Health Affairs},
	author = {Lillie-Blanton, Marsha and Hoffman, Catherine},
	year = {2005},
	pages = {398--408}
}

@article{carratala_health_2020,
  title={Health disparities by race and ethnicity},
  author={Carratala, Sofia and Maxwell, Connor},
  journal={Center for American Progress},
  volume={7},
  year={2020}
}

@article{nelson_unequal_2002,
  title={Unequal treatment: confronting racial and ethnic disparities in health care.},
  author={Nelson, Alan},
  journal={Journal of the national medical association},
  volume={94},
  number={8},
  pages={666},
  year={2002},
  publisher={National Medical Association}
}

@article{yearby_racial_2018,
  title={Racial disparities in health status and access to healthcare: the continuation of inequality in the United States due to structural racism},
  author={Yearby, Ruqaiijah},
  journal={American Journal of Economics and Sociology},
  volume={77},
  number={3-4},
  pages={1113--1152},
  year={2018},
  publisher={Wiley Online Library}
}

@article{orsi_blackwhite_2010,
  title={Black--white health disparities in the United States and Chicago: a 15-year progress analysis},
  author={Orsi, Jennifer M and Margellos-Anast, Helen and Whitman, Steven},
  journal={American journal of public health},
  volume={100},
  number={2},
  pages={349--356},
  year={2010},
  publisher={American Public Health Association}
}

@article{buchmueller_effect_2016,
  title={Effect of the Affordable Care Act on racial and ethnic disparities in health insurance coverage},
  author={Buchmueller, Thomas C and Levinson, Zachary M and Levy, Helen G and Wolfe, Barbara L},
  journal={American journal of public health},
  volume={106},
  number={8},
  pages={1416--1421},
  year={2016},
  publisher={American Public Health Association}
}

@article{campbell_black_2021,
  title={Black Lives Matter’s effect on police lethal use of force},
  author={Campbell, Travis},
  journal={Journal of Urban Economics},
  pages={103587},
  year={2023},
  publisher={Elsevier}
}

@article{luo_scandal_2022,
  title={Scandal, social movement, and change: Evidence from\# MeToo in Hollywood},
  author={Luo, Hong and Zhang, Laurina},
  journal={Management Science},
  volume={68},
  number={2},
  pages={1278--1296},
  year={2022},
  publisher={INFORMS}
}

@article{levy_effects_2019,
  title={The effects of social movements: Evidence from\# MeToo},
  author={Levy, Ro'ee and Mattsson, Martin},
  journal={Available at SSRN 3496903},
  year={2023}
}

@article{andrews_legitimacy_2016,
  title={The legitimacy of protest: explaining White Southerners' attitudes toward the civil rights movement},
  author={Andrews, Kenneth T and Beyerlein, Kraig and Tucker Farnum, Tuneka},
  journal={Social Forces},
  volume={94},
  number={3},
  pages={1021--1044},
  year={2016},
  publisher={Oxford University Press}
}

@article{webb_hooper_covid-19_2020,
	title = {{COVID}-19 and {Racial}/{Ethnic} {Disparities}},
	volume = {323},
	issn = {0098-7484},
	number = {24},
	journal = {JAMA},
	author = {Webb Hooper, Monica and Nápoles, Anna María and Pérez-Stable, Eliseo J.},
	year = {2020},
	pages = {2466--2467}

}

@techreport{pennebaker_development_2015,
  title={The development and psychometric properties of LIWC2015},
  author={Pennebaker, James W and Boyd, Ryan L and Jordan, Kayla and Blackburn, Kate},
  year={2015}
}

@inproceedings{serengil_lightface_2020,
  title={Lightface: A hybrid deep face recognition framework},
  author={Serengil, Sefik Ilkin and Ozpinar, Alper},
  booktitle={2020 innovations in intelligent systems and applications conference (ASYU)},
  pages={1--5},
  year={2020},
  organization={IEEE}
}

@article{yang_benchmarking_2021,
  title={Benchmarking commercial emotion detection systems using realistic distortions of facial image datasets},
  author={Yang, Kangning and Wang, Chaofan and Sarsenbayeva, Zhanna and Tag, Benjamin and Dingler, Tilman and Wadley, Greg and Goncalves, Jorge},
  journal={The visual computer},
  volume={37},
  pages={1447--1466},
  year={2021},
  publisher={Springer}
}

@article{snow_cultural_2018,
  title={The cultural outcomes of social movements},
  author={Van Dyke, Nella and Taylor, Verta},
  journal={The Wiley Blackwell companion to social movements},
  pages={482--498},
  year={2018},
  publisher={Wiley Online Library}
}

@book{bosi_consequences_2015,
  title={The consequences of social movements},
  author={Bosi, Lorenzo and Giugni, Marco and Uba, Katrin},
  year={2016},
  publisher={Cambridge University Press}
}

@article{younkin_colorblind_2018,
  title={The colorblind crowd? Founder race and performance in crowdfunding},
  author={Younkin, Peter and Kuppuswamy, Venkat},
  journal={Management Science},
  volume={64},
  number={7},
  pages={3269--3287},
  year={2018},
  publisher={INFORMS}
}

@article{jenq_beauty_2015,
  title={Beauty, weight, and skin color in charitable giving},
  author={Jenq, Christina and Pan, Jessica and Theseira, Walter},
  journal={Journal of Economic Behavior \& Organization},
  volume={119},
  pages={234--253},
  year={2015},
  publisher={Elsevier}
}

@book{giugni_how_1999,
  title={How social movements matter},
  author={Giugni, Marco and McAdam, Doug and Tilly, Charles},
  volume={10},
  year={1999},
  publisher={U of Minnesota Press}
}

@article{gorbatai_gender_2015,
  title={The narrative advantage: Gender and the language of crowdfunding},
  author={Gorbatai, Andreea and Nelson, Laura},
  journal={Haas School of Business UC Berkeley. Research Papers},
  pages={1--32},
  year={2015}
}

@article{chowkwanyun_racial_2020,
	title = {Racial {Health} {Disparities} and {Covid}-19 --- {Caution} and {Context}},
	volume = {383},
	issn = {0028-4793, 1533-4406},
	number = {3},
	journal = {New England Journal of Medicine},
	author = {Chowkwanyun, Merlin and Reed, Adolph L.},
	year = {2020},
	pages = {201--203}
}

@article{agarwal_antiracist_2022,
  title={Antiracist curriculum and digital platforms: Evidence from Black Lives Matter},
  author={Agarwal, Saharsh and Sen, Ananya},
  journal={Management Science},
  volume={68},
  number={4},
  pages={2932--2948},
  year={2022},
  publisher={INFORMS}
}

@article{wing2013strengthening,
  title={Strengthening the regression discontinuity design using additional design elements: A within-study comparison},
  author={Wing, Coady and Cook, Thomas D},
  journal={Journal of Policy Analysis and Management},
  volume={32},
  number={4},
  pages={853--877},
  year={2013},
  publisher={Wiley Online Library}
}

@article{reny2021opinion,
  title={The opinion-mobilizing effect of social protest against police violence: Evidence from the 2020 George Floyd protests},
  author={Reny, Tyler T and Newman, Benjamin J},
  journal={American political science review},
  volume={115},
  number={4},
  pages={1499--1507},
  year={2021},
  publisher={Cambridge University Press}
}

@article{moreland2020stay-at-home-timing,
  title={Timing of state and territorial COVID-19 stay-at-home orders and changes in population movement—United States, March 1--May 31, 2020},
  author={Moreland, Amanda and Herlihy, Christine and Tynan, Michael A and Sunshine, Gregory and McCord, Russell F and Hilton, Charity and Poovey, Jason and Werner, Angela K and Jones, Christopher D and Fulmer, Erika B and others},
  journal={Morbidity and Mortality Weekly Report},
  volume={69},
  number={35},
  pages={1198},
  year={2020},
  publisher={Centers for Disease Control and Prevention}
}

@article{perry2021SHO-inequality-pandemic,
  title={Pandemic precarity: COVID-19 is exposing and exacerbating inequalities in the American heartland},
  author={Perry, Brea L and Aronson, Brian and Pescosolido, Bernice A},
  journal={Proceedings of the National Academy of Sciences},
  volume={118},
  number={8},
  pages={e2020685118},
  year={2021},
  publisher={National Acad Sciences}
}

@article{montenovo2022SHO-inequality-determinants,
  title={Determinants of disparities in early COVID-19 job losses},
  author={Montenovo, Laura and Jiang, Xuan and Lozano-Rojas, Felipe and Schmutte, Ian and Simon, Kosali and Weinberg, Bruce A and Wing, Coady},
  journal={Demography},
  volume={59},
  number={3},
  pages={827--855},
  year={2022},
  publisher={Duke University Press}
}

@techreport{smith2016shared,
    author={Aaron Smith},
    title={Shared, collaborative and on demand: The new digital economy},
    institution = {Pew Research Center},
    year= {2016}
}

@article{madestam2013qje_teaparty_gather,
  title={Do political protests matter? evidence from the tea party movement},
  author={Madestam, Andreas and Shoag, Daniel and Veuger, Stan and Yanagizawa-Drott, David},
  journal={The Quarterly Journal of Economics},
  volume={128},
  number={4},
  pages={1633--1685},
  year={2013},
  publisher={MIT Press}
}

@article{engist2022_BLM_local_voteregistra,
  title={Do political protests mobilize voters? Evidence from the Black Lives Matter protests},
  author={Engist, Oliver and Schafmeister, Felix},
  journal={Public Choice},
  volume={193},
  number={3-4},
  pages={293--313},
  year={2022},
  publisher={Springer}
}

@article{chilton2020political,
  title={Political ideology and the law review selection process},
  author={Chilton, Adam and Masur, Jonathan and Rozema, Kyle},
  journal={American Law and Economics Review},
  volume={22},
  number={1},
  pages={211--240},
  year={2020},
  publisher={Oxford University Press}
}

@article{bailey2018social,
  title={Social connectedness: Measurement, determinants, and effects},
  author={Bailey, Michael and Cao, Rachel and Kuchler, Theresa and Stroebel, Johannes and Wong, Arlene},
  journal={Journal of Economic Perspectives},
  volume={32},
  number={3},
  pages={259--280},
  year={2018},
  publisher={American Economic Association 2014 Broadway, Suite 305, Nashville, TN 37203-2418}
}

\renewcommand \thesection{\Roman{section}.}
\renewcommand \thesubsection{\Roman{section}.\arabic{subsection}}
\renewcommand\thesubsubsection{\Roman{section}.\arabic{subsection}.\arabic{subsubsection}}

\setcounter{section}{0}
\setcounter{subsection}{0}
\setcounter{subsubsection}{1}

\newpage

\newgeometry{margin=2.6cm}

\begin{landscape}
\section{Figures}
\begin{figure}[htb]
    \centering
    \caption{A Example of Medical Fundraising Project in GoFundMe}
    \includegraphics[width=\textwidth]{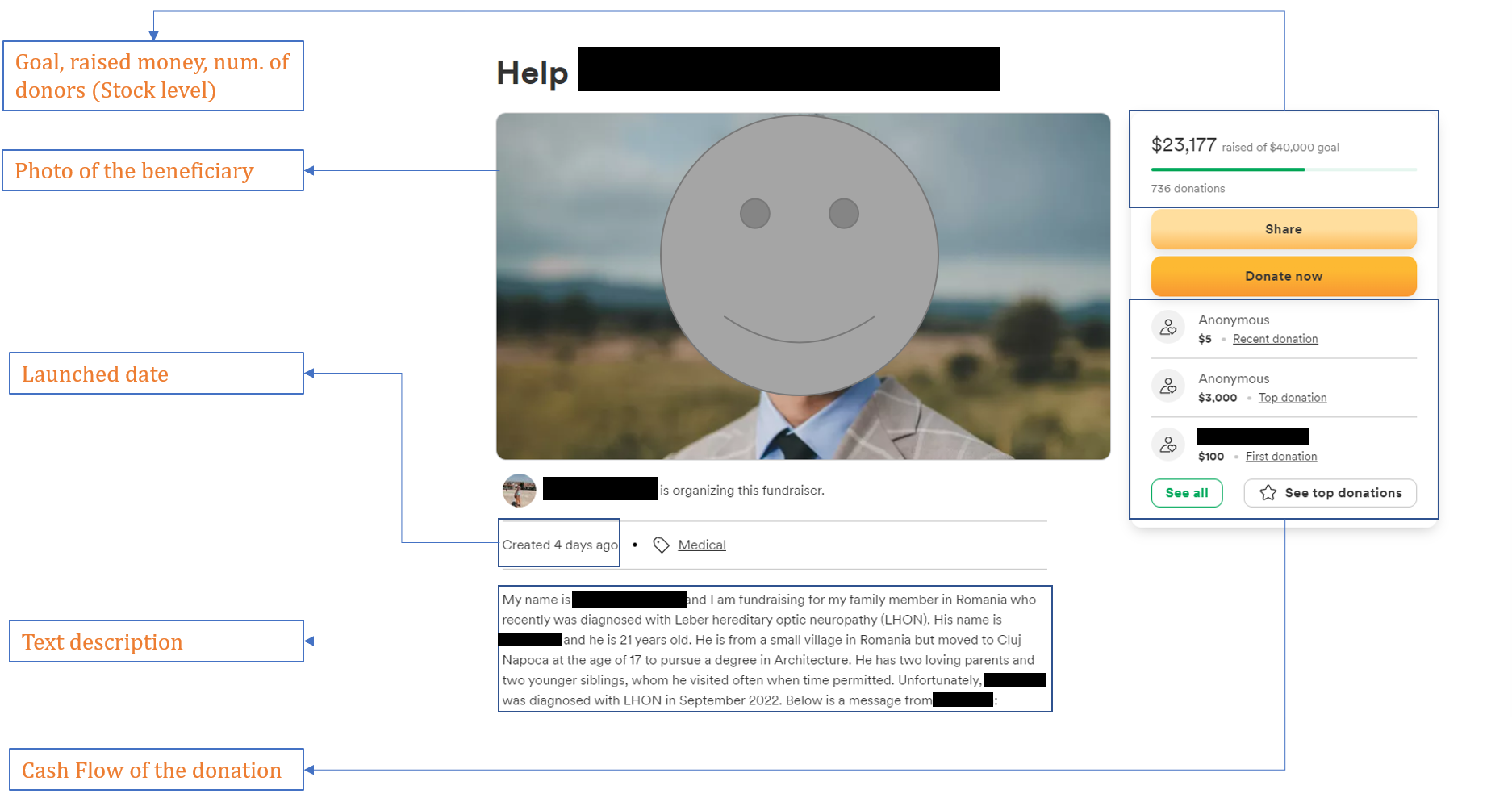}
    \vspace{1em}
    \label{fig_eg}
    \begin{minipage}[0.1cm]{1\textwidth}
    \small \textit{Note:} Figure \ref{fig_eg} presents a representative profile of a medical crowdfunding project  (recognizable faces and names are masked to preserve privacy). It displays the stock variables: the fundraising goal, the amount of money raised, and the number of donors. It also shows the flow of donations to this project. Additionally, the static characteristics of the project, including the launch date, text description, and a photo of the beneficiary, are also observable.
\end{minipage}
\end{figure}
    
\end{landscape}
\newpage

\newpage

\newpage

\newpage

\newpage

\newpage

\newpage

\restoregeometry
\newpage

\newpage

\section{Tables}
\begin{table}[htb]
\centering
\caption{Summary Statistics of Medical Crowdfunding Records}
\label{tab_su_GoFundMe}
\begin{tabular}{lrrrr}
\toprule
N = 362,953 & Mean & SD & Min & Max \\
\midrule
\multicolumn{5}{l}{\textit{\textbf{Panel A: Fundraising Outcomes}}} \\
Fund Goal (\$) & 12658.200 & 18103.780 & 250 & 160000 \\
Raised Money (\$) & 3354.647 & 5318.831 & 30 & 42670 \\
Raised Money/Goal & 0.384 & 0.413 & 0.002 & 3.126 \\
Fully or Over Funded & 0.101 & 0.301 & 0 & 1 \\
Average Number of Donors & 38.262 & 53.528 & 2 & 431 \\
Average Donation (\$) & 78.973 & 46.013 & 10 & 314.737 \\
 \midrule
\multicolumn{5}{l}{\textit{\textbf{Panel B: Racial Information in Photos}}} \\
Whether Include Black Faces & 0.101 & 0.301 & 0 & 1 \\
Total Faces & 1.996 & 1.963 & 1 & 20 \\ \midrule
\multicolumn{5}{l}{\textit{\textbf{Panel C: Characteristics of Project Description}}} \\
Total Words & 1561.980 & 1486.022 & 0 & 32434 \\
Positive Index & 3.726 & 2.291 & 0 & 100 \\
Negative Index & 1.556 & 1.283 & 0 & 100 \\
Authenticity Index & 20.020 & 26.343 & 0 & 99 \\
Male Index & 2.183 & 2.885 & 0 & 50 \\
Female Index & 2.348 & 3.074 & 0 & 33.330 \\
\midrule
\bottomrule
\end{tabular}
    \begin{minipage}[0.1cm]{1\textwidth}
    \vspace{1em}
    \scriptsize \textit{Note:} Table \ref{tab_su_GoFundMe} summarizes the crowdfunding data. We observe that 362,953 medical crowdfunding projects launched between Jan 2019 and Mar 2021 in the US. Panel A describes the eventual outcomes of these projects. On average, each project requires 12,659\$ and raised 3,355\$ finally; About 10\% of projects achieve their goal; Each project has received 38 donations on average and the mean value of the donation is 79\$. Panel B describes the racial information extracted from the beneficiaries' photos from each project's profile. Our algorithm can tell how many human faces are in one photo and how many of them are black faces. On average, each profile's photo has 2 faces and about 10\% of profiles contains black people's face. Panel C describes the characteristics of the text description of each profile. On average, people write 1,562 words to describe their situation. We also use LIWC (Linguistic Inquiry and Word Count) to quantify the other linguistic characteristics of the text description. The LIWC grades the text description by giving index (0 to 100) of the paragraph's positive/negative emotion, authenticity and tendency of gender.      
    \end{minipage}
\end{table}

\newpage
\begin{table}[p]
    \centering
    \caption{The Average Reduction of Racial Gap in Raised Funds before/after January 2020}
    \begin{tabular}{l*{3}{c}}
\hline\hline

          & (1)        & (2)    &     (3)         \\
\hline

& \multicolumn{3}{c}{\textit{Panel A}:\quad $y$ = Raised Funds (\$)}\\
\cmidrule{2-4}

$Black_j$ &   $-545.5^{***}$&   $-608^{***}$&   $-587^{***}$ \\
          &  (38.8)         &  (36.9)         &  (39.3)             \\
[1em]
$Black_j \times \textbf{1}(t>Jan. 2020)$&    $245^{***}$&    $146^{***}$&    $146^{***}$\\
          &  (0.005)         &  (0.005)         &  (0.004)             \\
\hline
& \multicolumn{3}{c}{\textit{Panel B}:\quad  $y$ = ln Raised Funds}\\
\cmidrule{2-4}
$Black_j$    &   $-0.237^{***}$&   $-0.243^{***}$&   $-0.234^{***}$\\
          & (-18.35)         & (-20.63)         & (-19.04)         \\
[1em]
$Black_j \times \textbf{1}(t>Jan. 2020)$&    $0.147^{***}$&    $0.118^{***}$&    $0.119^{***}$\\
          &   (8.23)         &   (7.33)         &   (7.20)         \\
\hline

Description \& Goal          &              &      Yes         &      Yes         \\

State FE \& Date FE         &              &      Yes         &      Yes         \\

Zip FE            &               &               &      Yes         \\      \\
[1em]
Observations     &   333106         &   332278         &   326853         \\
\hline\hline
\multicolumn{4}{l}{\footnotesize \textit{t} statistics in brackets}\\
\multicolumn{4}{l}{\footnotesize $^{*}$ $p<0.05$, $^{**}$ $p<0.01$, $^{***}$ $p<0.001$}\\
\end{tabular}
    \label{tab2}
    \vspace{1em}
        \begin{minipage}[0.1cm]{1\textwidth}
         \small \textit{Note:} Table \ref{tab2} reports the estimation results of the following diff-in-diff regression Equation 
        \begin{equation*}
    y_{j,t} = \beta_0 + \beta_1\textbf{1}(t>Jan. 2020)\times Black_{j} + \beta_2Black_j + X_{j}\Gamma + \delta_{t} + \sigma_{s} + \varepsilon_{j,t}
\end{equation*}
 $y_{j,t}$ is the outcome of project $j$ launching at date $t$. $\textbf{1}(t>Jan. 2020) = 1$ if the launching date $t$ is later than January, 2020. $Black_j = 1$ if project $j$ contains any black beneficiaries. $X_j$ is the control variables, including project $j$'s goal setting and features of its text description. $\delta_t$ is the month fixed effect. $\sigma_s$ is the state fixed effect of the beneficiaries' geographical location. \\
         
 Panel A reports the results when outcomes is the absolute value of the raised money and Panel B reports the estimates using the outcomes' log value. Column (1) presents the results without any controls; column (2) presents the results after controlling for goal, features of text description, state fixed effect and date fixed effect; column (3) in addition controls the zip code where the beneficiaries live. From panel A and B we learn that there is an initial racial gap in the amount of the eventually raised money between black and non-black people's projects: black people tend to raise 234\$ to 243\$ (6\% to 7\%) less than non-black people. However, we also learn that this racial gap significantly decreases for projects starting after January 2020. The racial gap decreases by 118\$ to 147\$ (1.9\% to 2.3\%).
    \end{minipage}
\end{table}

\newpage
\begin{table}[p]
    \centering
    \caption{DID estimation of the effect of the surge of BLM}
    \begin{tabular}{lcccc}
\hline \hline
 & \multicolumn{4}{c}{$y$ = Num. Daily Donation} \\ \cline{2-5} 
 & (1) & (2) & (3) & (4) \\ \cline{2-5} 
$Black_j$ & 0.093 & 0.084 & 0.056 &  \\
 & (1.42) & (1.29) & (0.82) &  \\
 &  &  &  &  \\
$Black_j\times   \textbf{1}(t\geq t_{May28})$ & 0.406*** & 0.419*** & 0.413*** & 0.313*** \\
 & (5.28) & (5.46) & (4.29) & (2.65) \\
 &  &  &  &  \\
$ln(Goal)_j$ & 0.597*** & 0.602*** & 0.602*** &  \\
 & (61.48) & (61.89) & (61.87) &  \\
 &  &  &  &  \\
$1(Covid)_j$ &  & 0.438*** & 0.438*** &  \\
 &  & (15.58) & (15.56) &  \\
 &  &  &  &  \\
 \hline
Description & Yes & Yes & Yes &  \\
Date FE & Yes & Yes & Yes &  \\
State FE & Yes & Yes & Yes &  \\
Project launch-date FE & Yes & Yes & Yes &  \\
$Black_j \times SAH_{st}$ &  &  & Yes & Yes  \\
Project FE &  &  &  & Yes \\
 &  &  &  &  \\
Control group mean value & 3.961 & 3.961 & 3.961 & 3.961 \\
Num. projects & 71488 & 71488 & 71488 & 71488 \\
Observations & 605888 & 605888 & 605888 & 605888 \\ \hline \hline
t statistics in brackets &  &  &  &  \\
\multicolumn{3}{l}{$^{*} p <   0.10$ $^{**} p < 0.05$  $^{***} p   < 0.01$} & \multicolumn{1}{l}{} & \multicolumn{1}{l}{}
\end{tabular}

    \label{tab:reg_mainDID}
    \begin{minipage}[0.1cm]{1\textwidth}
    \vspace{1em}
    \small \textit{Note:} Table \ref{tab:reg_mainDID} presents the results of the estimation of Equation \eqref{eq:main}. Columns 1 and 2 report the main specifications, with column 2 controlling for the fixed effects of COVID-19 related projects. Column 3 includes controls for the interaction between the indicator for Black individuals and the state-level Stay-at-Home (SAH) shock time dummy (marking the start and end of SAH orders). Column 4 accounts for project fixed effects. Throughout the main study period, 71,488 projects were actively receiving donations. Prior to the treatment period, projects related to non-Black individuals received on average 3.961 donations per day.   
    \end{minipage}
\end{table}

\newpage
\begin{landscape}

\begin{table}[p]
    \centering
    \caption{Donor Decomposition by Race}
    \scalebox{0.9}{
    \begin{tabular}{lccc}
\hline \hline
$y$ & Num. Daily Donation & Expected Num. of Black Donors & Expected Num. of non-Black Donors \\ \cline{2-4} 
 & (1) & (2) & (3) \\ \cline{2-4} 
$Black_j$ & 0.056 & 0.175*** & -0.119* \\
 & (0.82) & (18.06) & (-1.95) \\
 &  &  &  \\
$Black_j\times   \textbf{1}(t\geq t_{May28})$ & 0.413*** & 0.016 & 0.397*** \\
 & (4.29) & (1.16) & (4.68) \\
 &  &  &  \\
 \hline
Description \& Goal & Yes & Yes & Yes \\
Date FE & Yes & Yes & Yes \\
State FE & Yes & Yes & Yes \\
Project launch-date FE & Yes & Yes & Yes \\
$Black_j \times SAH_{st}$ & Yes & Yes & Yes \\
 &  &  &  \\
Mean: Black projects $(t\leq \text{May 27})$ & 3.955 & 0.572 & 3.383 \\
Mean: non-Black projects $(t\leq \text{May 27})$ & 3.968 & 0.395 & 3.572 \\
Num. projects & 71488 & 71488 & 71488 \\
Observations & 605888 & 605888 & 605888 \\
\hline \hline
\multicolumn{3}{l}{t statistics   in brackets} & \multicolumn{1}{l}{} \\
$^{*} p <   0.10$ $^{**} p < 0.05$  $^{***} p   < 0.01$ & \multicolumn{1}{l}{} & \multicolumn{1}{l}{} & \multicolumn{1}{l}{}
\end{tabular}

    }
    \label{tab:reg_race_decomposse}
    \begin{minipage}[0.1cm]{1\textwidth}
         \vspace{1em}
         \small \textit{Note}: Table \ref{tab:reg_race_decomposse}  provides estimations of Equation \eqref{eq:race_Black_effect} in column 2, and equation \eqref{eq:race_nonBlack_effect} in column 3. For comparison, equation \eqref{eq:main} is presented in column 1. Before the BLM movement, a project related to Black individuals was expected to receive 0.175 more donations from Black donors but 0.119 fewer donations from non-Black donors, compared to a project related to non-Black individuals. However, the BLM movement resulted in an increase of 0.413 donations per day to Black-related projects, with 0.397 of those donations coming from non-Black donors. Prior to the treatment, Black-related projects received 0.572 daily donations from Black donors and 3.383 from non-Black donors. In contrast, non-Black related projects received 3.95 daily donations from Black donors and 3.572 from non-Black donors.
    \end{minipage}
\end{table}
\end{landscape}

\newpage
\begin{table}[p]
    \centering
    \caption{Heterogeneity Analysis across Counties of Different Local Protest Intensity}
    \scalebox{0.85}{
    \begin{tabular}{@{}lccc@{}}
\hline \hline
 &    \multicolumn{3}{c}{ $y = $ Num. Daily Donation} \\ \cmidrule(l){2-4} 
 & (1) & (2) & (3) \\ \cmidrule(l){2-4} 
$\textbf{1}(t\geq t_{May28})\times Black_j   \times Q_1$ & 0.489** & 0.136 & -0.097 \\
 & {[}2.20{]} & {[}0.61{]} & {[}-0.40{]} \\
 &  &  &  \\
$\textbf{1}(t\geq t_{May28})\times Black_j   \times Q_2$ & 0.710*** & 0.615*** & 0.386* \\
 & {[}3.41{]} & {[}2.85{]} & {[}1.76{]} \\
 &  &  &  \\
$\textbf{1}(t\geq t_{May28})\times Black_j   \times Q_3$ & 0.280 & 0.167 & -0.006 \\
 & {[}1.30{]} & {[}0.76{]} & {[}-0.03{]} \\
 &  &  &  \\
$\textbf{1}(t\geq t_{May28})\times   Black_j$ & 0.103 & 0.418** & 0.459*** \\
 & {[}0.62{]} & {[}2.45{]} & {[}2.69{]} \\
 &  &  &  \\
$\textbf{1}(t\geq t_{May28})\times   Q_1$ & -0.563*** & -0.552*** &  \\
 & {[}-7.93{]} & {[}-7.76{]} &  \\
 &  &  &  \\
$\textbf{1}(t\geq t_{May28})\times   Q_2$ & -0.470*** & -0.467*** &  \\
 & {[}-6.47{]} & {[}-6.40{]} &  \\
 &  &  &  \\
$\textbf{1}(t\geq t_{May28})\times   Q_3$ & -0.216*** & -0.210*** &  \\
 & {[}-2.73{]} & {[}-2.65{]} &  \\
 &  &  &  \\
$ Black_j \times Q_1$ & -0.209 &  &  \\
 & {[}-1.11{]} &  &  \\
 &  &  &  \\
$ Black_j \times Q_2$ & -0.501*** &  &  \\
 & {[}-2.85{]} &  &  \\
 &  &  &  \\
$ Black_j \times Q_3$ & -0.334* &  &  \\
 & {[}-1.83{]} &  &  \\
 &  &  &  \\
 \hline
 Description \& Goal & Yes & Yes & Yes \\
County & Yes & Yes & Yes \\
County $\times$ Black  &  & Yes & Yes \\
County $\times \textbf{1}(t\geq t_{May28})$  &  &  & Yes \\
 &  &  &  \\
Observations & 605888 & 605888 & 605888 \\ \hline \hline
t statistics   in brackets &  &  &  \\
*p\textless{}0.10 ** p\textless{}0.05 *** p\textless{}0.01 &  &  &  \\
 &  &  & 
\end{tabular}
    }
    \label{tab:reg_county_DDD}
    \begin{minipage}[0.1cm]{1\textwidth}
         \vspace{1em}
         \small \textit{Note}: Table \ref{tab:reg_county_DDD} shows the results of the estimation from Equation \eqref{eq:DDD_county}, with various geographical fixed effects specifications. The projects in the counties with the highest number of gatherings (\textit{Quantile} 4) serve as the base level. The coefficients of $\textbf{1}(t\geq t_{May28})\times Black_{j}\times Q_{k}$, where $k = 1,2,3$, are the additional impact experienced by Black beneficiaries' projects beyond the overall effect represented by $\textbf{1}(t\geq t_{May28})\times Black_{j}$. The results show that Black beneficiaries' projects in counties of \textit{Quantile} 2 (with 4 - 18 gatherings) experienced the most substantial impact from the BLM movement.
    \end{minipage}
\end{table}

\newpage
\begin{table}[p]
    \centering
    \caption{OLS Estimates: Effect of Local Gatherings and National Protest Intensity}
    \scalebox{0.85}{
    \begin{tabular}{@{}lccccc@{}}
\hline \hline
 & \multicolumn{5}{c}{$y$ =   Num. Daily Donation} \\ \cmidrule(l){2-6} 

 & (1) & (2) & (3) & (4) & (5) \\ \cmidrule(l){2-6} 
$Black_j \times protest_{ct}$ & 0.412** & -0.54 & -0.46 & 0.210 & 0.183*** \\
 & (0.19) & (0.89) & (0.33) & (0.139) & (0.064) \\
 &  &  &  &  &  \\
$Black_j \times Protest_{-ct}$ & 0.0023*** & 0.0032* & 0.0035*** & 0.0016 & 0.002* \\
 & (0.0005) & (0.0015) & (0.00132) & (0.0013) & (0.0011) \\
 &  &  &  &  &  \\

 \hline
 Description \& Goal & Yes & Yes & Yes & Yes & Yes \\
County & Yes & Yes & Yes & Yes & Yes \\
County $\times$ Black & Yes & Yes & Yes & Yes & Yes \\
County $\times \textbf{1}(t\geq t_{May28})$ & Yes & Yes & Yes & Yes & Yes \\
 &  &  &  &  &  \\
Sample & Full & Quantile 1 & Quantile 2 & Quantile 3 & Quantile 4 \\
Observations & 605888 & 155369 & 178365 & 125383 & 146698 \\ 
\\

\hline \hline
t statistics in brackets &  &  &  &  &  \\
*   p\textless{}0.10  ** p\textless{}0.05  *** p\textless{}0.01 &  &  &  &  & 

\end{tabular}
    }
    \label{tab:reg_localprotest_OLS}
    \begin{minipage}[0.1cm]{1\textwidth}
         \vspace{1em}
         \small \textit{Note}: Table \ref{tab:reg_localprotest_OLS} presents the OLS estimates of Equation \eqref{eq:reg_localgather_OLS1} for measuring the effect of local protest rallies and the effect of the global national protests. Column (1) presents the results with a full sample. Column (2) - (5) presents the estimates for quantiles 1-4 respectively. Overall results suggest that for places with a moderate or little number of protests happening (\textit{Quantile} 1 and 2), the national protests matter more; while for places with a majority of protests happening (\textit{Quantile} 2 and 3), their own local protests matter.  
    \end{minipage}
\end{table}

\newpage

\begin{table}[p]
\centering
\caption{First Stage: Weekend Instrumental Variable on Protest Gatherings}
\label{tab:iv_1st_week}

\begin{tabular}{lcc}
\hline
\multicolumn{1}{l}{$y$} &\multicolumn{2}{c}{Aggregate Number of Protest Gathering At Date $t$}    \\ \cline{2-3}
& (1) & (2)\\
\cline{2-3}
\multicolumn{1}{l}{$\hat{\theta}_0$} & 137.0***  &  117.9*** \\
\multicolumn{1}{l}{}     & (0.0038) &   (0.2377)\\ \hline
                Week FE &   & Yes \\
                  &   &  \\ 
                  Observations &  184 &  184\\ 
                 $F$-statistics & 21.01  & 14.46 \\ 

                $R^2$ &  0.3325 &  0.7960\\ 
                \hline

                \multicolumn{3}{l}{\footnotesize Standard errors in brackets} \\
\multicolumn{3}{l}{\footnotesize * \(p<0.10\), ** \(p<0.05\), *** \(p<0.01\)} \\
\end{tabular}

\begin{minipage}[0.1cm]{1\textwidth}
         \vspace{1em}
         \small \textit{Note}: Table \ref{tab:iv_1st_week} presents the results for first stage regression of Equation \eqref{eq:iv_weenkend_1st_stage}. Column (1) presents results without week fixed effects, by comparing the number of protest gatherings for weekends v.s. weekdays directly. Column (2) presents the results with week fixed effects, by comparing the number of protest gatherings within the same week.  
    \end{minipage}
\end{table}

\newpage

\begin{landscape}
 
\begin{table}[]
    \centering
    
\caption{Reduced-Form Estimates: Weekend Instrumental Variable on Fundarising}
\label{tab:iv_reduced_week}
\scalebox{0.9}{
   
    \begin{tabular}{lcccccccccccr}
    \hline\hline
    \multicolumn{1}{l}{$y$} & \multicolumn{3}{c}{Num. Daily Donation Gap } & & \multicolumn{3}{c}{Num. Daily Donation for Black}  & &\multicolumn{3}{c}{Num. Daily Donation for n-Black } & \\
    \cline{2-13}
        & &  (1) & (2) & & & (3) &  (4)& & &(5) &(6) & \\
          \cline{2-4}  \cline{6-8}  \cline{10-12} \\
     $\hat{\alpha}_0$   & &  0.393*** & 0.319*** & & & 0.254* &  0.176& & &-0.129*** &-0.142*** &\\
        & &  (0.143) & (0.142) & & & (0.136) &  (0.134)& & &(0.045) &(0.045) &\\
\\
        \hline
         Description \& Goal & &  Yes & Yes & & & Yes &  Yes& & &Yes &Yes &\\
       Week FE & &   & Yes & & &  &  Yes& & & &Yes &\\
        & &   &  & & &  &  & & & & &\\
       Observations & &  706,944 & 706,944 & & & 79,109 &  79,109& & &627,835 &627,835 &\\
       $R^2$ & &  0.0066 & 0.0067 & & &  0.0061 &  0.0111& & &0.0049 &0.0061 &\\
      
        \hline
        \hline
        \multicolumn{4}{l}{\footnotesize Standard errors in brackets}\\
\multicolumn{4}{l}{\footnotesize * \(p<0.10\), * \(p<0.05\), * \(p<0.01\)}\\
    \end{tabular}

   }
\begin{minipage}[0.1cm]{1\textwidth}
         \vspace{1em}
         \small \textit{Note}: Table \ref{tab:iv_reduced_week} presents the results for the reduced-form regression specified by Equation \eqref{eq:iv_weenkend}, aiming to investigate the effect of weekend instrument on the fundraising. Columns (1) and (2) present the estimates of regressions strictly following Equation \eqref{eq:iv_weenkend}, checking how the racial disparity changes in response to our weekend instrument. Column (3) (4) and column (5) (6) presents the direct reduced-form effect on the fundraising outcome directly for black beneficiaries and non-black beneficiaries. The regression specification of columns (3) (4)  and (5) (6)  is: $ y_{j,t} = \alpha_0  \textbf{1}(t\geq t_{May28}) \times \textbf{\textit{h}}_t  + \alpha_1 \textbf{\textit{h}}_t+ X_{jt}\Gamma + \mu_{week} + \epsilon_{jt}$ for $Black_j=1$  and $Black_j=0$ respectively.
\end{minipage}
\end{table}
   
\end{landscape}

\newpage

\begin{landscape}
 
\begin{table}[]
    \centering
    
\caption{IV Estimates: Global Protest Intensity on Fundraising Using Weekends as Instrument Variables}
\label{tab:iv_main_week}
\scalebox{0.9}{
       \begin{tabular}{lcccccccccccr}
    \hline\hline
    \multicolumn{1}{l}{$y$} & \multicolumn{3}{c}{Num. Daily Donation Gap } & & \multicolumn{3}{c}{Num. Daily Donation for Black}  & &\multicolumn{3}{c}{Num. Daily Donation for n-Black } & \\
    \cline{2-13}
        & &  (1) &  & & & (2) &  & & &(3) & & \\
          \cline{2-12} \\
     $\hat{(\alpha/\theta)}$   & &  0.0027*** &  & & & 0.0015&  & & &-0.0012*** & &\\
        & &  (0.0013) &  & & & (0.00125) &  & & &(0.00039) & &\\
\\
        \hline
         Description \& Goal & &  Yes &  & & & Yes &  & & &Yes & &\\
       Week FE & & Yes  &  & & &  Yes&  & & & Yes& &\\
      & &   &  & & &  &  & & & & &\\
       Observations & &  706,944 &  & & & 79,109 &  & & &627,835 & &\\
       
        \hline
        \hline
        \multicolumn{4}{l}{\footnotesize Standard errors in brackets}\\
\multicolumn{4}{l}{\footnotesize * \(p<0.10\), * \(p<0.05\), * \(p<0.01\)}\\
    \end{tabular}

   }
   \begin{minipage}[0.1cm]{1\textwidth}
         \vspace{1em}
         \small \textit{Note}: Table \ref{tab:iv_main_week} presents the IV estimates of the effect of global protest intensity on fundraising outcome when using weekend IV. Column (1) present the estimates of regressions strictly following Equations \eqref{eq:iv_weenkend} and \eqref{eq:iv_weenkend_1st_stage}, checking how the racial disparity changes in response to protest intensity. Column (2) and (3) present the causal effect of global protest intensity on black beneficiaries' fundraising outcome and non-black beneficiaries' respectively.  
    \end{minipage}
\end{table}
   
\end{landscape}

\newpage

\begin{landscape}
\begin{table}[]
    \centering
        \caption{First Stage: Rain Occurrences on Protest Gathering}
    \label{tab:iv_1st_rain}

    \begin{tabular}{lcccc}
    \hline\hline
     & \multicolumn{4}{c}{$y$ = Number of Local Protest Rallies}\\
    \cline{2-5}
    Treatment & log Rain  & Rain  & log Rain & Rain \\
     &  (County Average) & (County Average) & (Population Center) & (Population Center)\\
    \cline{2-5}
         &  (1) &  (2) & (3) & (4)\\
         \cline{2-5}
        $\hat{\theta}$    &  -.0037*** &  -.0002** & -.00075* & (4) -.00008 \\
             &  (.0005) &  (.000094) & (.00045) & (.000064)\\
             \hline
           $\theta_c$ &  Yes &  Yes & Yes & Yes\\
        $\mu_t$ &  Yes &  Yes & Yes & Yes\\
         &   &   &  & \\
        $t$-statistics&   -7.19 &  -2.21 & -1.67  & -1.25\\
       $F$-statistics &  51.67 &   4.89 & 2.78 & 1.57\\
        Obs &  297,696 &  297,696 & 297,696 & 297,696\\
        $R^2$ &  0.0080 &  0.008 & 0.0080 & 0.0080\\
     
\hline\hline
\multicolumn{4}{l}{\footnotesize Standard errors in brackets}\\
\multicolumn{4}{l}{\footnotesize * \(p<0.10\), * \(p<0.05\), * \(p<0.01\)}\\
    \end{tabular}

   \begin{minipage}[0.1cm]{1\textwidth}
         \vspace{1em}
         \small \textit{Note}: Table \ref{tab:iv_1st_rain} presents the first stage regression of Equation \eqref{eq:iv_rain_1st_stage}, aiming to investigate how the local rain coincidence affects the local protest gathering. We present two measures -- county average precipitation v.s. precipitation of the county population center and two functional forms -- linear or log. Columns (1) - (4) present the results for them respectively and sequentially. Overall results show that rainfall precipitation significantly decreases the number of local protest rallies, especially when the log of average county rainfall is used as the instrument.  
    \end{minipage}
\end{table}
\end{landscape}

\newpage

\begin{landscape}
\begin{table}[]
    \centering
        \caption{Exogeneity Test: Rain Occurrences on Fundraising Before BLM movements}
    \label{tab:iv_exogeneity_rain}
    
    \begin{tabular}{lcccc}
    \hline\hline
     & \multicolumn{4}{c}{$y$ = Num. Daily Donation}\\
    \cline{2-5}
      Treatment & log Rain  & Rain  & log Rain & Rain \\
     &  (County Average) & (County Average) & (Population Center) & (Population Center)\\
    \cline{2-5}
         &  (1) &  (2) & (3) & (4)\\
         \cline{2-5}
        $Black_j \times $ Treatment    &  .071 &  -.0013 & .044 &  .00035 \\
             &  ( .060) &  (.0085) & (.0477) & (.0063)\\
         Treatment    &  .0062 &  .0035 & .0169 & .0030 \\
             &  (.022) &  (.0033) & ( .0186) & (.0025)\\
             \hline
 
        $\theta_c$ &  Yes &  Yes & Yes & Yes\\
        $\mu_t$ &  Yes &  Yes & Yes & Yes\\
         &   &   &  & \\
       $F$-statistics &  0.88 &   0.60 & 1.13 & 0.747\\
        Obs &  273,380 &  273,380 & 273,380 & 273,380\\
        $R^2$ &  0.0033 &  0.0033 & 0.0033 & 0.0033\\
       
\hline\hline
\multicolumn{4}{l}{\footnotesize Clustered errors in brackets}\\
\multicolumn{4}{l}{\footnotesize * \(p<0.10\), ** \(p<0.05\), *** \(p<0.01\)}\\
    \end{tabular}

    \begin{minipage}[0.1cm]{1\textwidth}
         \vspace{1em}
         \small \textit{Note}: Table \ref{tab:iv_exogeneity_rain} presents the results for examining whether rainfall occurrences have any impact on the fundraising gap prior to the event of the BLM movement. We present two measures -- county average precipitation v.s. precipitation of the county population center and two functional forms -- linear or log. Columns (1) - (4) present the results for them respectively and sequentially. Overall results show that rainfall precipitation has an insignificant effect on fundraising outcomes prior to May 28th.  
    \end{minipage}
\end{table}
\end{landscape}

\newpage

\begin{table}
    \centering
        \caption{IV Regression: The Causal Impact of Local Protest Gathering on Fundraising Gap}
    \label{tab:iv_main_rain}
    \scalebox{0.9}{
    
    \begin{tabular}{lccc}
    \hline\hline
     & \multicolumn{3}{c}{$y$ = Num. Daily Donation} \\
    \cline{2-4}
     &  IV - log Rain (County Average)  & Multiple IVs & OLS \\
    \cline{2-4}
         &  (1) &  (2) & (3) \\
       \cline{2-4}
        $ Black_j\times protest_{ct}$   &  0.39 &  0.87*** &  0.41**\\
           &  ( 0.52) &  ( 0.42) & (0.19) \\
         $protest_{ct}$    & 0.19 &  0.10 & 0.34***\\
             &  (0.20) &  (0.17) & (0.06) \\
             \hline

        $\theta_c$ &  Yes &  Yes & Yes \\
        $\mu_t$ &  Yes &  Yes & Yes \\
            &   &   &  \\
        Observation &  427917 &  427917 & 427917 \\
       Cragg-Donald Wald $F$-statistic &  603.724 &   199.219 & - \\
      
\hline\hline
\multicolumn{4}{l}{\footnotesize Clustered errors in brackets}\\
\multicolumn{4}{l}{\footnotesize * \(p<0.10\), * \(p<0.05\), * \(p<0.01\)}\\
    \end{tabular}

    }
    \begin{minipage}[0.1cm]{1\textwidth}
         \vspace{1em}
         \small \textit{Note}: Table \ref{tab:iv_main_rain} presents the rainfall IV estimates of Equation \eqref{eq:iv_rain_main} for measuring the causal effect of local protest gathering on the fundraising gap. Column (1) presents the single IV estimate from using log average county rainfall. Column (2) presents the multiple IV estimate that pools all four rainfall instruments in the same regression from Table \ref{tab:iv_1st_rain}. Column (3) presents the results from OLS estimates, in directing to Table \ref{tab:reg_localprotest_OLS}. 
    \end{minipage}
\end{table}

\newpage

\begin{landscape}

\begin{table}[]
    \centering
     \caption{The Role of Social Media: Weekend IV Estimates, Spillover Multiplier, and Global Protest Relevance By Intensity of Social Media}
    \label{tab:iv_socialmedia}
    \begin{tabular}{lccccccccccc}
    \hline\hline
   Social Media & \multicolumn{2}{c}{Facebook Connectivity} & & \multicolumn{2}{c}{Google Index} & & \multicolumn{2}{c}{Newspaper Coverage} & &  \multicolumn{2}{c}{Internet \& Digital Access.}\\
    \cline{2-12}
          & (1) & (2) &  & (3) & (4) & & (5) & (6) & & (7) & (8)\\
          \cline{2-12}
        & Low & High &  & Low & High & & Low & High & & Low & High\\
          \cline{2-3} \cline{5-6} \cline{8-9} \cline{11-12}
           Weekend IV Estimates      & .043 & .597*** &  & .092 & .475*** & &  .0916 & .562*** & &  .346* & .298\\
           & (.177) & (.222) &  & (.183) & (.223) & & (.213) & (.189)  & & (.198) & (.20)\\
           [1em]
         Spillover Multiplier  & 1.5 & 21.3 &  & 3.2 & 16.9 & & 3.26 & 20 & & 12.3 & 10.6\\
         [1em]
         Global Protest Relevance  & 33\% & 95\% &  & 68\% & 95\% & & 70\% & 95\% & & 91\% & 90.5\% \\
         [1em]
         \hline
         Observations  & 419,588 & 287,356 &  &  419,389 & 285,784 & &  376,745 &  328,734 & & 389,303 &  317,641\\
         
           \hline\hline
           \multicolumn{4}{l}{\footnotesize Clustered errors in brackets}\\
\multicolumn{4}{l}{\footnotesize * \(p<0.10\), * \(p<0.05\), * \(p<0.01\)}\\
    \end{tabular}
    \begin{minipage}[0.1cm]{1\textwidth}
         \vspace{1em}
         \small \textit{Note}: Table \ref{tab:iv_socialmedia} examines the role of social media in shaping the effect of global protest. We present the weekend IV estimates by the intensity of exposure to the social media reports on BLM themes. We examine three measures of social media: Facebook exposure, Google Search Index, and Newspaper Coverage. In addition, we also examine whether the internet\&digital infrastructure matter (without interaction with the BLM theme). We also present the corresponding spillover multiplier and the relevance of the global protest.  
    \end{minipage}
\end{table}
  
\end{landscape}
\appendix
\setcounter{page}{0}
\begin{center}
    \Large For Online Publication
\end{center}
 \section{Online Appendix}\label{sec:online_app_1}

\subsection{Robustness check}\label{app_robustness}
\textbf{Restricting to Projects Started Before BLM}

First, to address the concern that Black fundraisers might strategically react to the surge of the BLM movement, we restrict our sample to projects that were started from May 1 to May 25.  As we have discussed, the DID estimated effect based on these projects is purely owing to the changes in donors' behaviors during the treatment period. 

Table \ref{tab:reg_mainDID_robust_MayLaunch} reports the DID estimates of the surge of BLM's effect based on the restricted sample. Comparing column (1) and column (2) with Table \ref{tab:reg_mainDID}, one can tell the estimated effect from the restricted sample is still positive and economically significant. We lose the statistical significance owing to the reduction of data points. Besides, we also report the estimates after controlling the project fixed effect to absorb all the possible endogeneity from the beneficiaries' side. After controlling the project fixed effect, we only identify the treatment effect from projects that receive money prior to and during the treatment period. Column (4) in Table \ref{tab:reg_mainDID} and \ref{tab:reg_mainDID_robust_MayLaunch} report the results, and in our main analysis, we still get a significantly positive effect from the BLM's surge. These robust checks reinforce the argument that the effect that BLM brings to the Black people's projects is due to the BLM's effect on the donors' behaviors.

\textbf{RDiT Estimation}

The second robustness check is to use the RDiT estimation to estimate the surge of BLM's local causal effect within each racial group. Given the unexpected and unplanned nature of George Floyd’s killing, the RDiT provides an estimate that is comparable to RCT treatment effects. 

In this study, time is used as the ``running variable," measured as the number of days before (represented by negative values) and after (represented by positive values) the widespread protests that followed Floyd's killing. Figure \ref{fig:reg_RDiT_dailynumdonation} presents the RDiT estimation results with different bandwidths. Here we also control for the project fixed effect to address the possible endogeneity issue from the project-creation side. 

As one can observe, with a relatively short bandwidth in time, even after control for the project fixed effect, we still have enough power to identify the effect of the surge of BLM.\footnote{This can be attributed to the fact that the BLM's effect does not stay significant for more than 5 weeks according to the event study. Thus, the effect size stays significant while we keep the bandwidth within the first 5 weeks.} Our RDiT estimates shows that the surge of BLM has a significant positive effect on the number of donations to Black people's projects. This estimated effect stays robust with the reduction of the bandwidth. The estimated effect suggests that the surge of BLM causes Black people's projects to receive more donations right after the start of the spread of the movement. Besides, we also report the estimates for non-Black people. As is clear, while Black people experience more donation volume, the donations to non-Black people do not shrink. Namely, the premium that Black people undergo is not in the cost of the cutback for non-Black people.

\subsection{Testfying Other Confounding Events: Empirical Details}\label{app_assumptiontest}

\textbf{Confounder 1: Pandemic}

The first example of the threat is the pandemic per se. We observe the leap of the new Covid-19 cases in mid-March and mid-June, 2020 in the US. Previous research showed that Black people were more susceptible to Covid-19 and their health status is more adversely affected, leading to a greater demand for medical services (\cite{chowkwanyun_racial_2020,webb_hooper_covid-19_2020}). 

The pandemic may then contaminate our estimation of the surge of the BLM movement's effect if it disproportionately affects Black people as a ``sudden" shock around the start point of the explosion of the Floyd protests. In particular, a simultaneous issue may happen in which large gatherings due to BLM protests might cause an outburst of Covid-19 cases among Black people. Therefore, social movements themselves did not change donation outcomes, but the following surge in Covid cases did.

To address these concerns, we begin by investigating whether the pandemic had a sudden impact on Black individuals, similar to how the surge of the BLM movement took effect. We examine whether there were any significant changes in the infection or death rates across different races around the beginning of our treatment period. If we find that the change in the infection or death rate was smooth and not sudden, then it is reasonable to assume that the pandemic did not affect Black people as a sudden shock. Additionally, we use a regression discontinuity design to test for any discontinuous change in the Covid infection ratio at the start of our treatment period. If we find no evidence of such a discontinuity or sudden change, it is unlikely that the pandemic caused swift surges in any mediator that contributed to the increase in donations to Black causes. 

Shown in Figure \ref{fig:covid_black_other_ratio}, the Covid newly-infected (or death) ratio between Black and non-Black smoothly decreased over time. In particular, the infection ratio or death ratio across races experienced a very smooth change around the start point of our treatment period. Figure \ref{fig:effects_blm_covid_varyingB} reports the test if there is any discontinuous change in the Covid infection ratio around the start point of the treatment period with RDD regressions. There is no such discontinuous change both statistically and economically. 

Overall, there is no discontinuity of the Covid pattern around the neighborhood of the surge of the Floyd protests. Also, our main analysis shows that controlling for the Covid-related project has a negligible effect on the estimation of the BLM's effect. Hence, it is hard to believe that Covid cases would cause swift surges in any mediator that will finally contribute to the upticks in the donation to Black people.  

Besides, we also add a robust check in our main analysis, where we control for the dummy of Covid-related projects for both Black and non-Black people.\footnote{By analyzing the text description of the projects, we can tag the projects asking for money due to the Covid.} If Black people got more donations because they launched over-represented Covid-related projects than non-Black people, then the dummy would absorb this premium.

\textbf{Confounder 2: Stay-at-home orders}

Another crucial threat to our identification comes from the mandatory stay-at-home orders (SAH) across states. U.S. states began implementing these policies in March 2020 and rescinded these orders from late April. Different states implemented and canceled their staying-home order at different times. On average, the state governments implemented SHA around late March 2020, and they canceled SAH around mid-May 2020 (\cite{moreland2020stay-at-home-timing}).

Both the start and the end of the SAH can be threats to our identification assumptions. Black people are overrepresented in positions that require physical presence or direct interaction with customers and have limited access to remote work. Thus, SAH shocks can affect Black people disproportionately (\cite{perry2021SHO-inequality-pandemic,montenovo2022SHO-inequality-determinants}). For example, the rescinding of the SAH, which happened very close to the Floyd protests, might increase the Black donors' budgets particularly. These Black donors might then give more money to Black beneficiaries. Thus, part of our estimated effect of the BLM movement is from the cancellation of the SAH.

To address this possible simultaneous problem, we rely on our daily level data and show there is no abrupt change in the fundraising outcomes right after the announcement/cancellation of SAH. We employ a RDiT strategy but with varying bandwidths of the post-shock period to explore whether there exists an effect of the SAH's enactment or cancellation. For project $j$ at date $t$, and for each bandwidth pick $B$, we have the following regression:
\begin{equation}
    y_{j,t} = \gamma_0 Black_j \times 1(t>t^{p}_s) + \gamma_1 Black_j + X_{j}\Gamma + \tau_{t}+X_{jt}\beta+\varepsilon_{jts}
    \label{eq:sho}
\end{equation}
Here we have $t \in [t^p_{s}-B,t^p_{s}+B]$. $t^{p}_s$ denotes either the enactment of the SAH policies or the cancellation of them for state $s$. $\gamma_0$ represents the ``effect" of SAH on the donation gap between Black people and non-Black people. 

The challenge to excluding the effect of the SAH is the fact that they happen at a time tightly close to the start time of our treatment. Many states just rescinded the SAH a few days before the surge of the Floyd protests. It is then difficult to distinguish the potential SAH effect from the BLM effect. However, by leveraging the daily-level data, we are able to solve this issue based on the bandwidth-indexed regressions above. If the SAH did not take effect on the donations, then we expect the estimated $\hat{\gamma}_0$ to show a pattern that (I) When the bandwidth $B$ does not overlap with the treatment period, we shall observe negligible effect from the SAH shock; (II) When $B$ covers the dates after May 28, 2020, we shall get a significant estimated $\hat{\gamma}_0$. However, the significance here is not from the SAH's effect but from the BLM's effect.

Figure \ref{fig:reg_robust_SAH} plots the relationship between our estimates $\hat{\gamma}_0$ and our choice of bandwidth $B$ from Equation \eqref{eq:sho}.  Panel (a) shows the effect of SAH's enactment and panel (b) shows the effect of SAH's cancellation. 

The estimated $\hat{\gamma}_0$ shows the pattern which indicates that the SAH did not take effect on the donations. (I) When the bandwidth $B$ is small and does not overlap with the treatment period, the ``effect" of SAH enactment/cancellation policies is statistically indistinguishable from zero. It suggests that the SAH has a trivial effect on the donations; (II) When $B$ covers the dates after May 28, 2020,  $\hat{\gamma}_0$ is significantly positive. With the larger $B$ we include the sample affected by the BLM movements in our analysis and mistakenly attribute the effect of the BLM to the "effect" of SAH enactment/cancellation policies. For the SAH enactment, when $B$ is larger than 62 days (the average duration between SAH enactment and the treatment start),  $\hat{\gamma}_0$ starts to increase and become increasingly significant. For the SAH cancellation, the $\hat{\gamma}_0$ becomes significant when $B$ is larger than 10 days since the cancellations of SAH happened around May 15th of 2022. Taken together, the enactment/cancellation of the SAH is not likely to take effect on the donations. 

Though SAH is not expected to contaminate our estimation of the causal role of the surge of BLM on the number of daily donations, we also control for it in our main analysis. Column (3) in Table \ref{tab:reg_mainDID} and \ref{tab:reg_mainDID_robust_MayLaunch} control for the interaction between the indicator of Black people and the time dummy of the SAH shocks. To control for the shock of the SAH enactment, we define the time dummy indicating dates before the SAH start; As for the shock of the SAH end, we define the time dummy indicating dates after the SAH end. Also, by controlling for the SAH shock, our estimates does not show significant change.

In general, both pandemic and the SAH are unlikely to be the confounders that break out identification assumptions. The estimated effect from Table \ref{tab:reg_mainDID} and Figure \ref{eq:main_eventstudy} is the causal effect that the surge of BLM has on the daily donation number of Black people's project from May 28 to September 1, 2020.

\subsection{OLS Estimates for Local Rallies and Global Gatherings}\label{sec_app_olsprotest}

The OLS regression to study how the local rallies and global gatherings may affect donors' behaviors is the following:
\begin{equation}\label{eq:reg_localgather_OLS}
\begin{split}
    y_{jct}  = &  \beta_0 Black_j \times protest_{ct} + \beta_1 Black_j \times Protest_{-ct} + \beta_2 Black_j\times\textbf{1}(t\geq t_{May28}) \\
    & + \beta_3 Black_j + X_{jt}\Gamma_0  + \tau_{t}+\varepsilon_{jct}
\end{split} 
\end{equation}

$protest_{ct}$ represents the number of gatherings that occurred in county $c$ at day $t$. $Protest_{-ct}$ is the total number of gatherings with that occurred in the U.S. other than county $c$ at the same time at day $t$. $X_j$ is the observed characteristics of the project \textit{j}, and also other fixed effects (time fixed effects, county fixed effects and the interaction with Black indicator) that could be considered. \footnote{To address the concern that donations might not respond to the BLM gatherings on the same day, we also use the 3-day prior moving average of $protest_{ct}$ and $Protest_{t}$ as alternative measurements of the shocks of local and national gatherings, which returns consistent estimates as our primary specification.}

In this analysis, $\beta_0$ and $\beta_1$ are the coefficients of primary interest, representing the effect of local rallies and global gatherings. Table \ref{tab:reg_localprotest_OLS} reports the estimates of Equation \eqref{eq:reg_localgather_OLS}. Column (1) covers the estimates using projects from across the nation, while columns (2) through (5) report the estimates for projects from counties that witnessed the number of protest rallies ranging from the smallest quantile (\textit{Quantile} 1) to the largest quantile (\textit{Quantile} 2).

The primary conclusion is that local gatherings seem to have minimal impact on projects in counties with few such events. This is particularly true for projects in Quantile 2, which show the largest additional effect in the heterogeneity analysis in subsection \ref{sec:heter_DDD} Equation \eqref{eq:DDD_county}. 

The secondary conclusion is that the BLM movement mainly exerts its influence through its broad global impact. This is seen in projects in \textit{Quantile} 1 and 2, where we note insignificant effects from local gatherings but significant effects from the national BLM intensity. Additionally, projects in \textit{Quantile}  3 and 4 show effects related to local gatherings and negligible effects from the national shock. However, the overall conclusion remains the same, as these projects were initiated in areas with a high frequency of gatherings, and the ``local effect" from these events is essentially the global effect on a national level.\footnote{Note that $Protest_{t}$ essentially captures the effect of sizable gatherings outside county \textit{c}. Consequently, counties in \textit{Quantile} 3 and \textit{Quantile} 4, characterized by a high frequency of gatherings, show a smaller effect. However, these local gatherings essentially reflect the national intensity of the BLM movement, particularly since these counties experienced the majority of the gatherings. Moreover, these local gatherings serve as the national shocks in the counties belonging to \textit{Quantile} 1 and \textit{Quantile} 2. Hence, the local gatherings in \textit{Quantile} 3 and \textit{Quantile} 4 fundamentally function as national shocks.}

\subsection{Construction of Social Media Coverage}\label{sec_app_socialmedia}
In this subsection, we show our procedure to gauge social media exposure intensity to ``BLM" of a county.

\noindent\textbf{Google Search Trends}\\
On a given keyword, Google Search Trends reports cross-DMA (Nielsen's Digital Metropolitan Area) indexes. To translate into a county/zip level number, we get access to the cross-walk data linking each zip code in the U.S. with the DMA code from a third-party researcher.\footnote{\url{https://gist.github.com/clarkenheim/023882f8d77741f4d5347f80d95bc259}} The method to construct the cross-walk data is to calculate the center point of every zip code geo boundary, plot those points on a DMA boundary map, and find the containing DMA of each zip centroid. Currently, this is the only free source that provides zipcode-to-DMA crosswalks. Nielsen's original cross-walk data is restricted to public users since a 2011 court decision found that Nielsen's DMA maps are copyright-protected.\footnote{See \url{https://pub.bna.com/ptcj/0806446Aug29.pdf}}

We query Google Search API on the keyword ``Black Lives Matter" during the period from April to October, 2020. Then using the cross-walk linking DMAs with zip/counties, we construct the county-level Google search trend index on ``Black Lives Matter" through taking an average of DMA-level index while weighting by the population size of the county.

\noindent\textbf{News Coverage}\\
Our data to construct newspaper coverage measure is the NewsLibrary  data  base (\url{newslibrary.com}). We use an automated script to search for all the newspaper coverages that contain at least one of the keywords ``Black Lives Matter" ``Floyd" ``protest" and ``racial justice". We have located 10,937 articles containing  one of our specified keywords since May 25, 2020. We have information regarding the name of the newspaper the article is posted on, the title of the article, the major content, and the location where it was posted. Our measure of newspaper coverages on ``BLM" calculates the total number of articles published  in each state.

\newpage

 \section{Online Appendix: Figures and Tables}\label{sec:online_app_2}

 \setcounter{table}{0}
\setcounter{figure}{0}

\begin{figure}[htb]
    \centering
    \caption{Average amount of donation by launch time}
    \includegraphics{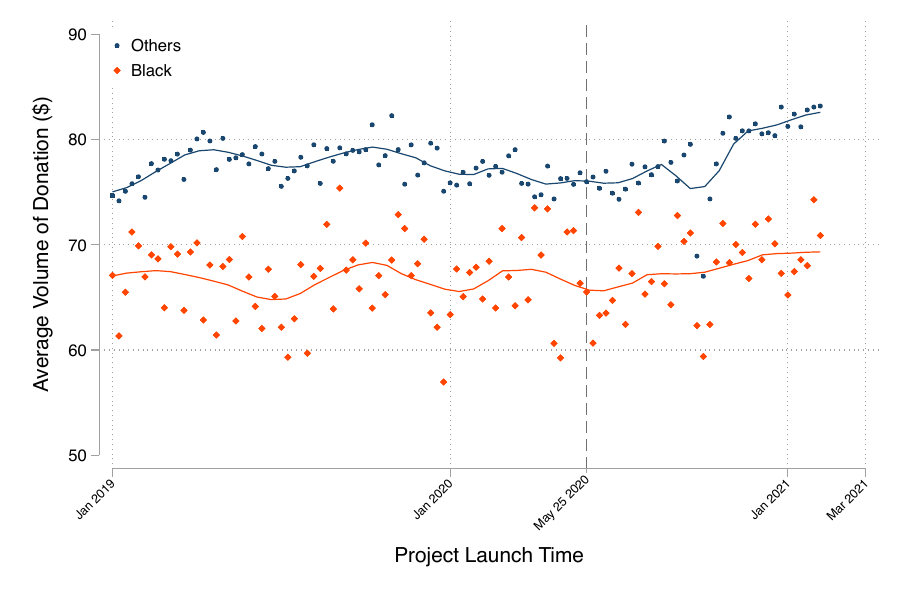}

    \label{fig:avgdonation_t}
    \vspace{1em}
    \begin{minipage}[0.1cm]{1\textwidth}
    \small \textit{Note:} Figure \ref{fig:avgdonation_t} presents the average donation amount for projects launched at various times. The x-axis represents the launch date of the projects, while the y-axis represents the average donation amount. Each data point signifies the mean value of the average donation for projects initiated in a given week. Projects involving Black beneficiaries are denoted by orange dots, and other projects are indicated by navy dots. The curves illustrate the non-parametric fitted lines for each group. The figure shows that both the relative gap between Black and non-Black beneficiaries' projects and the absolute level of the average donation amount remain steady regardless of the projects' launch dates.
    \end{minipage}
\end{figure}

\newpage
\begin{figure}[htb]
    \centering
    \caption{Goal setting by launch time}
    \includegraphics{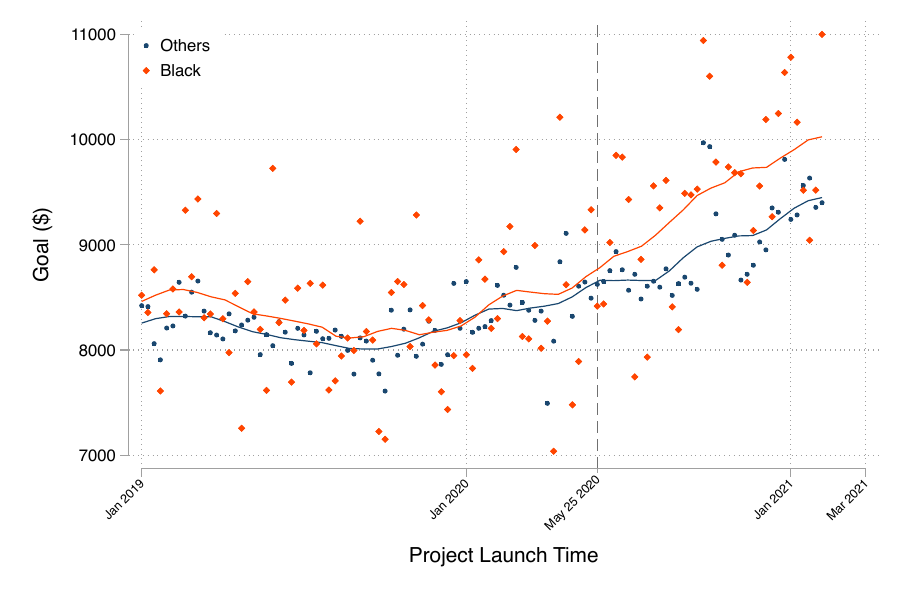}
    
    \label{fig:goal_t}
    
    \begin{minipage}[0.1cm]{1\textwidth}
    \small \textit{Note:} Figure \ref{fig:goal_t} presents the fundraising goals for projects launched at different times. The x-axis signifies the project launch date, while the y-axis represents the fundraising goal. Each data point denotes the average fundraising goal for projects launched in that week. Projects with Black beneficiaries are indicated by orange dots, and the other projects by navy dots. The curves illustrate the non-parametric fitted lines for each group. As can be seen from the figure, the disparity in fundraising goals between projects with Black beneficiaries and others remains relatively unchanged regardless of the projects' launch dates.
\end{minipage}

\end{figure}

\newpage
\begin{figure}[htb]
    \centering
    \caption{Number of Launched Projects by Days}
    \includegraphics[width=\textwidth]{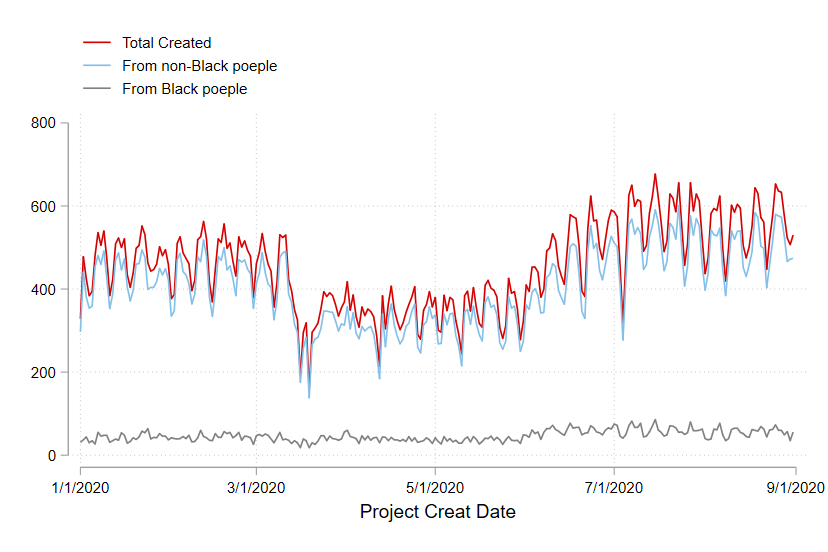}
    \label{fig:num_project_date}
    \vspace{1em}
    \begin{minipage}[0.1cm]{1\textwidth}
    \small \textit{Note:} Figure \ref{fig:num_project_date} illustrates the daily count of new projects created by each racial group over time. The x-axis represents the date, while the y-axis signifies the number of projects launched on that day. The blue curve corresponds to non-Black individuals, the gray curve to Black individuals, and the red curve indicates the total number of projects. It can be observed that the ratio of new projects created by Black individuals to those created by non-Black individuals remains relatively stable over time. Notably, during the surge period of the BLM movement (post late-May 2020), the number of new projects launched by Black individuals does not increase in comparison to those launched by non-Black individuals.
    \end{minipage}
\end{figure}

\newpage
\begin{figure}[htb]
    \centering
    \caption{The Daily Average Amount (\$) of Donation}
    \includegraphics{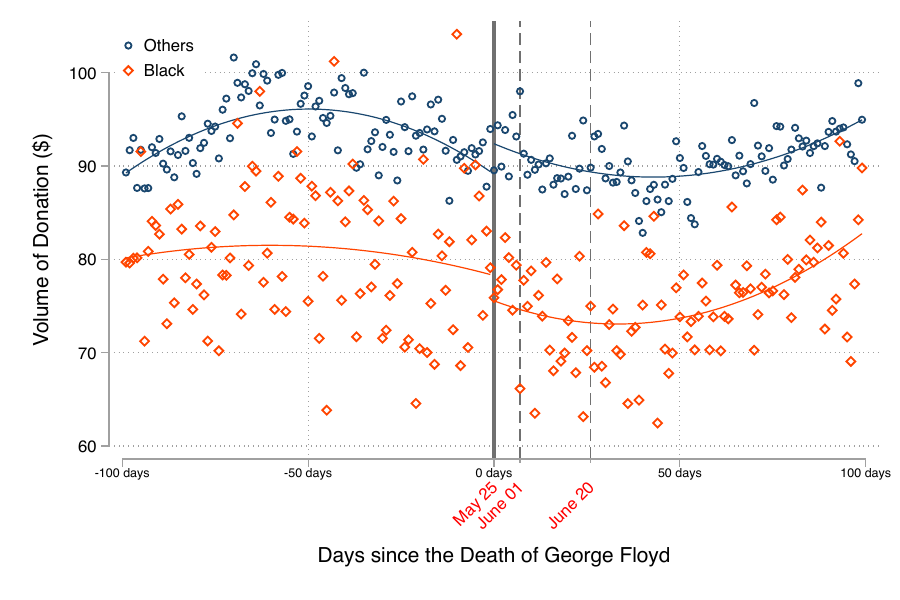}
    \label{fig:donation_volume}
    \vspace{1em}
    \begin{minipage}[0.1cm]{1\textwidth}
    \small \textit{Note:} Figure \ref{fig:donation_volume} illustrates the average donation amount per day, relative to the date of George Floyd's death. The x-axis represents the date relative to this event. Each data point represents the total donation amount received by Black/non-Black individuals on date $t$, divided by the total number of donations given to accounts associated with Black/non-Black individuals on that day. Interestingly, unlike the pattern of donation counts, there's no discernible change in the average donation amount post George Floyd's death. In general, each donor donated less to projects associated with Black beneficiaries as compared to other beneficiaries, and this pattern remained consistent after the event.
    \end{minipage}
    
\end{figure}

\newpage

\begin{figure}[htb]
    \centering
    \caption{RDiT Estimation of the effect of the surge of BLM}
    \includegraphics[width = \textwidth]{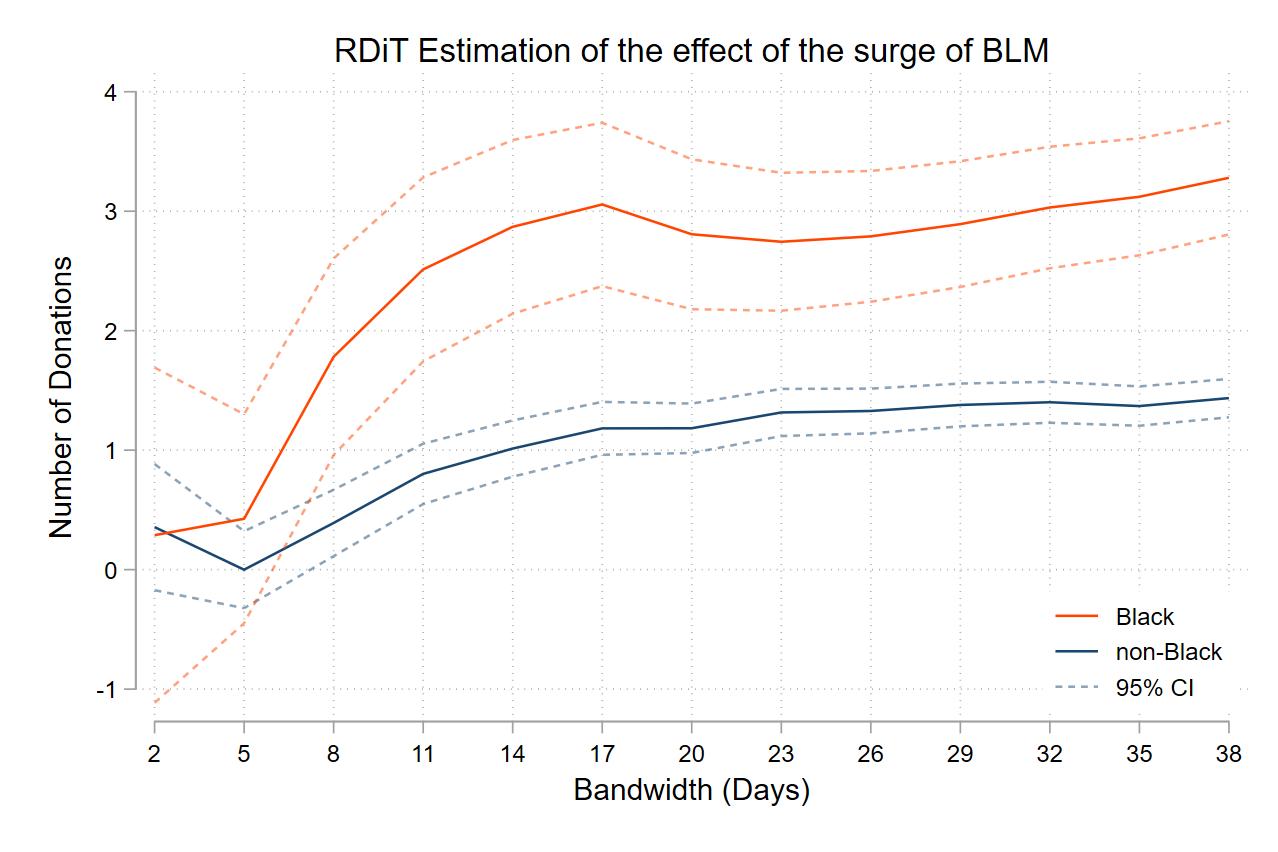}
    \label{fig:reg_RDiT_dailynumdonation}
    \vspace{1em}
    \begin{minipage}[0.1cm]{1\textwidth}
    \small \textit{Note:} Figure \ref{fig:reg_RDiT_dailynumdonation} illustrates the Regression Discontinuity in Time (RDiT) estimates ($\Hat{\alpha_1}$) of the effect of the BLM movement's surge on each racial group's projects over different bandwidths ($B$) around May 28, 2020. We use the regression specification:
    $$
        y_{j,t} = \alpha_0\textbf{1}(t\geq t_{May28}) \times (t-t_{May28})+\alpha_1\textbf{1}(t\geq t_{May28}) + \delta_{t} + X_{jt}\beta + \varepsilon_{j,t},
    $$ where $t \in [t_{May28}-B,t_{May28}+B]$ signifies the days relative to May 28, 2020, with positive values indicating the post-shock period, and $\delta_{j}$ is the project fixed effect. The surge in BLM movement a substantial positive effect on the number of donations to projects associated with Black individuals (the orange curve), and this estimated impact remains robust despite a decrease in the bandwidth. Conversely, the donation volume to non-Black projects (the navy curve) doesn't exhibit any decline.
    \end{minipage}
\end{figure}

\newpage
\begin{figure}[htb]
    \centering
    \caption{Time-series of Covid Infection Cases and Death Cases}
    \includegraphics[scale=0.8]{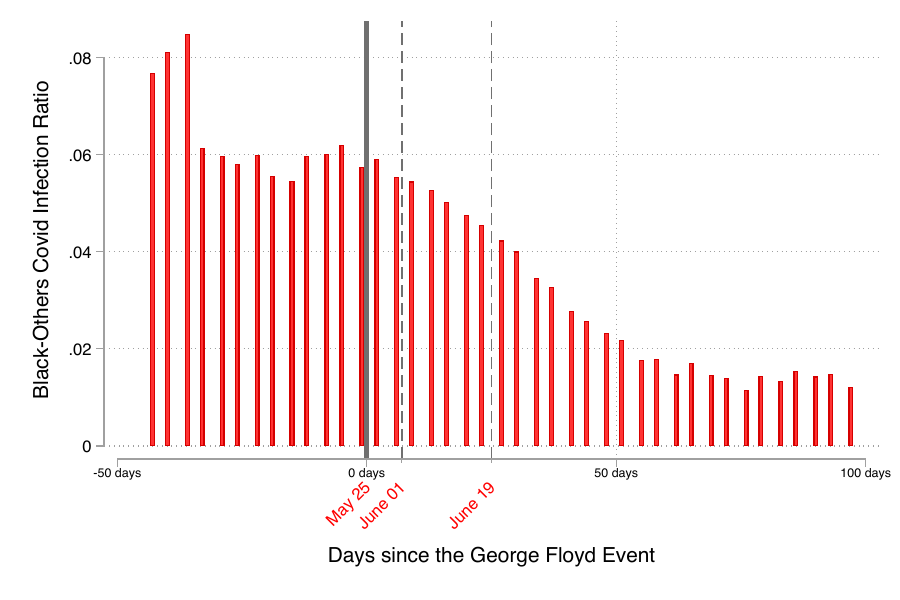}
    \includegraphics[scale=0.8]{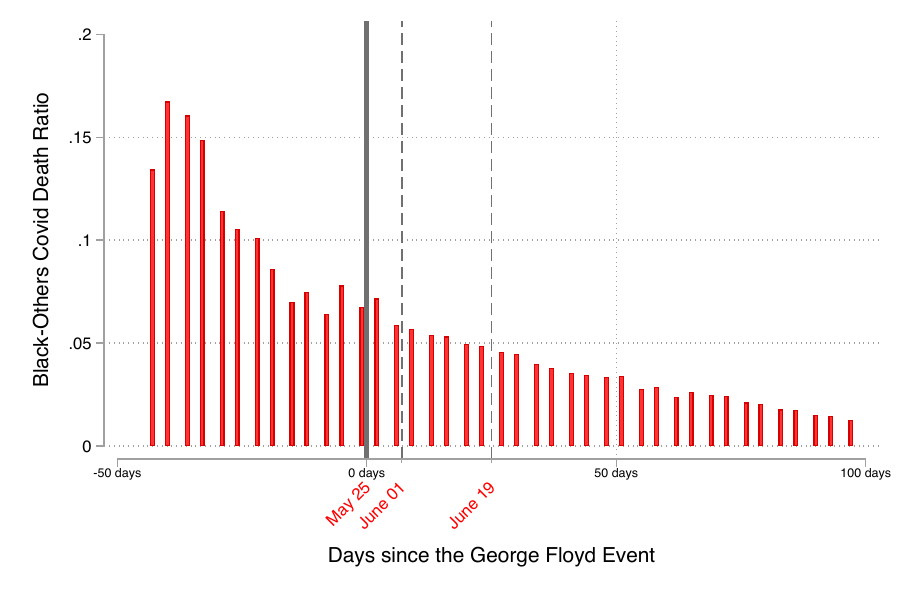}
    
    \label{fig:covid_black_other_ratio}
    \vspace{1em}
    \begin{minipage}[0.1cm]{1\textwidth}
    \small \textit{Note:} Figure \ref{fig:covid_black_other_ratio} presents the proportion of Black individuals among newly infected COVID-19 cases and new COVID-19-related deaths. The x-axis represents the date relative to May 28, 2020. The height of the bars indicates the proportion of Black individuals. Notably, in both panels, we do not see a significant increase or decrease in the share of Black people among new COVID-19 cases or deaths.
    \end{minipage}
\end{figure}

    
    

\newpage
\begin{figure}[htb]
    \centering
    \caption{The Effect of Post George Floyd Event on the Portion of the Black People Infected Covid-19}
    \includegraphics{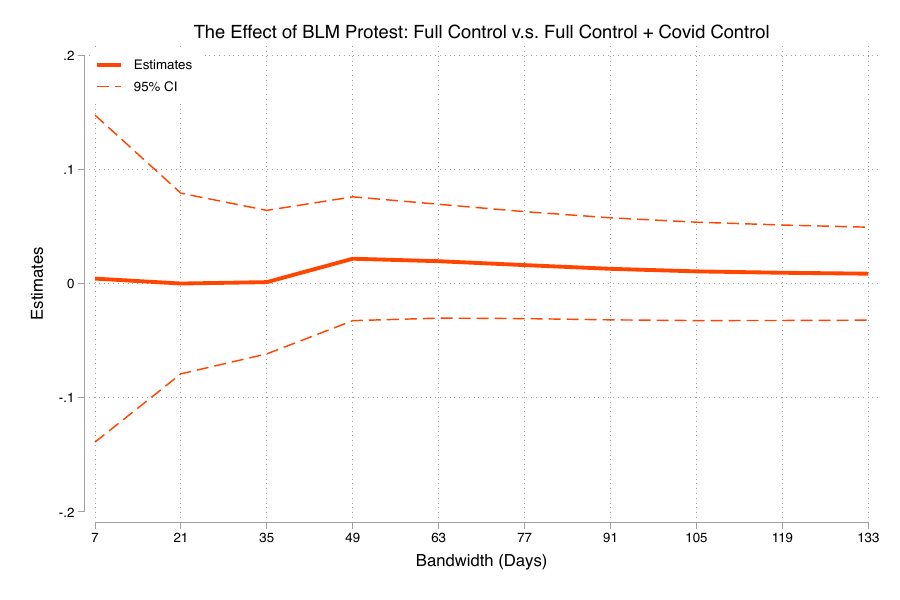}
    
    \label{fig:effects_blm_covid_varyingB}
    \vspace{1em}
    \begin{minipage}[0.1cm]{1\textwidth}
     \small \textit{Note:} Figure \ref{fig:effects_blm_covid_varyingB} presents the Regression Discontinuity in Time (RDiT) estimates ($\Hat{\alpha_1}$) of the impact of being post-George Floyd event on the proportion of Black people among newly infected COVID-19 cases, over various bandwidths ($B$) around May 27, 2020. We utilize the following regression specification:
      $$
      y_{s,t} = \alpha_0 \textbf{1}(t\geq t_{May28}) \times  (t- t_{May28}) + \alpha_1 \textbf{1}(t\geq t_{May28}) + \varepsilon_{j,t}.
      $$
      Here, $y_{s,t}$ denotes the proportion of Black individuals among those infected by COVID-19 in state $s$ on day $t$. $t \in [ t_{May28}-B,t_{May28}+B]$ indicates the days relative to May 28, 2020, with positive values marking the post-shock period. Both statistically and economically, we do not observe a discontinuous change in the number of Black people's COVID-19 cases.
    \end{minipage}
\end{figure}

\newpage

\begin{figure}[htb]
    \centering
    \caption{The "Effect" of Stay-at-Home Orders}
    \includegraphics[width = .8\textwidth]{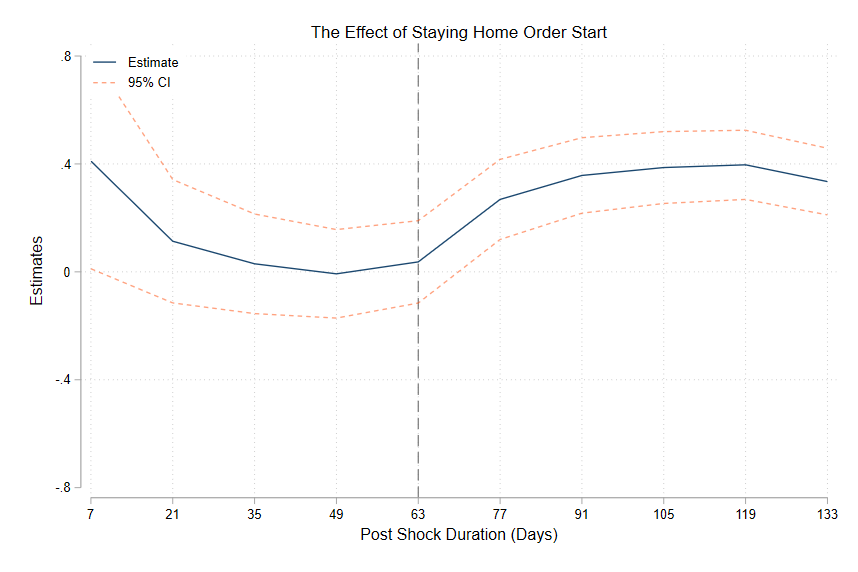}
    \includegraphics[width = .8\textwidth]{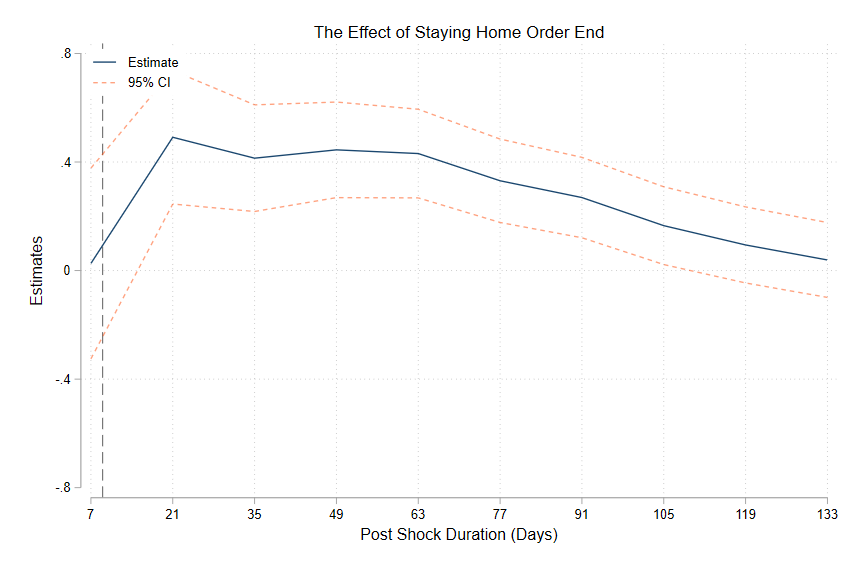}
    \label{fig:reg_robust_SAH}
    \vspace{1em}
    \begin{minipage}[0.1cm]{1\textwidth}
     \small \textit{Note:} Figure \ref{fig:reg_robust_SAH} depicts the estimated effects ($\Hat{\gamma_0}$) of the initiation and cancellation of Stay-at-home Orders over various bandwidths ($B$) as per Equation \eqref{eq:sho}. The bandwidth begins at 7 days (non-overlapping with the BLM period) and progressively extends to coincide with the BLM period. The vertical dashed lines mark the cutoff points when the bandwidths overlap with the BLM period. The pattern in the estimated $\Hat{\gamma_0}$ suggests that the Stay-at-home Orders did not significantly influence donation patterns.
    \end{minipage}
\end{figure}

\newpage

\begin{landscape}
 \begin{figure}
     \centering
     \caption{The Effect of Floyd Movement in History}
     \includegraphics[scale=0.6]{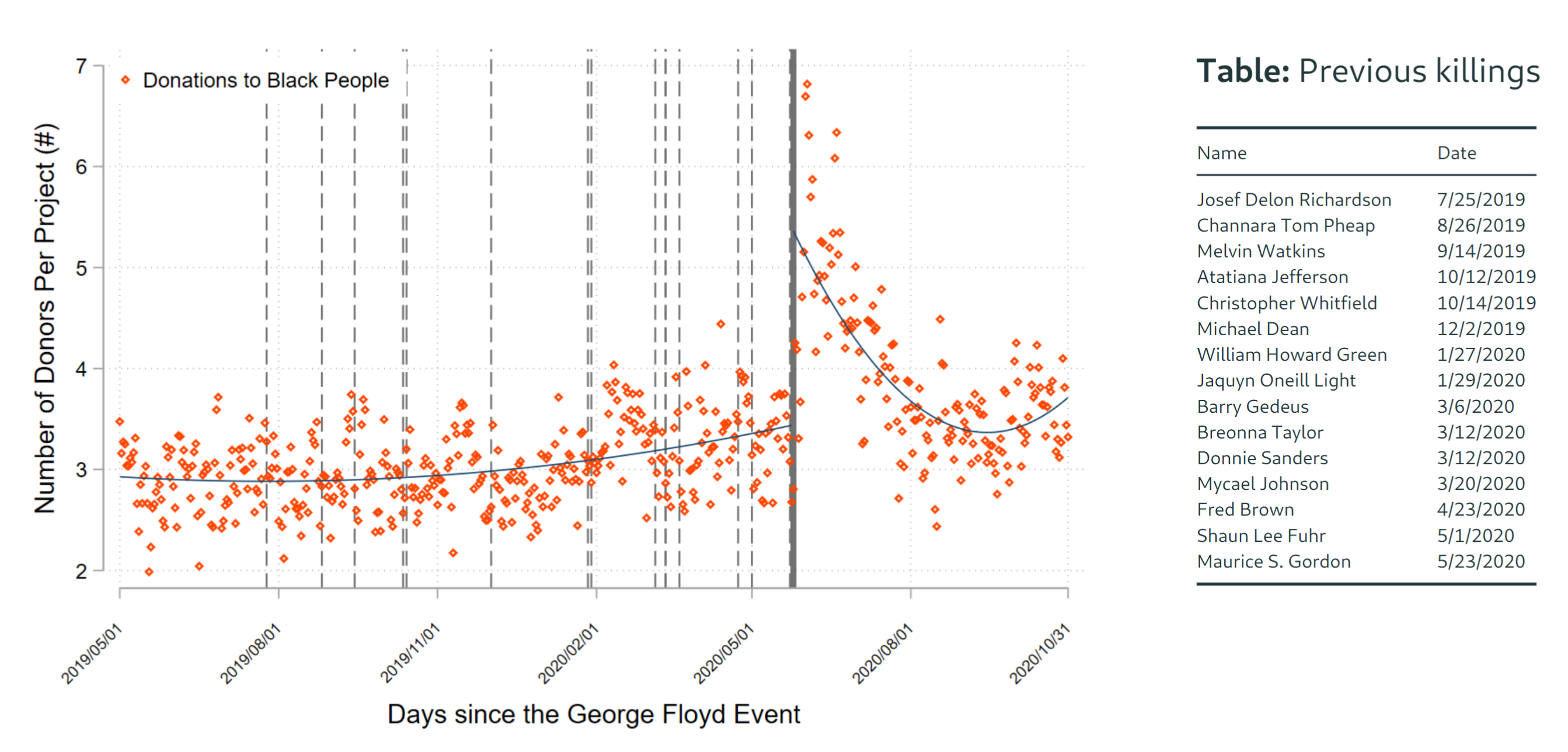}
     \label{fig:previous_killings}
 \end{figure}
   \vspace{1em}
    \begin{minipage}[0.1cm]{1\textwidth}
     \small \textit{Note:} Figure \ref{fig:previous_killings} presents number of daily donations to Black project, tracing back to our earliest data, from May, 2019 to October, 2020. During this period, there are 15 police injustice killings and the corresponding BLM protests. We mark the timing of these events using dashed vertical lines. 
    \end{minipage}
\end{landscape}

\newpage

\begin{landscape}
   \begin{figure}[ht]
       \centering
    \caption{Heterogeneous Effects of BLM By County's Characteristics}
    
    \subfigure[Pre-BLM Inequality]{\includegraphics[width=0.6\textwidth]{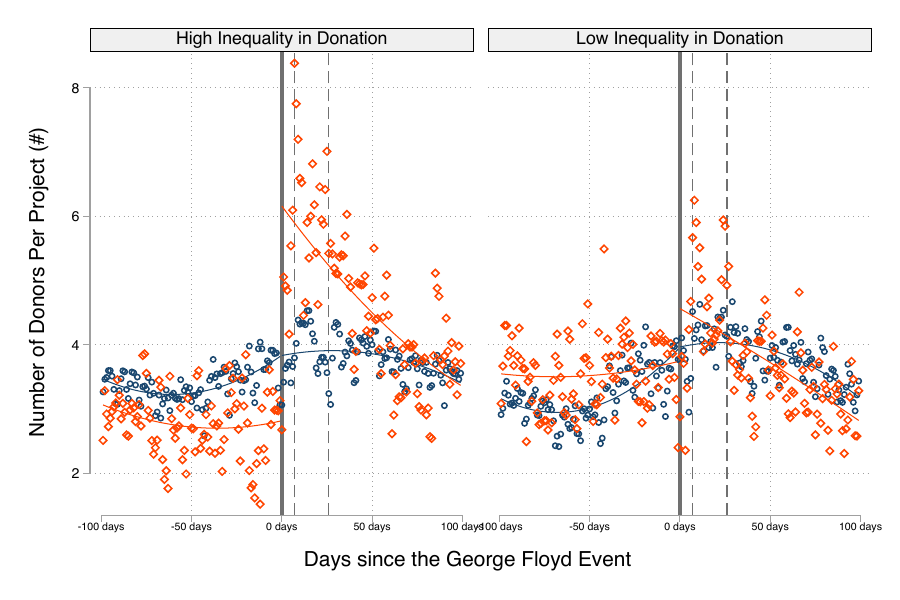}}
    \subfigure[Implicit Attitude Test]{\includegraphics[width=0.6\textwidth]{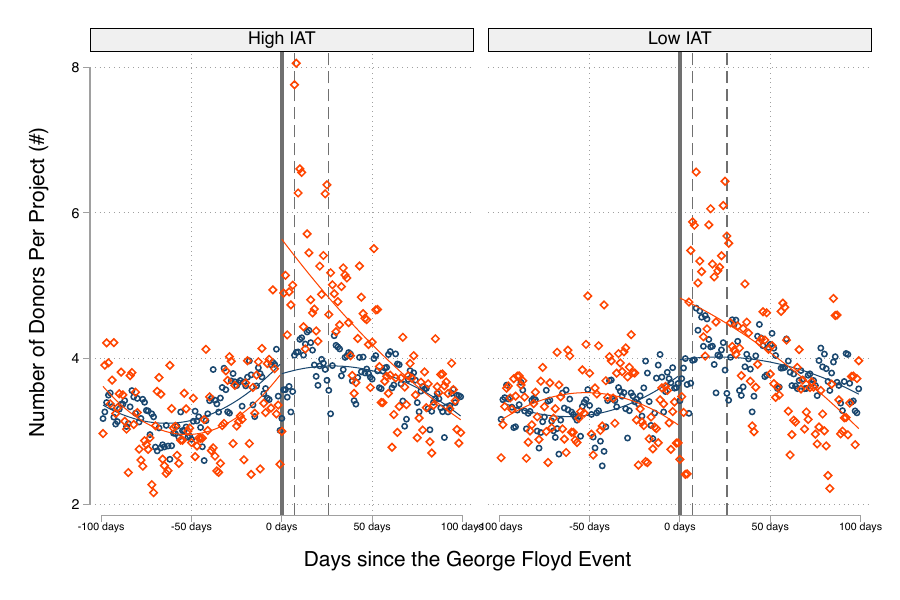}}
    \subfigure[Black Population Ratio]{\includegraphics[width=0.6\textwidth]{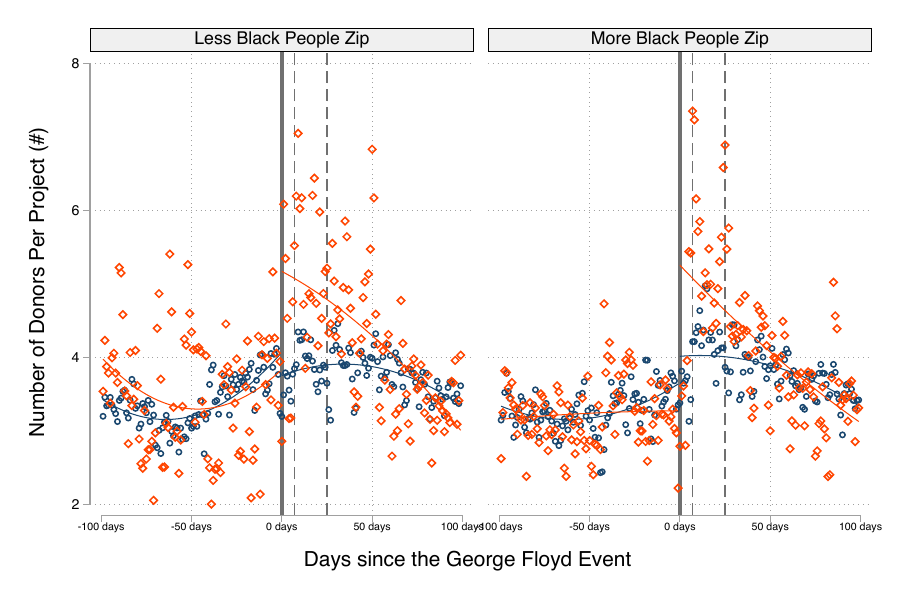}}
    \subfigure[Income]{\includegraphics[width=0.6\textwidth]{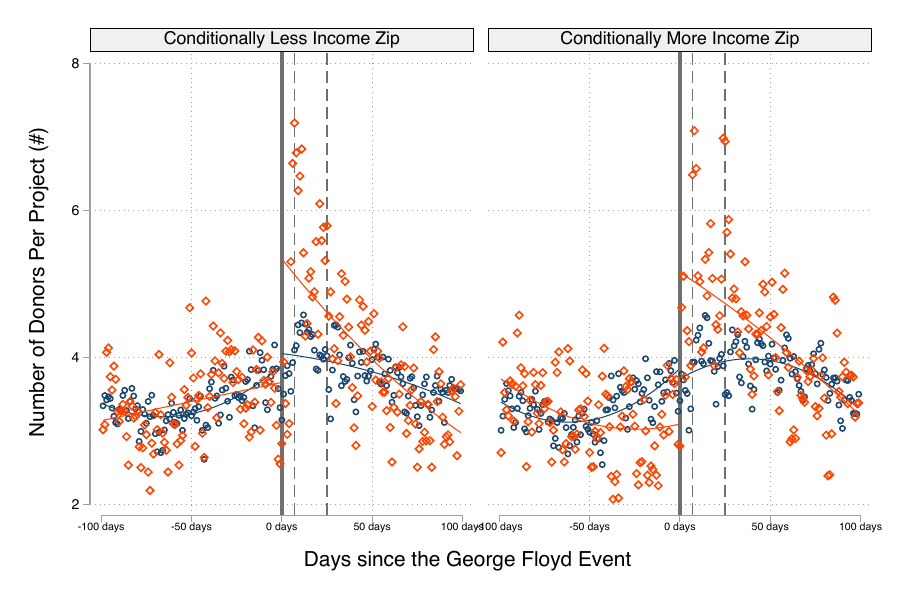}}
    \label{fig:heterogeneity_BLM}
   \end{figure}
\end{landscape}

\begin{landscape}
   \begin{figure}
       \centering
    \caption*{Heterogeneous Effects of BLM By County's Characteristics (Continued)}
     \subfigure[PA vs. Other States]{\includegraphics[width=0.6\textwidth]{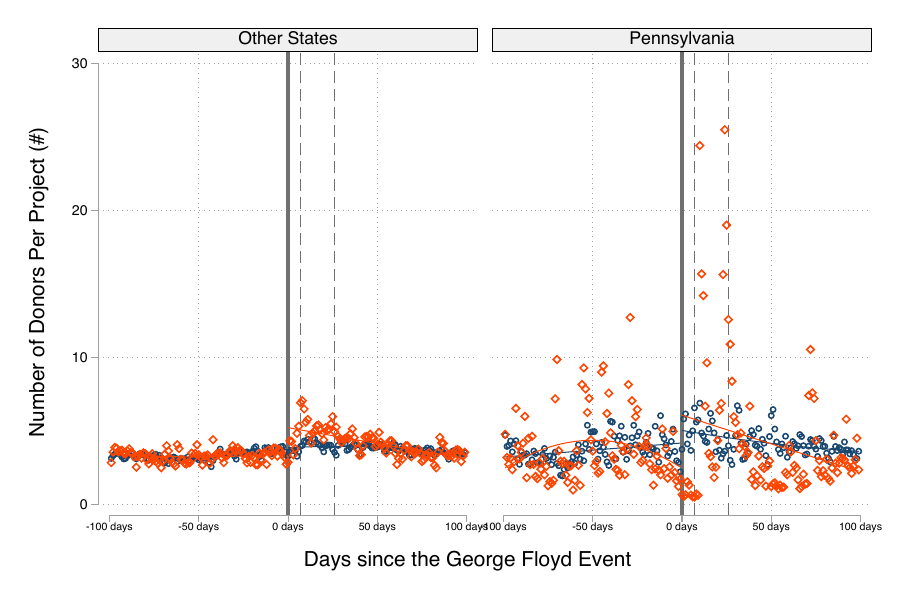}}
    \subfigure[Democratic or Republican]{\includegraphics[width=0.6\textwidth]{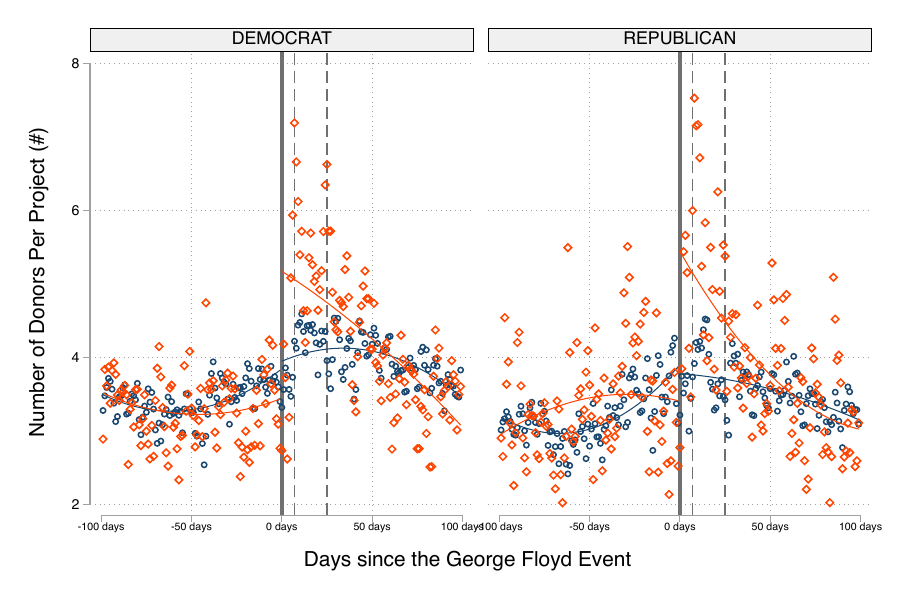}}
    
    \vspace{1em}
    \begin{minipage}[0.1cm]{1\textwidth}
    \small \textit{Note:} In Figure \ref{fig:heterogeneity_BLM}, we divide our sample into counties with higher and lower level by the following six heterogeneities: (a) pre-BLM racial disparity in fundraising, (b) prejudice against black people (measured from 2019's Implicit Attitude Test), (c) the black population ratio, (d) the average income of black
people, (e) PA v.s. other States, and (f) political tendency (2019's voting outcome: Republican or Democratic). The x-axis of this figure is the date relative to May 25, 2020. Each data point represents the average number of donations per active project received on that day. The left panel presents the pattern for counties with a higher level (above the median). The right panel presents the pattern for counties with a lower level(below the median).
    \end{minipage}
   \end{figure}
\end{landscape}
\vspace{1em}

\begin{figure}[htb]
\caption{Geographical Distribution of the Protest Rallies}
    \centering
    \includegraphics[width=0.72\textwidth]{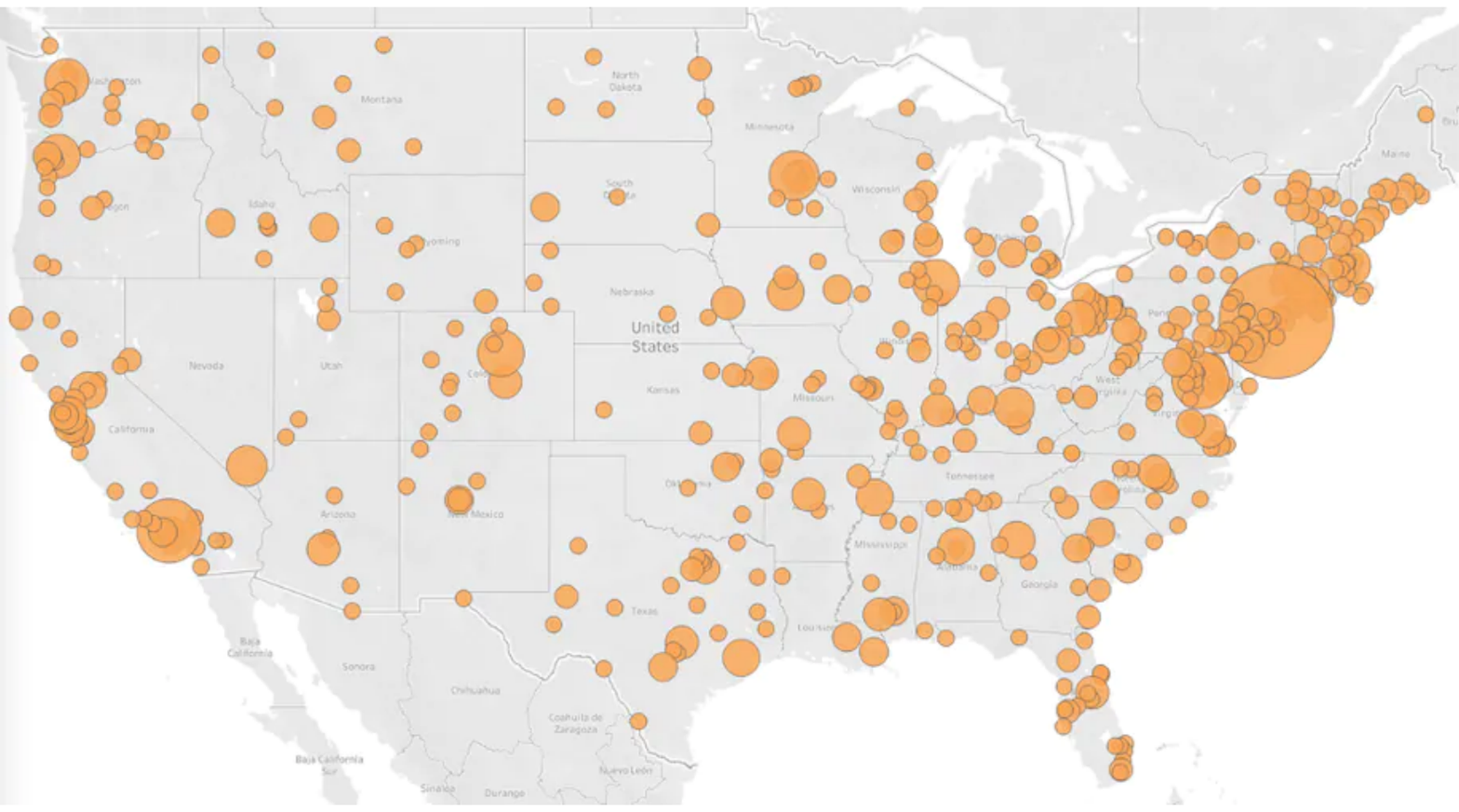}
    \includegraphics[width=0.75\textwidth]{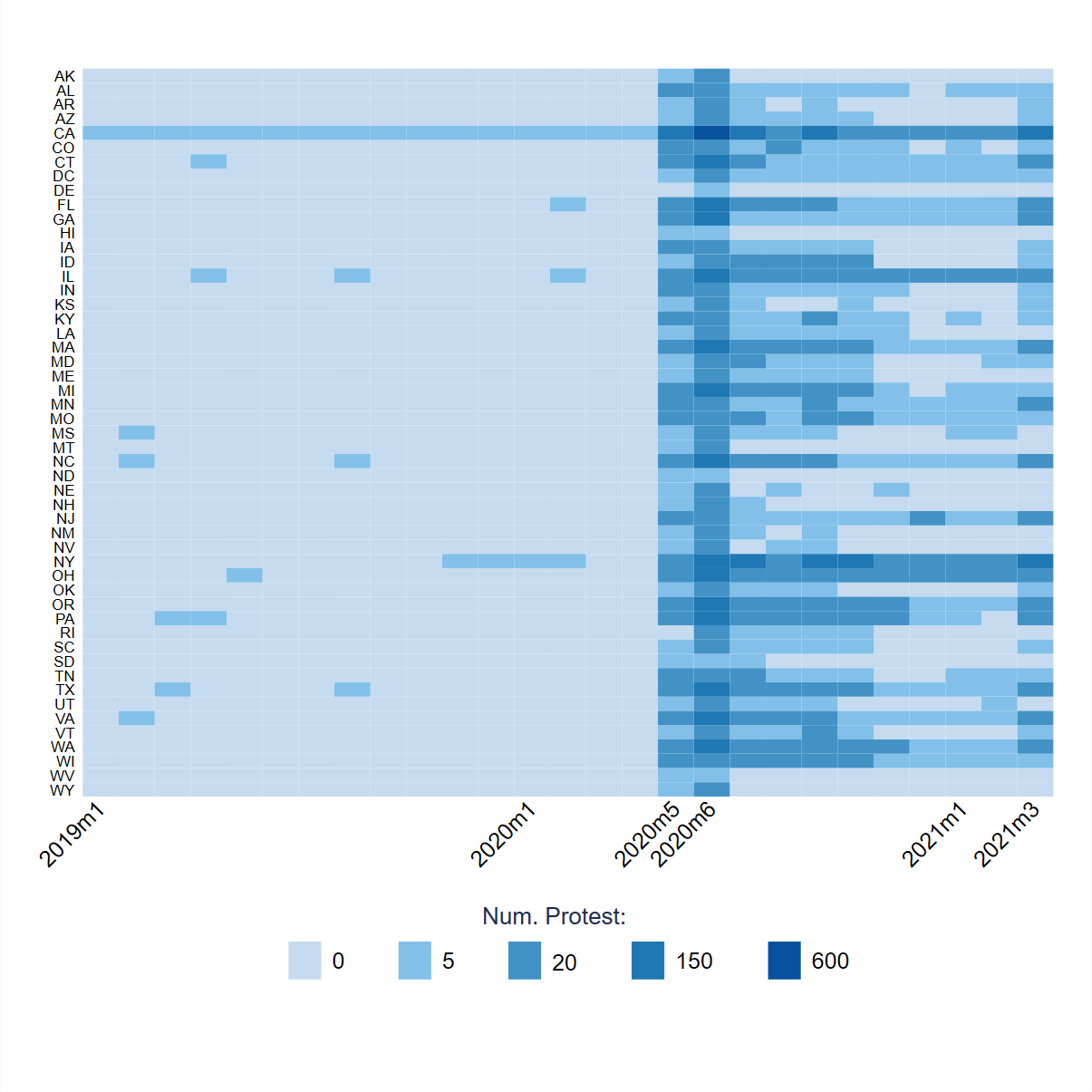}
    \label{fig:geodist_protest}
    \vspace{1em}
    \begin{minipage}[0.1cm]{1\textwidth}
     \small \textit{Note:} Figure \ref{fig:geodist_protest} presents protests in solidarity with George Floyd/in support of Black Lives Matter, May 28-June 4, 2020. Circle size reflects the number of separate events. Data are from Crowdcounting.org; map by Gabriel Perez-Putnam. Figure below presents the panel view of protest intensity by states over time, Jan 2019 - March 2021.
    \end{minipage}
\end{figure}

\newpage

\begin{figure}[htb]
    \centering
    \caption{Projects' distribution across counties with different num. of gatherings}
    \includegraphics[width = \textwidth]{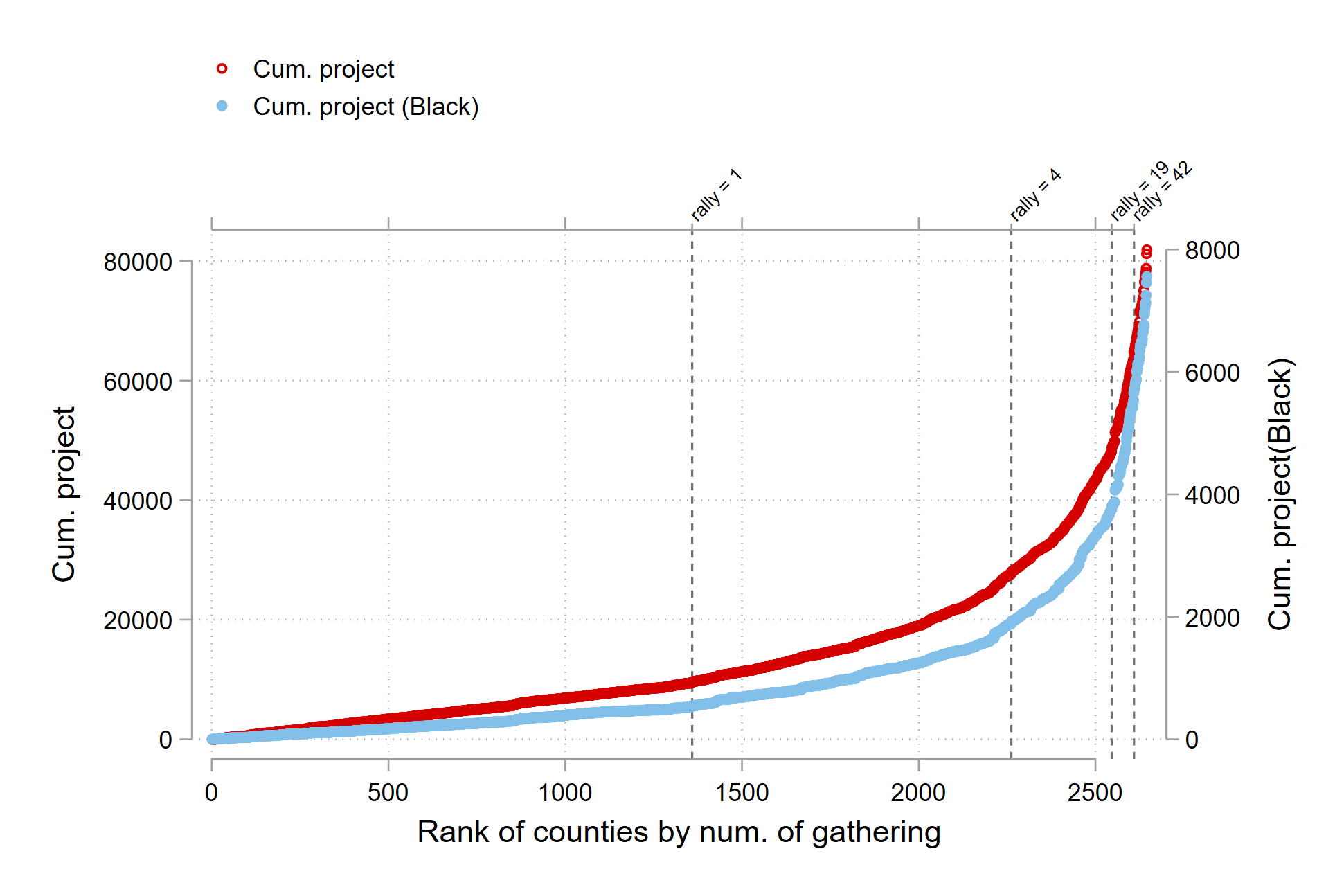}
    \label{fig:protest_cum_project}
    \vspace{1em}
    \begin{minipage}[0.1cm]{1\textwidth}
     \small \textit{Note:} Figure \ref{fig:protest_cum_project} presents the distribution of projects across counties with varying levels of BLM-related gatherings from May 26 to September 1, 2020. The x-axis represents the ranking of the counties based on the total number of BLM-related gatherings. The
y-axis indicates the cumulative number of projects. The vertical dashed lines depict the number of gatherings at each rank. The red curve corresponds to all projects, while the blue curve represents projects associated with the Black community.
    \end{minipage}
\end{figure}

\newpage
\begin{table}[p]
    \centering
    \caption{The Average Change of Racial Gap in Goal before/after January 2020}
    \begin{tabular}{l*{3}{c}}
\hline\hline

          & (1)        & (2)    &     (3)         \\
\hline

& \multicolumn{3}{c}{\textit{Panel A}:\quad $y$ = Goal (\$)}\\
\cmidrule{2-4}

$Black_j$ &   $742.3^{***}$&   136.1&   73\\
          &  (173)         &  (114)         &  (121)             \\
[1em]
$Black_j \times \textbf{1}(t>Jan. 2020)$&    $1257^{***}$&    321.0&    339.0 \\
          &  (256)         &  (166.8)         &  (172.9)             \\
\hline
& \multicolumn{3}{c}{\textit{Panel B}:\quad  $y$ = ln Goal}\\
\cmidrule{2-4}
$Black_j$    &  $ 0.0335^{***}$&   $3.22e-15^{***}$&   $-2e-15^{***}$\\
          & (0.01)         & (5.4e-16)         & (5.2e-16)         \\
[1em]
$Black_j \times \textbf{1}(t>Jan. 2020)$&    $0.072^{***}$&    -2.7e-16&    2.3e-16\\
          &   (0.014)         &   (7.0e-16)         &   (6.6e-16)         \\
\hline
Description        &              &      Yes         &      Yes         \\

State FE \& Date FE         &              &      Yes         &      Yes         \\

Zip FE            &               &               &      Yes         \\      \\
[1em]
Observations     &   333322         &   332496        &   327088        \\
\hline\hline
\multicolumn{4}{l}{\footnotesize \textit{t} statistics in brackets}\\
\multicolumn{4}{l}{\footnotesize $^{*} (p<0.05)$, $^{**} p<0.01$, $^{***} p<0.001$}\\
\end{tabular}
    \label{tab3}
    \vspace{1em}
        \begin{minipage}[0.1cm]{1\textwidth}
         \small \textit{Note:} Table \ref{tab3} reports the estimation results of the following diff-in-diff regression Equation 
       \begin{equation*}
    y_{j,t} = \beta_0 + \beta_1\textbf{1}(t>Jan. 2020)\times Black_{j} + \beta_2Black_j + X_{j}\Gamma + \delta_{t} + \sigma_{s} + \varepsilon_{j,t}
\end{equation*}
Here $y_{j,t}$ is the goal of project $j$ launching at date $t$. $\textbf{1}(t>Jan. 2020) = 1$ if the launching date $t$ is later than January, 2020. $Black_j = 1$ if project $j$ contains any black beneficiaries. $X_j$ is the control variables, including project $j$'s goal setting and features of its text description. $\delta_t$ is the month fixed effect. $\sigma_s$ is the state fixed effect of the beneficiaries' geographical location. \\

         Panel A reports the results when the outcome is the absolute value of the projects' goal and Panel B reports the estimates using the log value of the projects' goal. Column (1) presents the results without any controls; column (2) presents the results after controlling for goal, features of text description, state fixed effect and date fixed effect; column (3) in addition controls the zip code where the beneficiaries live. We learn that though there is statistical significant gap between black and non-black beneficiaries in goal setting but the size of the gap is negligible in economic view. The initial gap is less than 0.1\% after adding control variables. Also, this gap does not decrease for projects starting after BLM.
    \end{minipage}
\end{table}

\newpage
\begin{table}[p]
    \centering
    \caption{The Average Reduction of Racial Gap in Total Number of Donors and Average Amount of Donations before/after January 2020}
    \begin{tabular}{l*{3}{c}}
\hline\hline
          & (1)        & (2)    &     (3)         \\
\hline
& \multicolumn{3}{c}{\textit{Panel A}:\quad $y$ = Total Number of Donors}\\
\cmidrule{2-4}
$Black_j$   &   $-0.092^{***}$&   $-0.087^{***}$&   $-0.099^{***}$\\
          &  (-8.68)         &  (-8.80)         &  (-9.72)         \\
[1em]
$Black_j \times \textbf{1}(t>Jan. 2020)$&    $0.129^{***}$&    $0.103^{***}$&    $0.103^{***}$\\
          &   (8.81)         &   (7.72)         &   (7.59)         \\

\hline
& \multicolumn{3}{c}{\textit{Panel B}:\quad $y$ = Average Amount of Donation}\\
\cmidrule{2-4}

$Black_j$   &  $-11.324^{***}$&  $-12.164^{***}$&  $-10.640^{***}$\\
          & (-27.62)         & (-30.72)         & (-25.35)         \\
[1em]
$Black_j \times \textbf{1}(t>Jan. 2020)$&    $0.406$         &    $0.111$         &    $0.063$         \\
          &   (0.73)         &   (0.21)         &   (0.11)         \\
[1em]
\hline

Description \& Goal      &              &      Yes         &      Yes         \\

State FE \& Date FE         &              &      Yes         &      Yes         \\

Zip FE            &               &               &      Yes         \\      \\
[1em] 
Observations    &   339092         &   338245         &   332843         \\
\hline\hline
\multicolumn{4}{l}{\footnotesize \textit{t} statistics in brackets}\\
\multicolumn{4}{l}{\footnotesize $^{*}$ $p<0.05$, $^{**}$ $p<0.01$, $^{***}$ $p<0.001$}\\
\end{tabular}
    \label{tab4}
    \vspace{1em}
        \begin{minipage}[0.1cm]{1\textwidth}
         \small \textit{Note:}  Table \ref{tab4} reports the estimation results of the following diff-in-diff regression Equation 
       \begin{equation*}
    y_{j,t} = \beta_0 + \beta_1\textbf{1}(t>Jan. 2020)\times Black_{j} + \beta_2Black_j + X_{j}\Gamma + \delta_{t} + \sigma_{s} + \varepsilon_{j,t}
\end{equation*}
Here $y_{j,t}$ is either the total number of donors or the average amount of donation received by project $j$ launching at date $t$. $\textbf{1}(t>Jan. 2020) = 1$ if the launching date $t$ is later than January, 2020. $Black_j = 1$ if project $j$ contains any black beneficiaries. $X_j$ is the control variables, including project $j$'s goal setting and features of its text description. $\delta_t$ is the month fixed effect. $\sigma_s$ is the state fixed effect of the beneficiaries' geographical location. \\

          Panel A reports the results when outcomes is total number of donors and Panel B reports estimates for average amount of donation. Column (1) presents the results without any controls; column (2) presents the results after controlling for goal, features of text description, state fixed effect and date fixed effect; column (3) in addition controls the zip code where the beneficiaries live. 
    \end{minipage}
\end{table}

\begin{landscape}
\newpage
\begin{table}[p]
    \centering
    \caption{Observed Characteristics of Profiles before and during Floyd Protests}
    \begin{tabular}{@{}lcccc@{}}
\toprule
 $y$ & $ln(\text{Length of Text Des.})$ & $\textbf{1}(\text{Topic: Cancer})$ & $\textbf{1}(\text{Topic: Surgery})$ & $ln(\text{Fund Goal})$ \\ \cline{2-5}
 & (1) & (2) & (3) & (4) \\ \cline{2-5}
$Black_j \times\textbf{1}(t\geq t_{May28})$ & -0.003  & 0.0004  & 0.035   & -0.016  \\
                               & (0.017) & (0.010) & (0.012) & (0.029) \\
$\textbf{1}(t\geq t_{May28})$               & -0.021  & -0.0153 & 0.060   & 0.132   \\
                               & (0.005) & (0.004) & (0.004) & (0.009) \\
$Black_j$                      & -0.043  & -0.054  & -0.039  & 0.101   \\
                               & (0.112) & (0.007) & (0.009) & (0.020) \\ \midrule
State FE \& Date FE         &       Yes        &      Yes         &      Yes      & Yes    \\
[1em]
Observations                            & 110,894 & 70,108  & 70,108  & 110,894 \\ \bottomrule
\multicolumn{4}{l}{\footnotesize \textit{t} statistics in brackets}\\
\multicolumn{4}{l}{\footnotesize $^{*}$ $p<0.05$, $^{**}$ $p<0.01$, $^{***}$ $p<0.001$}\\
\end{tabular}
    \label{tab_observed_charcts}
    \begin{minipage}[0.1cm]{1\textwidth}
         \vspace{1em}
         \small \textit{Note:}  Table \ref{tab_observed_charcts} reports the estimation results of the following diff-in-diffs regression Equation 
         \begin{equation*}
          y_{j,t} = \beta_0 + \beta_1\textbf{1}(t\geq t_{May28})\times Black_{j} + \beta_2Black_j + \sigma_{s} + \varepsilon_{j,t}
          \end{equation*}
        Here $y_{j,t}$ is the observed characteristics of project $j$ launching at date $t$. $\textbf{1}(t\geq t_{May28}) = 1$ if the launching date $t$ is later than May 28, 2020. $Black_j = 1$ if project $j$ contains any Black beneficiaries. $\sigma_s$ is the state fixed effect of the beneficiaries' geographical location. \\
         
        $ln(\text{Length of Text Des.})$ is the log value of the length of the text description; \textbf{1}(Topic: Cancer) and \textbf{1}(Topic: Surgery) are the likelihood that the description is in a topic related to cancer or surgery. We apply an LDA algorithm, which generates these two topics and the related likelihood value for every profile's description in a data-driven style; $ln(\text{Fund Goal})$ is the log value of the goal set of the project. We report the standard deviation in the bracket. From the first row of this table, one can tell that the Black fundraisers do not edit their profiles differently during the surge of the BLM. 
    \end{minipage}
\end{table}
\end{landscape}

\newpage
\begin{table}[p]
    \centering
    \caption{DID estimation of the effect of the surge of BLM \\
    {\small(for projects launched from May 1 to May 24, 2020)}}
     \begin{tabular}{lcccc}
\hline \hline
 & \multicolumn{4}{c}{$y$ = Num. Daily   Donation} \\ \cline{2-5} 
 & (1) & (2) & (3) & (4) \\ \cline{2-5} 
$Black_j$ & 0.060 & 0.053 & -0.155 &  \\
 & (0.30) & (0.27) & (-0.64) &  \\
 &  &  &  &  \\
$Black_j\times   \textbf{1}(t\geq t_{May28})$ & 0.329 & 0.344 & 0.201 & 0.303 \\
 & (1.54) & (1.61) & (0.95) & (0.93) \\
 &  &  &  &  \\
$ln(Goal)_j$ & 0.915*** & 0.917*** & 0.918*** &  \\
 & (24.86) & (24.88) & (24.89) &  \\
 &  &  &  &  \\
 $1(Covid)_j$ &  & 0.431*** & 0.431*** &  \\
 &  & (5.57) & (5.58) &  \\
 &  &  &  &  \\
 \hline
Description & Yes & Yes & Yes &  \\
Date FE & Yes & Yes & Yes &  \\
State FE & Yes & Yes & Yes &  \\
Project launch-date FE & Yes & Yes & Yes &  \\
$Black_j \times SAH_{st}$ &  &  & Yes & Yes \\
Project FE &  &  &  & Yes \\
 &  &  &  &  \\
Control group mean value & 6.597 & 6.597 & 6.597 & 6.597 \\
Num. projects & 5644 & 5644 & 5644 & 5644 \\
Observations & 68167 & 68167 & 68167 & 68167 \\ \hline \hline
t statistics in brackets &  &  &  &  \\
\multicolumn{3}{l}{$^{*} p <   0.10$ $^{**} p < 0.05$  $^{***} p   < 0.01$} & \multicolumn{1}{l}{} & \multicolumn{1}{l}{}
\end{tabular}

     \label{tab:reg_mainDID_robust_MayLaunch}
    \begin{minipage}[0.1cm]{1\textwidth}
    \vspace{1em}
    \small \textit{Note:} Table \ref{tab:reg_mainDID_robust_MayLaunch} shows the estimation results of Equation \eqref{eq:main}, restricting to the projects initiated between May 1 and May 24, 2020.  Columns 1 and 2 report the main specifications, with column 2 controlling for the fixed effects of COVID-19 related projects. Column 3 includes controls for the interaction between the indicator for Black individuals and the state-level Stay-at-Home (SAH) shock time dummy (marking the start and end of SAH orders). Column 4 accounts for project fixed effects.  5,644 projects, which launched from May 1 to May 24, 2020, were actively receiving donations. Prior to the treatment period, projects related to non-Black individuals received on average 6.597 donations per day. 
    \end{minipage}
\end{table}

\end{document}